\documentclass[12pt,hidelinks]{article}
\usepackage[body={17.5cm, 22cm},right=2cm]{geometry}
\usepackage{color}
\usepackage{amssymb,graphicx}
\usepackage{amsmath}
\usepackage{epsfig}
\usepackage[bookmarks=true,pdfborder={0 0 0}]{hyperref} 
\usepackage{cite}

\newcommand{\keV}{{\, {\rm keV}}}

\newcommand{\GeV}{{\, {\rm GeV}}}

\usepackage{float}
\allowdisplaybreaks

\definecolor{myorange}{RGB}{199,146,32}
\definecolor{myred}{RGB}{204,0,0}

\begin{document}
\setcounter{page}{0}
\thispagestyle{empty}

\parskip 3pt

\font\mini=cmr10 at 2pt

\begin{titlepage}
~\vspace{1cm}
\begin{center}

{\LARGE \bf Axions from Strings: the Attractive Solution}

\vspace{0.6cm}

{\large
Marco~Gorghetto$^a$,
Edward~Hardy$^b$,  and Giovanni~Villadoro$^{c}$}
\\
\vspace{.6cm}
{\normalsize { \sl $^{a}$ 
SISSA International School for Advanced Studies and INFN Trieste, \\
Via Bonomea 265, 34136, Trieste, Italy }}

\vspace{.3cm}
{\normalsize { \sl $^{b}$ Department of Mathematical Sciences, University of Liverpool, \\ Liverpool, L69 7ZL, United Kingdom}}

\vspace{.3cm}
{\normalsize { \sl $^{c}$ Abdus Salam International Centre for Theoretical Physics, \\
Strada Costiera 11, 34151, Trieste, Italy}}

\end{center}
\vspace{.8cm}
\begin{abstract}

We study the system of axion strings that forms in the early Universe if the Peccei-Quinn symmetry is restored after inflation. Using numerical simulations, we establish the existence  of an asymptotic solution to which the system is attracted independently of the initial conditions. We study in detail the properties of this solution, including
the average number of strings per Hubble patch, the distribution of loops and long strings, the way that different types of radiation are emitted, and the shape of the spectrum of axions  produced. 
We find clear evidence of logarithmic violations of the scaling properties of the attractor solution.  We also find that, while most of the axions are emitted with momenta of order Hubble, most of the axion energy density is contained in axions with energy of order 
the string core scale, at least in the parameter range available in the simulation. 
While such a spectrum would lead to a negligible number density of relic axions from strings when extrapolated to the physical parameter region,
we show that the presence of small logarithmic corrections 
to the spectrum shape could completely alter such a conclusion.
A detailed understanding of the evolution of the axion spectrum is therefore crucial for a reliable estimate of the relic axion abundance from strings. 
\end{abstract}

\end{titlepage}

\tableofcontents

\section{Introduction} \label{sec:intro}

The QCD axion \cite{Peccei:1977hh,Wilczek:1977pj,Weinberg:1977ma,Kim:1979if,Shifman:1979if,
Zhitnitsky:1980tq,Dine:1981rt}  is the simplest and most robust of the known solutions of the Standard Model (SM) Strong CP problem, and it also  automatically forms a component of cold dark matter (DM) \cite{Preskill:1982cy,Abbott:1982af,Dine:1982ah}. Consequently, a QCD axion that makes up the entire measured DM relic abundance is one of the best motivated scenarios for physics beyond the SM. In addition, numerous  experiments aimed at detecting axions are currently running or in development. These will be sensitive to a substantial proportion of the relevant parameter space, and, if an axion is discovered, they could measure its mass and couplings precisely  (see e.g.~\cite{
Graham:2015ouw,Irastorza:2018dyq}).

The dynamics by which QCD axion DM is produced, and its final relic abundance, depends on the cosmological history of the Universe (see e.g.~\cite{Sikivie:2006ni,Marsh:2015xka}). If the Peccei-Quinn (PQ) symmetry that gives rise to the axion was broken after inflation, the axion field had initially random fluctuations over the present day observable Universe. Instead, if the PQ symmetry was broken during inflation and never subsequently restored, the axion field was initially homogeneous. In this case the axion relic abundance is incalculable because it depends on the local value of the axion field after 
inflation.\footnote{There is only an upper bound on the axion mass for which it can make up the full relic abundance, coming from isocurvature constraints   \cite{Lyth:1989pb,Turner:1990uz,Linde:1991km}.}  

In this paper we study the class of models in which PQ breaking happens after inflation. This includes theories in which inflation happened at a scale above the axion decay constant $f_a$, and also those with inflation at a lower scale but which were reheated to a temperature above $f_a$.\footnote{The precise boundary between the regimes depends on the details of thermalisation during reheating.} In this case, assuming a standard cosmological history,  the relic abundance is calculable in terms of the axion mass due to the random initial conditions.\footnote{As we will discuss later the relation between the relic abundance and the mass of the axion might be affected for extreme choices of the model parameters. }.
As a result, there is in principle a unique calculable prediction for the axion mass if it is to make up the complete DM density in such models, which would be extremely valuable for experimental axion searches.

However, in this scenario the mechanisms by which DM axions are produced are complex, and calculating the relic abundance is challenging   \cite{Sikivie:1982qv,Vilenkin:1982ks,Vilenkin:1984ib,Davis:1985pt,Davis:1986xc}. The random initial axion field after PQ symmetry breaking leads to the formation of axion strings \cite{Kibble:1976sj,Kibble:1980mv,Vilenkin:1981kz}. These are topologically stable field configurations that wind around the $U$(1) vacuum manifold of the broken PQ symmetry as a loop in physical space is travelled.
Interactions between strings are thought to result in the length of string per Hubble volume, measured in Hubble lengths, remaining approximately of ${\cal O}(1)$ as the Universe expands \cite{Albrecht:1984xv,Bennett:1987vf,Allen:1990tv,Vincent:1996rb,Martins:2000cs,Vilenkin:2000jqa}. To maintain such a regime the string network must release energy. This dominantly happens through the production of axions, which form a potentially significant fraction of the total relic abundance  \cite{Vilenkin:1986ku,Davis:1989nj,Dabholkar:1989ju,Battye:1993jv,Battye:1994au,Yamaguchi:1998gx,Yamaguchi:1999yp,Hagmann:2000ja}. 

The string system persists until the temperature of the Universe drops to around the QCD scale, when the axion mass turns on and becomes cosmologically relevant, leading to the formation of domain walls. The subsequent dynamics depends on the anomaly coefficient between QCD and the PQ symmetry, $N_W$, which is equal to the number of minima that are generated in the axion potential \cite{Sikivie:1982qv,Georgi:1982ph}. $N_W =1$ corresponds to the scenario in which the domain walls are automatically unstable and decay, destroying the string network and releasing further DM axions in the process \cite{Sikivie:1982qv,Georgi:1982ph,Chang:1998tb}. If $N_W >1$ the domain walls are generically stable, and the model is not phenomenologically viable, unless further explicit breaking of the PQ symmetry is introduced \cite{Gelmini:1988sf,Larsson:1996sp,Zeldovich:1974uw,Hiramatsu:2012sc} (which might reintroduce the strong CP problem) or the ${\mathbb Z}_{N_W}$ symmetry is gauged~\cite{DiLuzio:2017tjx}. In the following we set $N_W=1$ so that the PQ breaking scale $v_{\rm PQ}=f_a$, however, since we will focus on the early evolution of the string
network, the general case can be recovered from our results by simply replacing $f_a$ with $v_{\rm PQ}=N_W f_a$.
After the QCD crossover the comoving axion number density is conserved, and the coherent axion field behaves as cold DM.

In our present work we consider the string network before the axion mass turns on. An understanding of this stage of its evolution is crucial, both to calculate the relic abundance of axions produced at such times and to set the appropriate initial conditions when analysing the system once the axion mass becomes relevant. The important properties of the network, which we aim to determine, include: the average length of string per Hubble volume, and the way that this is distributed in loops of different lengths; the rate of energy release into axions; and the spectrum of axions emitted. 
Due to the complex evolution and interactions of the strings, and later the domain walls, an accurate analytic calculation of the axion relic abundance appears hopeless in this scenario, and instead some form of numerical simulation is required. The most direct approach, which we will follow, is to simulate a complete UV theory that gives rise to the axion numerically on a discrete lattice \cite{Hagmann:1990mj} (more details are given in Section~\ref{sec:sim}). 

The properties of the axion strings are critically affected by the fact that they come from a global symmetry. As we discuss further in Sections~\ref{sec:sim} and \ref{sec:scaling},  this means that the energy per unit length of an isolated string is logarithmically divergent. In the early Universe the divergence is cut off by the distance to a string of opposite orientation, or for a small loop its diameter, both of which are typically of order of the Hubble length $H^{-1}$ \cite{Vilenkin:1982ks}. 
Consequently, the theoretical expectation for the string tension $\mu$, defined as the average string energy per unit length $\mu=E/L$, in this system is roughly
\begin{equation} \label{eq:muintro}
\mu \approx \pi f_a^2 \log\left(\gamma \frac{m_r}{H} \right) ~, 
\end{equation}
where $m_r^{-1}$ is the size of the string core and $m_r$ corresponds to the mass scale of the degrees of freedom that UV complete the axion effective theory (in many models $m_r \sim f_a$). Meanwhile, $\gamma$ is a numerical factor that to a first approximation is expected to be of order $1$, and which is likely to be time dependent due to, for example, the number of strings per Hubble volume changing.
 Because of the expansion of the Universe, the majority of the DM axions produced by strings are emitted shortly before the network is destroyed, at the time of the QCD crossover. At this point  $\log (m_r / H) \approx 70 $ is enormous for $m_r \sim f_a$, changing the expected string tension significantly. 
Further, the large scale separation between $m_r$ and $H$ also suppresses the coupling between strings and axions by the same logarithmic factor \cite{Dabholkar:1989ju}, and is expected to render emission of heavy modes associated to the theory's UV completion irrelevant.

However, the huge scale separation in the physically relevant regime presents an immediate problem in attempting to study the system using numerical simulations. To resolve the dynamics of the strings, a 2 dimensional slice of the lattice perpendicular to a string must contain at least a few grid points inside the string core, and to capture the interactions and dynamics of strings a few Hubble volumes must be simulated (we show this by analysing the systematic errors in simulations in Appendix~\ref{sec:app2}). Given the computational power available, the largest grids that can be simulated have $N^3 \sim 1000^3$ lattice points. Consequently, the maximum scale separations that can be directly studied correspond to $\log\left(m_r/H\right) \approx 6$. In this system, the tension of strings, and their couplings to axion and heavy degrees of freedom, are far from the physically relevant values. Indeed, even if these could be adjusted to the physical values by modifying the UV theory this would not be sufficient. For example, the properties of the string network depend on whether collapsing string loops rebound and oscillate many times before disappearing, and if strings that approach each other recombine. In order to accurately capture such dynamics, processes on all scales between Hubble and the string core size must be resolved.

Making physically relevant predictions about the system at the time of QCD crossover therefore requires that results from simulations are extrapolated over a vast difference in scale separations. What makes such an extrapolation not obviously hopeless
is the possible existence of an attractor in the evolution, an understanding of which would
allow a controlled extrapolation to be made.
A key point of our work is that this is an extremely delicate process. In particular, we stress that a careful analysis of which features of the string network are being assumed to remain constant, or change, between $\log (m_r/H ) \approx 6$ and $\log (m_r/H) \approx 70$ is required, and that naive extrapolations can lead to misleading results.

The calculation of the axion relic abundance by strings and domain walls has been the subject of extensive prior investigation. However, there is still substantial disagreement about the qualitative and quantitative dynamics of strings, and  predictions of the resulting axion DM density differ dramatically. Previous work has been based on numerical simulations of global strings, and also theoretical analysis and  numerical simulations of local strings, which are thought to reproduce the dynamics of global strings in the limit of large scale separation. 
To enable comparison, we postpone a discussion of the literature until we have presented our results. Instead, here we simply note that there is currently an order of magnitude disagreement about the average numbers of strings per Hubble volume at the time of the QCD crossover, which introduces a similar uncertainty on the axion number density produced; and also an uncertainty on the form of the energy spectrum of the axions produced, which can change the axion number density by almost two orders of magnitude. 

Turning to the structure of this paper: In Section~\ref{sec:sim} we discuss the theory that we numerically simulate, and the technical challenges that this involves. In Section~\ref{sec:scaling} we demonstrate the existence of an attractor solution, which is approached independently of the system's initial conditions, and analyse its properties. In Section~\ref{sec:spectrum} we study the energy released by the string network, and the spectrum of axions emitted. There we also discuss the impact on the axion relic abundance, and the process of extrapolating to the physical point in parameter space. In Section~\ref{sec:concl} we summarise our results and the possibilities for future development.  
Additional technical details about our simulations may be found in Appendix~\ref{sec:app1}, and an extensive analysis of the systematic errors is given in Appendix~\ref{sec:app2}. In Appendix~\ref{sec:app_mask} we present a detailed analysis of the distribution of energy into different components, in  Appendix~\ref{sec:app_init} we provide further evidence for the existence of an attractor solution, and in Appendix~\ref{sec:app_scaling} we give details of how we fit the parameters of the string network. Finally, in Appendix~\ref{sec:app_eft} we analyse whether the properties of the global strings that we simulate are converging to those of local strings.

\section{Axion Strings and Simulations} \label{sec:sim}

We consider a complex scalar field $\phi$ taken to have the $U(1)_{\rm PQ}$ invariant Lagrangian
\begin{equation} \label{eq:lag1}
\mathcal{L}=|\partial_\mu\phi|^2
- V(\phi)\,, \qquad \text{with} \quad V(\phi)=\frac{m_r^2}{2 f_a^2} \left(|\phi|^2 -\frac{f_a^2}{2}\right)^2 \,, 
\end{equation}
in a spatially flat Friedmann-Robertson-Walker background. 
The metric is $ds^2=dt^2-R^2(t)dx^2$, and the Universe is assumed to expand as in radiation domination, so the scale factor $R(t) \propto t^{1/2}$, and the Hubble parameter $H \equiv \dot{R}/{R}=1/(2t)$.  

The potential $V(\phi)$ leads to $\phi$ getting a $U(1)_{\rm PQ}$ breaking vacuum expectation value (VEV) $\left|\left<\phi \right>\right|^2  = f_a^2/2$. 
We decompose
\begin{equation} \label{eq:decom}
\phi(x)=\frac{r(x)+ f_a}{\sqrt{2}}\,e^{i \frac{a(x)}{f_a}}\,,
\end{equation}
into the radial field $r(x)$, which has a mass $m_r$, and the axion field $a(x)$, which has a period $2\pi f_a$. Since we focus on the properties of the system at temperatures above the QCD crossover, the PQ breaking axion potential generated by QCD can be neglected and the axion is massless (at lower temperatures this must be added to eq.~\eqref{eq:lag1}).

The average Hamiltonian density $\rho_{\rm tot}=\langle T_{00}\rangle$ of the complex field $\phi$ is
\begin{equation} \label{eq:ham1}
\rho_{\rm tot} = \ \left\langle|\dot{\phi}|^2+|{\nabla}\phi|^2+V(\phi)\right\rangle ~,
\end{equation}
where $\dot{\phi}=d\phi/dt$, ${\nabla}$ is the gradient with respect to the physical spatial coordinates $R(t) x$, and $\langle A\rangle\equiv \lim_{V\to\infty}\frac{1}{V}\int_V d^3x\ A$ is the spatial average of $A$. After decomposing $\phi$ as in eq.~\eqref{eq:decom},
\begin{equation} \label{eq:rhoar}
\begin{aligned}
\rho_{\rm tot} =& \left\langle \frac{1}{2}\dot{a}^2+\frac{1}{2}|{\nabla} a|^2 \right\rangle + \left\langle\frac{1}{2}\dot{r}^2+\frac{1}{2}|{\nabla} r|^2+ V(r) \right\rangle \\
& + \left\langle\left(\frac{r^2}{2f_a^2}+\frac{r}{f_a}\right)\left(\dot{a}^2+|{\nabla} a|^2\right)\right\rangle ~,
\end{aligned}
\end{equation}
where $V(r) = \frac{m_r^2}{8 f_a^2} r^2 \left(r + 2 f_a \right)^2$. The terms on the first line of eq.~\eqref{eq:rhoar} correspond to the kinetic and potential parts of the axion and radial modes' energies, and the term on the last line is the interaction energy between the two. In the small field limit $\left|r\right|/f_a\to 0$, the Hamiltonian can be approximated as the sum of that from decoupled axions and radial modes (i.e. by the first line of eq.~\eqref{eq:rhoar}). However, away from this limit interaction terms between the two fields are not negligible and make the axion-radial system strongly coupled.

Note that the field's  equation of motion 
\begin{equation} \label{eq:eom}
\ddot{\phi} +3 H\dot{\phi}- {\nabla}^2\phi+\phi \frac{m_r^2}{f_a^2}\left(|\phi|^2 -\frac{f_a^2}{2}\right)=0  \,,
\end{equation}
does not depend on the ratio $m_r/f_a$ directly. Indeed the dependence on the two scales $f_a$ and $m_r$ can be reabsorbed by rescaling respectively the field $\phi\to \phi f_a$ and the space-time coordinates 
$t\to t/m_r$ and $x\to x/m_r$. Therefore, up to a trivial field rescaling, the physics is only sensitive to the ratio $m_r/H=2m_r t$.

The equations of motion in eq.~\eqref{eq:eom} admit solitonic string-like solutions.
As mentioned in the Introduction, these are topologically non-trivial configurations that contain loops in space where the axion field $a$ wraps  the fundamental domain $[0,2\pi f_a]$ non-trivially. 
The prototype of such solutions is a static, infinite, string lying along the $z$-axis. In cylindrical coordinates $(\rho,\theta,z)$ this is given by 
\begin{equation} \label{eq:protostring}
\phi(x)=\frac{f_a}{\sqrt{2}} g(m_r \rho)e^{i\theta}~,
\end{equation}
where $g$ is a profile function that satisfies $g(\rho)=C\rho+{\cal O}(\rho^3)$ for $\rho\to0$ and $g(\rho)= 1-\rho^{-2}+{\cal O}(\rho^{-4})$ for $\rho\to\infty$. The string core is defined as the region in which $\phi$ is close to the maximum of its potential, i.e. when $r/f_a\sim -1$, which corresponds to points at a distance less than $m_r^{-1}$ away from the centre of the string $\rho=0$. In this part of space the axion-radial mode system is strongly coupled, and all of the terms in eq.~\eqref{eq:rhoar} 
  contribute to the string energy density. 
However, for a single string configuration the axion energy density diverges logarithmically for $\rho\to \infty$ due to the angular gradient $\frac{1}{2}\langle|\nabla a|^2\rangle$. Consequently, the total string energy is dominantly in this component, and is mostly stored away from the string core. In the early Universe this leads to a string tension of the form given in eq.~\eqref{eq:muintro}.

To analyse the dynamics of the string system in the early Universe, we numerically integrate the equations of motion, eq.~\eqref{eq:eom}, in $3+1$ dimensions. Starting from suitable initial conditions, for example $\phi$ random with sufficiently large fluctuations, axion strings automatically form and evolve. 
In doing so, we are assuming that solutions of \emph{classical} equations of motion capture the physics of strings and axion radiation. This is justified because strings are themselves intrinsically classical and the relevant part of the axion radiation has large
occupation number.

\begin{figure}[t]
\begin{center}
\includegraphics[height=6.cm]{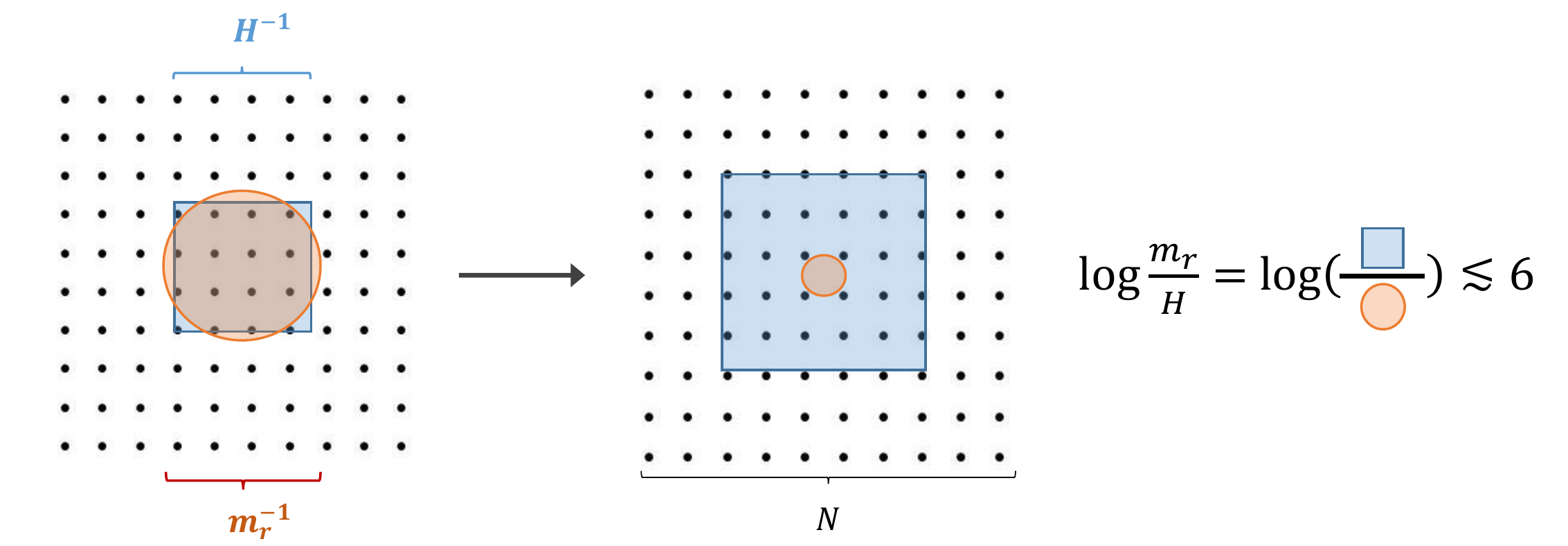} 
\caption{\label{fig:core_picture}
An illustration of how the size of a string core, shaded red, and a Hubble volume, shaded blue, evolve relative to the lattice points in our simulations, where $N$ is the number of lattice points in a spatial dimension. Requiring that the simulation contains at least a few Hubble volumes and that a string core contains at least $\sim 1$ lattice point constrains the maximum scale separation that can be studied.}
\end{center} 
\end{figure}

The complex scalar $\phi$ is discretised on a lattice with approximately $N^3=1250^3$ grid points, and evolved in fixed steps of conformal time $\tau \sim \sqrt{t}$. Our simulation is carried out with periodic boundary conditions, and in comoving coordinates $x$, so that the comoving distance between grid points remains constant and the physical distance between grid points grows $\sim \sqrt{t}$. Further details of the algorithms used are given in Appendix~\ref{sec:app1}. As the system is evolved forward, the number of Hubble lengths contained in the box side decreases $\sim 1/\sqrt{t}$ and the number of lattice points inside a string core also decreases $\sim 1/\sqrt{t}$, as shown in Figure~\ref{fig:core_picture}. The maximum accessible scale separation corresponds to an upper bound on the final time that can
be simulated. Other possible sources of numerical uncertainty include the time step used in the simulation and the way that the contribution of the string energy is excluded from the calculation of the energy in free axions, and a full analysis is given in Appendix~\ref{sec:app2}.

The simulations that we carry out with $m_r / H \sim 1$ can be interpreted in terms of a theory with $m_r \approx f_a $ at a time when $H \approx f_a$, however this is not a physically relevant part of parameter space. First, for such models the scale separation simulated is far from that at the time of the QCD crossover, when the majority of axions produced by the string network are emitted. Additionally, this regime does not correspond to a system that is  realisable even at early times: when $H \sim f_a$, the temperature of the Universe is $T \sim \sqrt{f_a M_{\rm Pl}}$, where $M_{\rm Pl}$ is the Planck mass, and a physical theory will be in the unbroken, PQ symmetric, phase. However, by studying the potential eq.~\eqref{eq:lag1} without including finite temperature effects, we can use the results obtained to extrapolate the properties of the string system to low temperatures, when such corrections are actually negligible. In particular, at the time of the QCD crossover $T\sim \GeV$ and finite temperature corrections to the potential of the radial scalar are irrelevant.

Alternatively, simulations at a scale separation $m_r / H \sim 1$ can be interpreted as studying a model with $m_r \ll f_a$ at a time when the Hubble parameter is $H\ll f_a$. Axion theories with a light radial mode are less commonly considered, but in this case the simulations are directly analysing a physically realisable point in parameter space. In particular, taking $m_{r} \sim 10^{-18} \GeV$, the Hubble parameter is that at the time of the QCD crossover (although such a light scalar is excluded by fifth-force experiments).\footnote{In such models the PQ phase transition still happens at $T \sim f_a$, and finite temperature corrections to the potential of the radial mode are not important, provided that it is weakly coupled to states in the thermal bath.} Meanwhile, theories in an intermediate regime, with a radial degree of freedom that has a mass $10~\keV \lesssim m_r \lesssim f_a $, are not experimentally excluded by evolution of stars and fifth-force experiments. These correspond to scale separations at the time of the QCD crossover that are larger than can be reached with simulations, so that extrapolation is still required, but which are smaller than if $m_r \approx f_a$.

The Lagrangian in eq.~\eqref{eq:lag1} includes only one heavy scalar degree of freedom, and it is clearly not the most generic that can arise in UV-complete axion theories. However, when $H \ll m_r$ the dynamics of axion strings and radiation  are expected to be largely independent of which massive degrees of freedom are included, since these  only get excited when strings interact over distances of order $m_r^{-1}$, e.g. when loops shrink or long strings intersect. The energy of a global string configuration remains logarithmically divergent in more complicated theories, since this comes from the axion angular gradient, which is always present. In principle one could also include interactions with other fields that $\phi$ is coupled to --- e.g. SM fields --- in eq.~\eqref{eq:lag1}, but away from the string cores the couplings of axions to these are suppressed by powers of $f_a$, and are negligible. Meanwhile the radial mode can have order 1 couplings to SM fields, however it is expected to decouple from the dynamics of the strings and axions at large scale separations. At early times, when the temperature is high, interactions of strings with the visible sector thermal plasma could modify their dynamics \cite{Harari:1987ht}, however these effects will also be negligible at temperatures around the QCD crossover.

As well as the physical system, the literature has often used a deformed theory in which 
the mass of the radial mode in eq.~\eqref{eq:lag1} is replaced with a time dependent one
\begin{equation} \label{eq:fatstringdef}
m_r\left(t\right) =m_{i} \frac{R(t_i)}{R(t)}= m_{i} \sqrt{\frac{t_i}{t}} ~,
\end{equation}
where $m_{i}$ is the mass of the radial mode at the initial time $t_i$. This is equivalent to a theory with a quartic coupling $\lambda \sim m_r^2(t)/f_a^2$ that decreases with time, and is known as the \emph{fat string} scenario. The size of the string cores increases as $m_r^{-1}(t) \sim t^{1/2}$, and the number of lattice points inside a string core remains constant throughout a simulation. The maximum scale separation that can be simulated is unchanged compared to the physical potential, however the time taken to reach a particular scale separation (starting from $H\sim m_r$) is increased, and simulations can be run for longer before arriving at the upper bound. As a result, energy left over from the initial conditions is redshifted more, and there is more time available for properties of the string system to reach their asymptotic behaviour (there are additional benefits that will be seen in Section~\ref{sec:spectrum}). Despite these advantages, we stress that by making the potential time dependent the equations of motion of the system are changed by an order 1 amount. Consequently, although the dynamics of axion strings might remain qualitatively similar to those of the physical Lagrangian this is not guaranteed, and the numerical values of the parameters of the scaling solution are not expected to be the same in the two cases. We perform simulations using both techniques and discuss their advantages and disadvantages.

In both the fat string and the physical scenarios, the axions produced in the simulations are massless, and their energy densities redshift as $\sim 1/R\left(t\right)^4$. Meanwhile, the radial modes produced are highly non-relativistic. In simulations with the physical Lagrangian their energy density redshifts as $\sim 1/R\left(t\right)^3$. In simulations of the fat string Lagrangian the scalar mass decreases with time, and the energy density of these states redshifts as $\sim 1/R\left(t\right)^4$, the same as axions.

\section{The Scaling Solution} \label{sec:scaling}
It has long been claimed 
 that a system of axion  strings is driven towards a particular  solution, which is independent of its initial conditions~\cite{Kibble:1976sj,Kibble:1980mv,Vilenkin:1981kz}. Indeed, this feature is crucial for making predictions about the properties of the string system at late times, and in particular of the axion relic abundance from strings, that do not depend on the dynamics of the system at early times, which are model dependent.
 
The existence of such an attractor solution is simple to motivate qualitatively. Strings can lose the energy stored in their length by radiating axions and radial modes. Therefore bends in strings with curvature larger than the Hubble scale tend to straighten, and closed string loops smaller than the horizon are expected to disappear, emitting radiation. Additionally, long strings (or equivalently string loops larger than the horizon) can interact when they enter each other's horizon through a process called recombination: when strings cross they can recombine into a new configuration with a lower tension, and similarly a region of high curvature in a long string can split off forming an isolated loop. The net effect is a reduction of the total string length and 
the production of smaller loops, string segments with larger curvatures, and radiation. The rate at which such processes occur
depends on the density of strings within the horizon. Below some critical density recombination is inefficient. In this case, the number of strings in each Hubble patch increases as the Universe expands and new strings enter the horizon. On the other hand, above a critical density recombination becomes efficient, reducing the number of strings within the horizon. As a result, the density of strings is pushed towards a particular (not necessarily time independent) value. Other statistical properties of the network, such as the distribution of the string density in loops of different length, are expected to converge similarly to a common behaviour.

We define the average number of strings per Hubble patch $\xi(t)$ as
\begin{equation} \label{eq:xidef}
\xi(t)\equiv \lim_{L\to \infty } \frac{{\ell}_{\rm tot}(L)\,t^2}{L^3} \,,
\end{equation}
where $\ell_{\rm tot}(L)$ is the total length stored in strings in a volume $L^3$. Hence the energy density of strings is
\begin{equation} \label{eq:rhos}
\rho_{s}(t)=\xi(t) \frac{\mu_{\rm eff}(t)}{t^2} ~,
\end{equation}
where, given eq.~(\ref{eq:muintro}),  the effective string tension $\mu_{\rm eff}(t)$ is expected to be
\begin{equation}\label{eq:mueff}
\mu_{\rm eff}(t)=\mu_0 \log\left(\frac{m_r\,\gamma(t)}{H \sqrt{\xi(t)}} \right)\,,
\end{equation}
with\footnote{Eq.~(\ref{eq:mueff}) strictly applies for non-relativistic strings. As we will see
relativistic effects turn out to be small.} $\mu_0=\pi f_a^2$.  The factor $m_r/(H\sqrt\xi)$ is anticipated  to capture the main time dependence of $\mu_{\rm eff}(t)$ since the logarithm is cut-off by the average distance between strings ($\propto t/\sqrt{\xi}$). The remaining time dependence is encoded in the factor $\gamma(t)$, which takes into account the non-trivial shape of the strings and
is expected either to be constant or to have at most a very mild time dependence. Indeed, it will be a non-trivial check of the string network's properties that the energy density in strings extracted from the simulations is well reproduced by eqs.~(\ref{eq:rhos}) and (\ref{eq:mueff}).

The existence of the scaling law (\ref{eq:rhos}) with constant $\xi(t)=\xi_0$ can easily be understood for local strings.
For these the string tension is localised on the core and is constant 
$\mu_{\rm eff}(t)=\mu$. Neglecting the core size the problem only has one scale, 
$H=1/2t$, which completely fixes the scaling law for the energy density $\rho_s(t)=\xi_0 \mu/t^2$. 
The presence of a single scale suggests that during the scaling
regime all the properties of the string configuration should be scale invariant. 

On the other hand, for the global case several properties of the strings, including their tension and their coupling to axions~\cite{Dabholkar:1989ju}, depend logarithmically on the core size $m_r$, therefore
logarithmic corrections to the scaling law can be expected. To account for these effects 
we leave an explicit time dependence both in $\xi(t)$ and in $\gamma(t)$,
besides the one contained in $m_r/H$ inside $\mu_{\rm eff}(t)$.

In the rest of the section we will establish the existence of the (approximate) scaling solution for axion string
networks, and study its properties in detail. In particular we will present results from
numerical simulations demonstrating the presence of the attractor, its independence
from the initial conditions, the behaviour of the parameter $\xi(t)$, and the distribution of
loops and long strings during the scaling regime.

\subsection{The Attractive Solution} \label{sec:attrac}
The existence of the attractor can be tested by studying if different statistical properties of the
string network converge to the same values independently of the initial conditions of the field.
As a representative example, here we focus on the evolution of the average number 
of strings per Hubble patch $\xi(t)$. In Appendix~\ref{sec:app_init} we present additional results showing that the number density, the total and the instantaneous spectrum of axions emitted from the string network also clearly converge, regardless of the initial conditions. The convergence is particularly evident in the instantaneous emission spectrum, which depends only on the string configuration at a particular moment and has no memory of earlier times.

We set the initial conditions in two different ways. In the first, we just generate
sets of random fields. In the second, we construct initial conditions with a fixed 
number of strings by evolving random configurations until the total string length in the box 
reaches 
a required value, and then we reset the clock rescaling Hubble 
(more details about the procedure can be found in Appendix~\ref{sec:app1}).
Besides allowing us to start simulations with a predetermined density of strings, the second method produces initial field configurations that have less primordial background radiation 
(although this would redshift away anyway) and more suitable for a cleaner study of the instantaneous spectrum. In Appendix~\ref{sec:app_init} we show that the properties and evolution of the string network are independent of the way in which the initial conditions are set.

\begin{figure}[t] 
\begin{center}
\includegraphics[height=6.77cm]{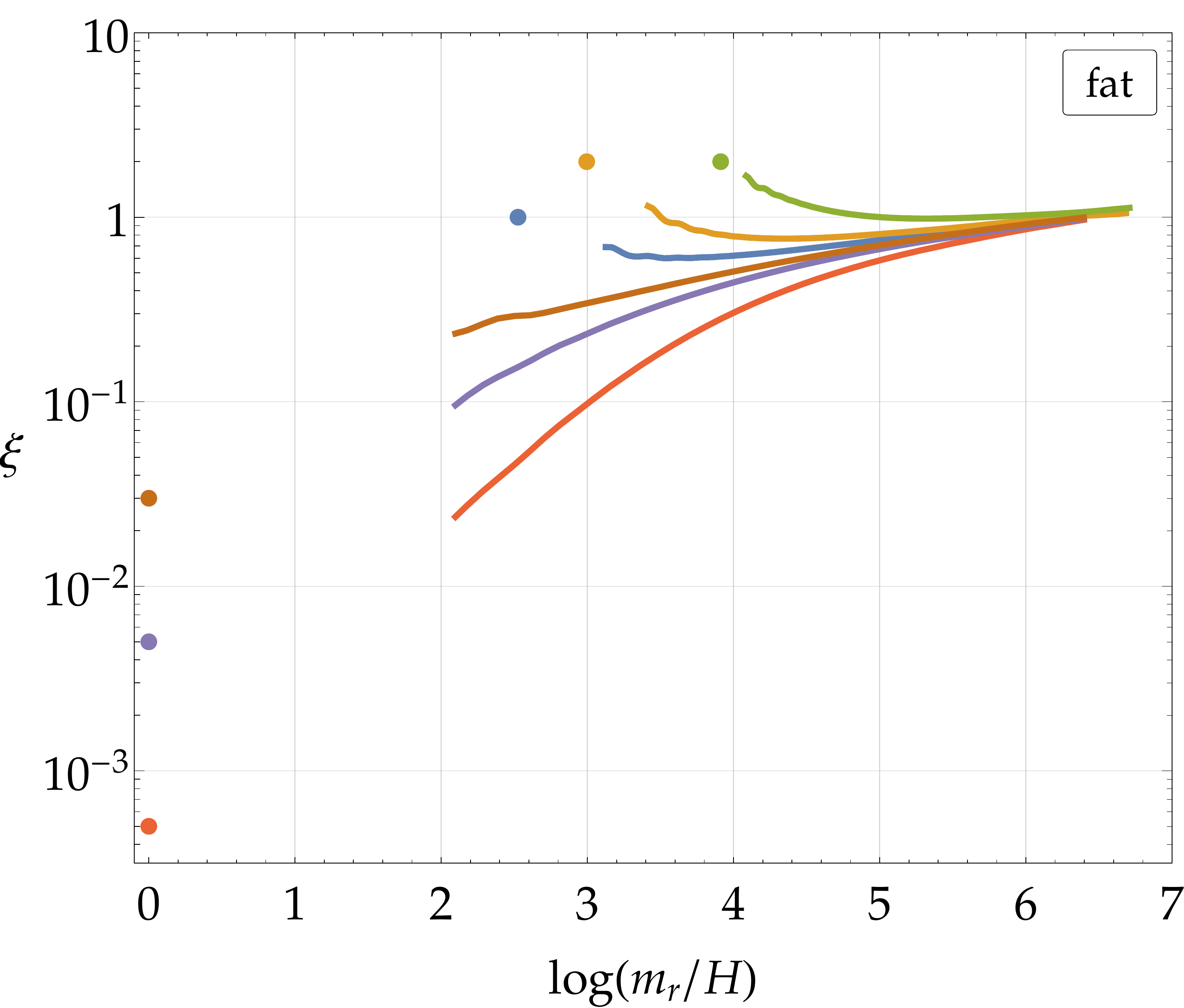}\hspace{0.5cm}
\includegraphics[height=6.77cm]{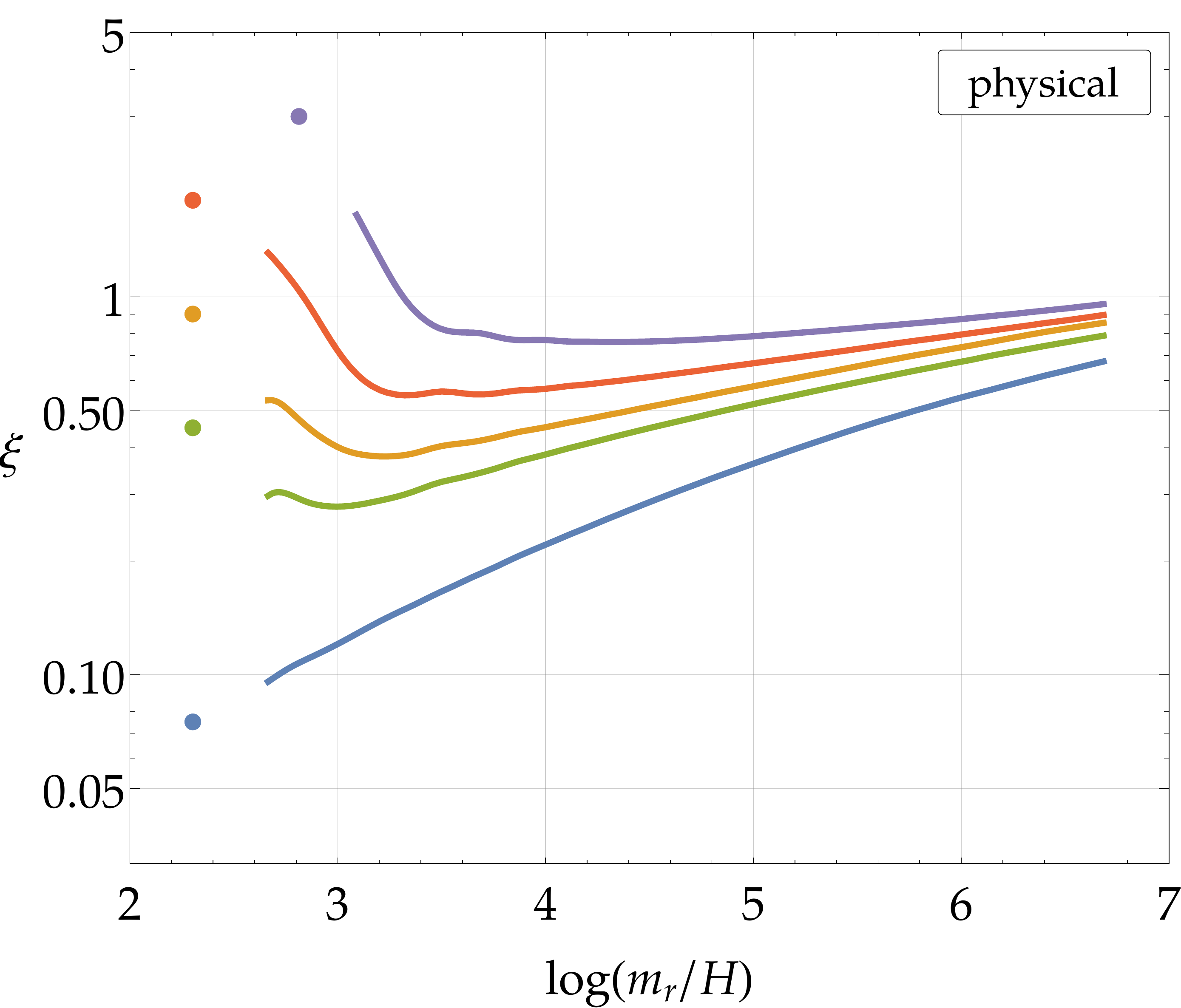}
\caption{\label{fig:attractor} 
The evolution of the average number of strings per Hubble patch $\xi(t)$ as a function of time (here represented by Hubble) using the 
fat string trick (for which $m_r\propto t^{-1/2}$, left) and for the physical case ($m_r=$~const, right) for different initial conditions. Each curve corresponds to the average of many simulations with the same initial value of $\xi$. The position of the coloured dots indicates the initial time and the value of $\xi$ for each simulation set. The number of simulations has been taken large enough that the statistical uncertainties are smaller than the thickness of the curves.} \end{center}
\end{figure}

In Figure~\ref{fig:attractor} we show the evolution of $\xi$ with time in simulations, 
starting from initial conditions with different numbers of strings using the second method described above.
$\xi$ 
 has been computed at different time shots from its definition eq.~\eqref{eq:xidef}, using the algorithm described in
Appendix~\ref{sec:app1}.
 As discussed in Section~\ref{sec:sim}, it takes a longer time to get to the same value of 
 $ \log(m_r/H)$ in the fat string scenario than for the physical theory. As a result the attractor regime is reached at smaller values of the log in the fat string case (here and in the rest of the paper we  sometimes use the short-hand notation ``log'' to refer to $\log(m_r/H)$).

In both the fat string and the physical models, the convergence towards a common value of $\xi$ is manifest. 
In the fat string case, the initial values of $\xi$ span
more than three orders of magnitude and, by the end of the simulations, they lead to the same value of $\xi$ to within 10\%.
For the physical case the convergence is a little slower but it is still clear. 
In simulations starting at $H=m_r$ it is not possible to initially have more than one string per Hubble patch. As a result, to achieve initial conditions with a clear overdensity of strings we started such simulations
later, when $H<m_r$. The corresponding data in the Figure~\ref{fig:attractor} have the initial points (the coloured dots) at larger values of  $\log(m_r/H)$.

The network of global strings was first studied using field theoretic computer simulations in \cite{Yamaguchi:1998gx,Yamaguchi:1999yp}, and more recently over a longer time range in \cite{Hiramatsu:2010yu}.  
Ref.~\cite{Fleury:2015aca,Klaer:2017qhr} showed evidence for the existence 
an attractor for the fat string system, respectively in two and three dimensions.
Our simulations have a similar time range and constitute an independent check of the convergence of $\xi$ in the fat string system starting from a wide range of initial conditions, and a demonstration of the attractor's existence for the physical case. Further, in Appendix~\ref{sec:app_init}, we show that other properties of the network, including the spectrum of axions emitted, also converge.

\subsection{Scaling Violation} \label{sec:scalvio}
\begin{figure}[t] 
\begin{center}
\includegraphics[height=6.77cm]{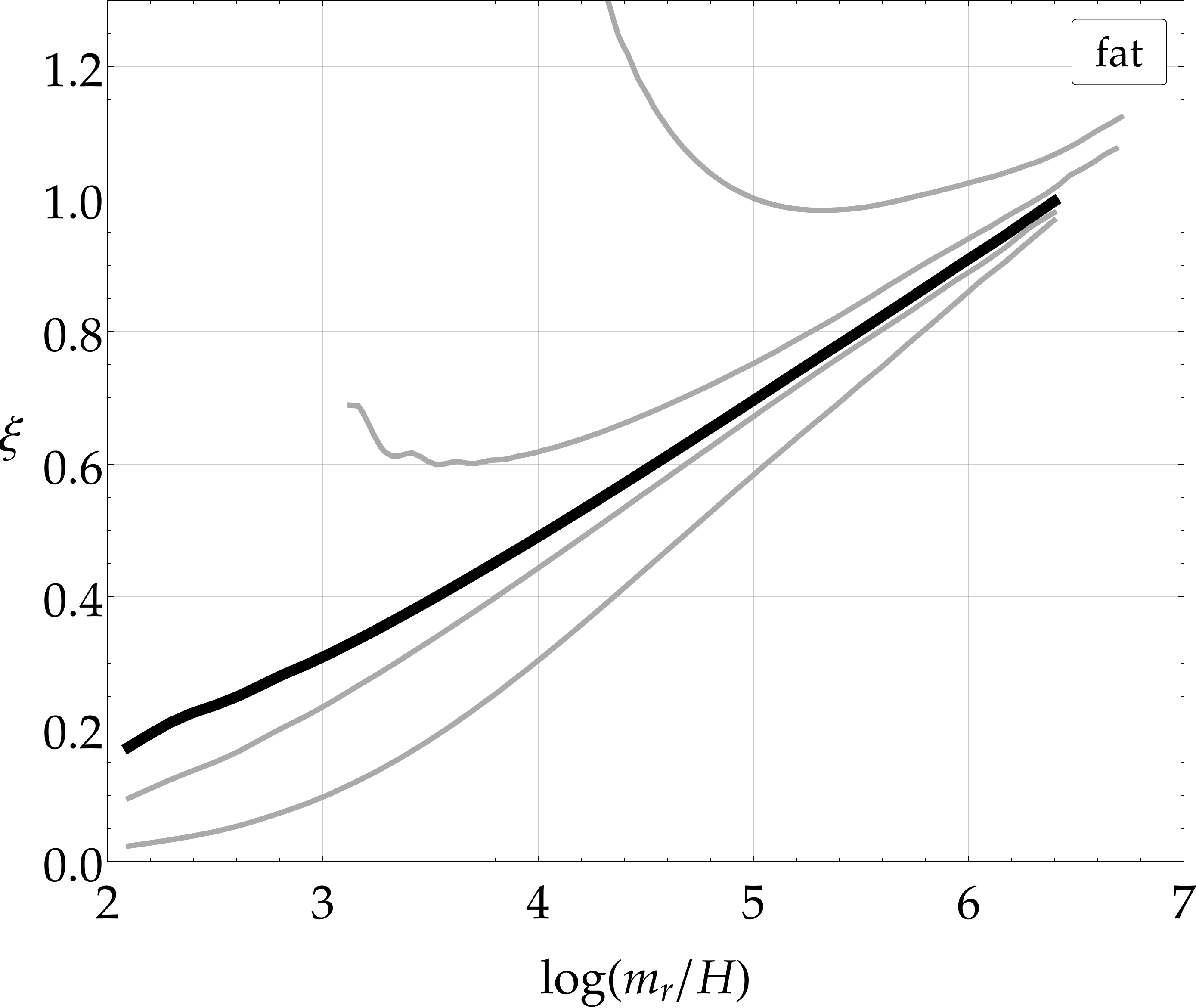}\hspace{0.5cm}
\includegraphics[height=6.77cm]{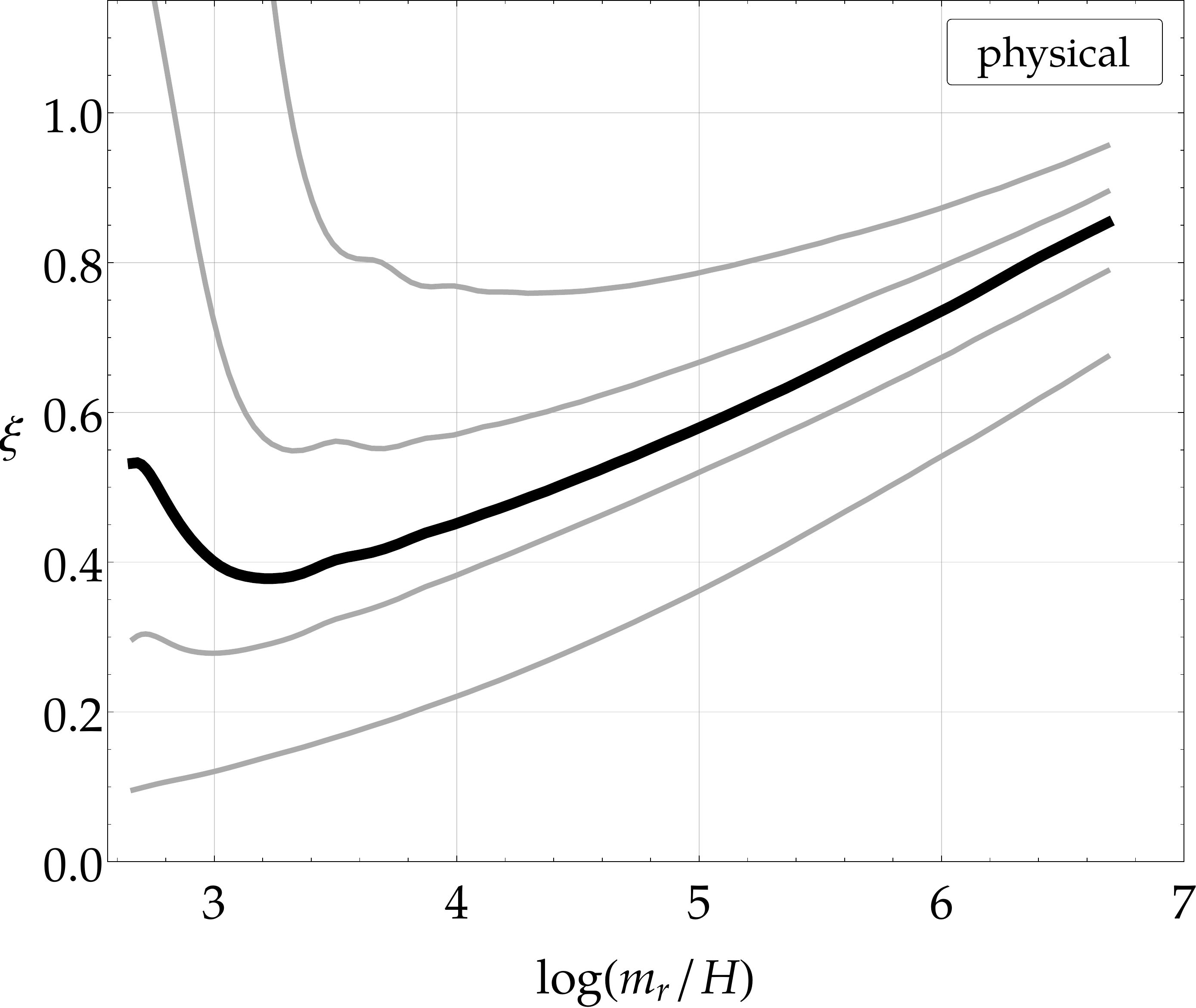} 
\caption{\label{fig:xit} The growth of $\xi(t)$ with time ($\log(m_r/H)$) for different initial conditions using the fat string trick (left) and in the physical case (right). The black curves correspond to the initial conditions that are the closest to the attractor solution.}\end{center}
\end{figure}

Having shown that the attractor solution exists, we now turn to study its properties. One prominent feature is that,
although different boundary conditions converge to a common value of $\xi$,
this value does not seem to be constant in time. To see the change more clearly, 
in Figure~\ref{fig:xit} we show the plots of Figure~\ref{fig:attractor} on a linear scale. 
A growth of  $\xi$ linear in $\log(m_r/H)$ is evident both for the fat string and for the physical cases. 
The fact that simulations with an overdensity of strings first rapidly
evolve to smaller values of $\xi$, converging to the attractor, and then start increasing again is particularly convincing. 
This strongly indicates that the growth is an intrinsic property of the scaling solution,
rather than the sign that the attractor has not yet been reached.

The behaviour shown in Figure~\ref{fig:xit} is compatible with the asymptotic form 
\begin{equation} \label{eq:xifit}
\xi(t)=\alpha \log\left( \frac{m_r}{H}\right)+\beta\,.
\end{equation}
In particular, at late times $\beta$ is subleading, and the value of $\xi(t)$ is dominated by 
the logarithmic term. 
 From Figure~\ref{fig:xit} it can be seen that the coefficient $\alpha$ is universal, independent
of the initial conditions. 
Indeed, the derivatives
$t\,\xi'(t)=\partial\xi/\partial[\log(m_r/H)]$ of the curves tend towards a common value $\alpha$ more rapidly than $\xi$ itself.

We can use the convergence of the slopes to a constant value
to
select the optimal initial conditions, i.e. those for which the scaling regime is reached 
 the earliest. The corresponding lines are those plotted in solid black in Figure~\ref{fig:xit}, and curves starting from
different boundary conditions reach the same constant slope
at later times. Considering only simulations that reach the scaling regime 
(i.e. those that show a sufficiently large region of constant slope) we extract estimates 
for~$\alpha$
\begin{equation} \label{eq:alphafit}
\alpha_{\rm fat}= 0.22(2)\,, \qquad  \alpha_{\rm phys}=0.15(5)\,.
 \end{equation}  
Here the errors clearly have no statistical significance, but rather they represent our educated, 
conservative, guesses about the uncertainty.
Plots showing the behaviour of the slope for
different boundary conditions, and more details about the fit, can be found in the Appendices~\ref{sec:app_init} and \ref{sec:app_scaling}.
We do not report the values of $\beta$, since the uncertainty on these from different initial conditions is larger, and they
are irrelevant for the physically interesting values of the log. 

As mentioned, a logarithmic violation of the scaling behaviour is not 
completely unexpected, since several properties of the string network, 
including the string tension and coupling to axions, have a similar dependence.
If such behaviour is maintained at later times, as seems plausible, the average number
of strings per Hubble patch will grow substantially for values of the log relevant to
the QCD axion. For example, if $m_r \sim f_a$ 
eq.~\eqref{eq:alphafit} would imply $\xi=10.5\pm 3.5$ at the physically relevant separation, 
$\log(m_r/H)\approx 70$.
This value is an order of magnitude larger than that found
in refs.~\cite{Yamaguchi:1998gx,Yamaguchi:1999yp,Hiramatsu:2010yu} from numerical simulations 
on smaller grids, and the value that is usually assumed in rough estimates of the axion abundance produced by strings. 
 Meanwhile, the extrapolated value of $\xi$ that we obtain is only a factor of 2 larger than that recently obtained in 
ref.~\cite{Klaer:2017qhr}, which used the fat string trick and a different UV completion of the core.  
 We stress that although the fat string system shows the same qualitative linear growth with the log as physical strings, the numerical parameters are somewhat different in the two cases. This is not surprising, and in extrapolating to the physical scale separation it is important to study the physical system, not just the fat string case.

The logarithmic enhancement in $\xi$ was first observed and studied in the 2+1 dimensional
simulations of ref.~\cite{Yamaguchi:1998iv}, but it was missed in the 3+1 dimensional ones of
ref.~\cite{Yamaguchi:1998gx,Yamaguchi:1999yp,Hiramatsu:2010yu}, 
perhaps due to the use of smaller grids (which limited the time range
of their simulations) and the choice of overdense initial conditions. Indeed, the combination
of these two factors can produce a fake plateau at the intermediate values of the logs 
that were analysed (see e.g. the top curve of the right hand plot in Figure~\ref{fig:xit}). 

Conversely, the later simulations of the fat string case in ref.~\cite{Fleury:2015aca}, 
made on larger grids, also observe a logarithmic increase,  
in agreement with our results. Finally, in the recent analysis of ref.~\cite{Klaer:2017qhr}, 
which use fat strings and a different UV completion of the core to partly include 
the effects of a large scale separation, the value of $\xi$
increases with the log. The growth rate is not clear and the authors
suggest that part of it may be due to spurious $H/m_r$ effects. However, the results of this reference for purely global strings are in agreement with our numerical fit, and correspond to initially overdense networks.

We will resist the temptation to interpret the log-increase of $\xi(t)$
in terms of the reduction of the string-axion coupling or the increase of the 
string tension. In fact a similar growth also seems to be present for local strings, 
and can be seen in Figure~3 of ref.~\cite{Klaer:2017qhr} and is hidden in Figure~7 
of ref.~\cite{Hindmarsh:2017qff}.\footnote{We have carried out preliminary 
simulations of local strings to study this effect, and these appear to confirm such a growth.}
Neglecting gravity, the only way that local strings can maintain a scaling regime is through the production of heavy modes associated to the core scale. The latter then cannot be neglected and its presence allows for a violation of the scaling law $\rho_s=\xi_0\mu/t^2$ argued before. 
As we will discuss below, the way that global strings lose energy is not
so different from the one above, which might explain the similarity in the way that the scaling is violated.

\subsection{Long \emph{vs} Short: the scale-invariant distribution of loops} \label{sec:longvshort}
In order to further characterise the attractor solution, and to better understand its properties, 
we also study how the total string length per Hubble patch $\xi$ is distributed over different
loop sizes. If we call $dn_\ell/d\ell$ the loop number density, i.e. the 
number of loops per unit volume and per unit of loop length, then the quantity
\begin{equation}
\xi_\ell\equiv  t^2 \int_0^\ell d\ell'\, \ell' \,\frac{dn_{\ell'}}{d\ell'} \,,
\end{equation}
represents the contribution to $\xi$ from loops of size smaller than $\ell$  and, in particular, $\xi_\infty = \xi$. 

We have performed a large number of simulations
using the fat string trick to acquire enough statistics to study $\xi_\ell$ as a function of time and loop size $\ell$. The initial conditions were fixed as for the thick line in Figure~\ref{fig:xit} left, so that the system started close to the scaling solution. 

\begin{figure}[t]\begin{center}
\includegraphics[scale=0.45]{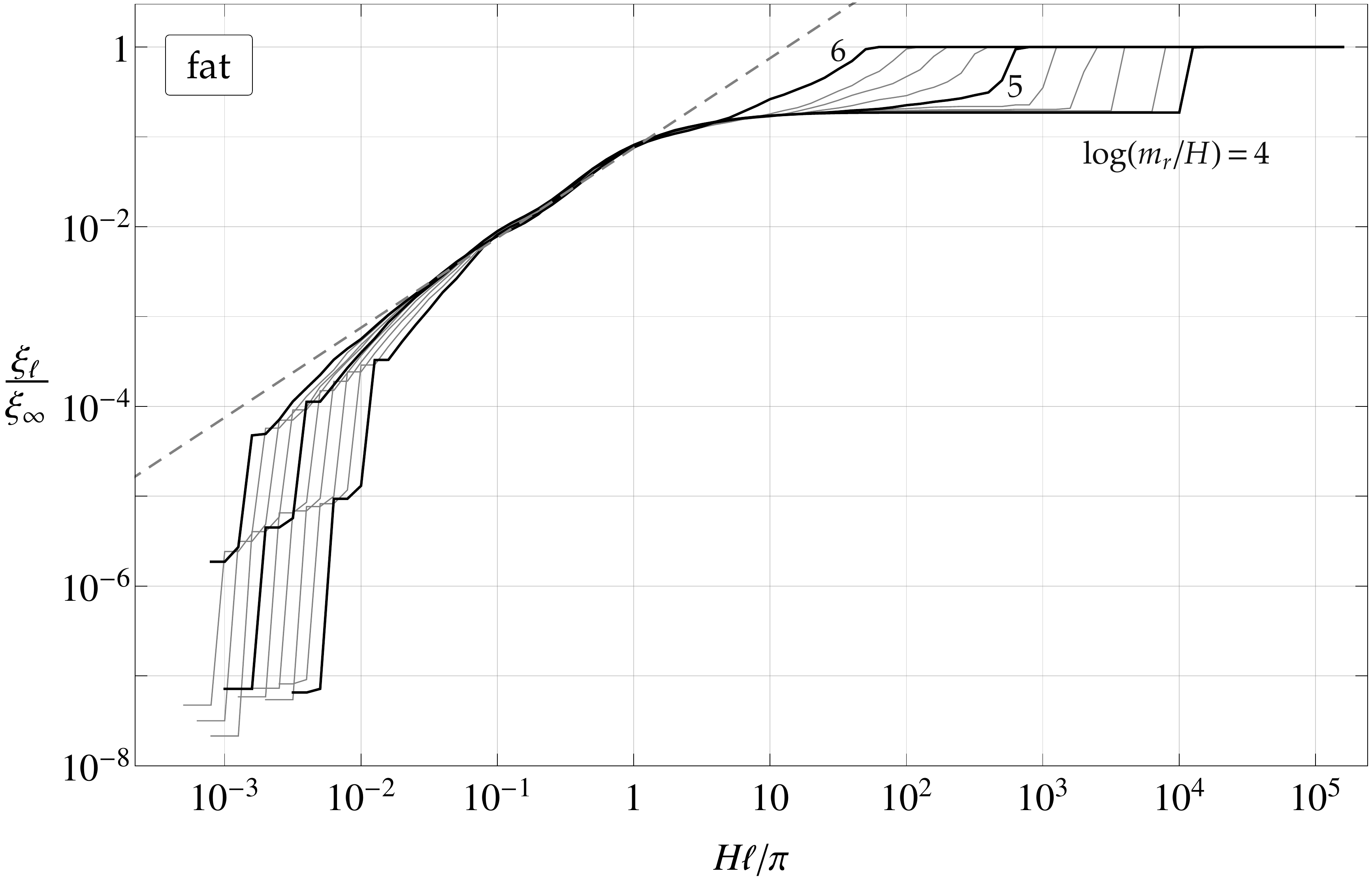}
\caption{ \label{fig:loopdistr}
The fraction of the total string length  $\xi_\ell/\xi_\infty$
that is contained in loops
smaller than $\ell$ for different time shots.}\end{center}
\end{figure}

The results for the ratio $\xi_l/\xi_\infty$, which gives the percentage of string length contained in loops of size smaller than $l$, are plotted in
Figure~\ref{fig:loopdistr}, and reveal several features of the string network.
Most of the string length, more than 80\%, is contained in loops much larger
than Hubble, of order of the full box size (in fact it seems that most of the string length is contained
in a single loop wrapped around the simulation box multiple times). This leads to the abrupt increase
in the right part of the plot, which saturates $\xi_\ell$ to its asymptotic value $\xi$.
Less than 20\% of the total contribution to $\xi$ is contained in loops of size of order $H^{-1}$ and smaller, which results in the appearance of the 
 plateau with $\xi_\ell /\xi_\infty \approx 0.2$ at $H\ell\gg 1$.\footnote{The fact that only approximately 10\% of the string length is contained in sub-Hubble loops was mentioned in \cite{Klaer:2017qhr}, which matches our more detailed analysis.}  
On the left of the plot the UV cut-off corresponding to the smallest possible loops, 
of order the core size, is also visible. 
As the Universe expands the physical size of the simulation box in units of Hubble shrinks, 
and as a result $\xi_\ell$ saturates its asymptotic value $\xi$ at smaller and smaller values 
of $H\ell$. At the same time, the value of $m_r/H$ grows so the UV cut-off moves to the left.

The lines in Figure~\ref{fig:loopdistr} corresponding to different times approximately overlap for values of $\ell$ sufficiently far from the UV and IR cutoffs. Therefore, since $\xi_{\infty}$ grows logarithmically with time, the corresponding growth in $\xi_\ell$ is homogeneous in $\ell$. This signals that the logarithmic increase of $\xi$ is equally distributed
over all scales, and that the ratio between long and short strings stays constant in time (see also
Figure~\ref{fig:hubbleTest} in Appendix~\ref{sec:app2}). The fact that $\xi_{\ell}/\xi_\infty$ remains constant in time for $\ell \lesssim  10 \pi / H$ also shows that the number of loops of a particular length per Hubble patch does not change, apart from this logarithmic increase. As loops shrink and disappear (or recombine with other strings) they are replaced at the same rate by larger loops themselves shrinking, or by new loops being produced from interactions of long strings, which is an indication that the attractor solution has been reached. When, at the final times, the Hubble scale becomes of order of the box size  
there is no longer a sharp distinction between long and short strings.

Another feature of the loop distribution is evident from the plot: for  loop lengths smaller than Hubble $\xi_\ell \propto \ell$ (the dashed line in the plot), so $d\xi_\ell/d\ell\propto \ell dn_\ell/d\ell=\,$const. This means that the number of loops in each logarithmic interval of length is constant, over almost two orders of magnitude. Equivalently, 
the 10\% of the full string length contained in sub-Hubble loops is equally distributed over all loop sizes on a linear scale. This approximate power law seems to become a better fit
as time progresses, suggesting that it is an intrinsic property of the attractor solution,
and further confirming that the attractor regime has been reached within the time range of
the simulation.\footnote{At late times a similar power law behaviour is also present in field theoretic
simulations of local strings, see ref.~\cite{Hindmarsh:2008dw}.}

\begin{figure}[t]\begin{center}
\includegraphics[scale=0.18]{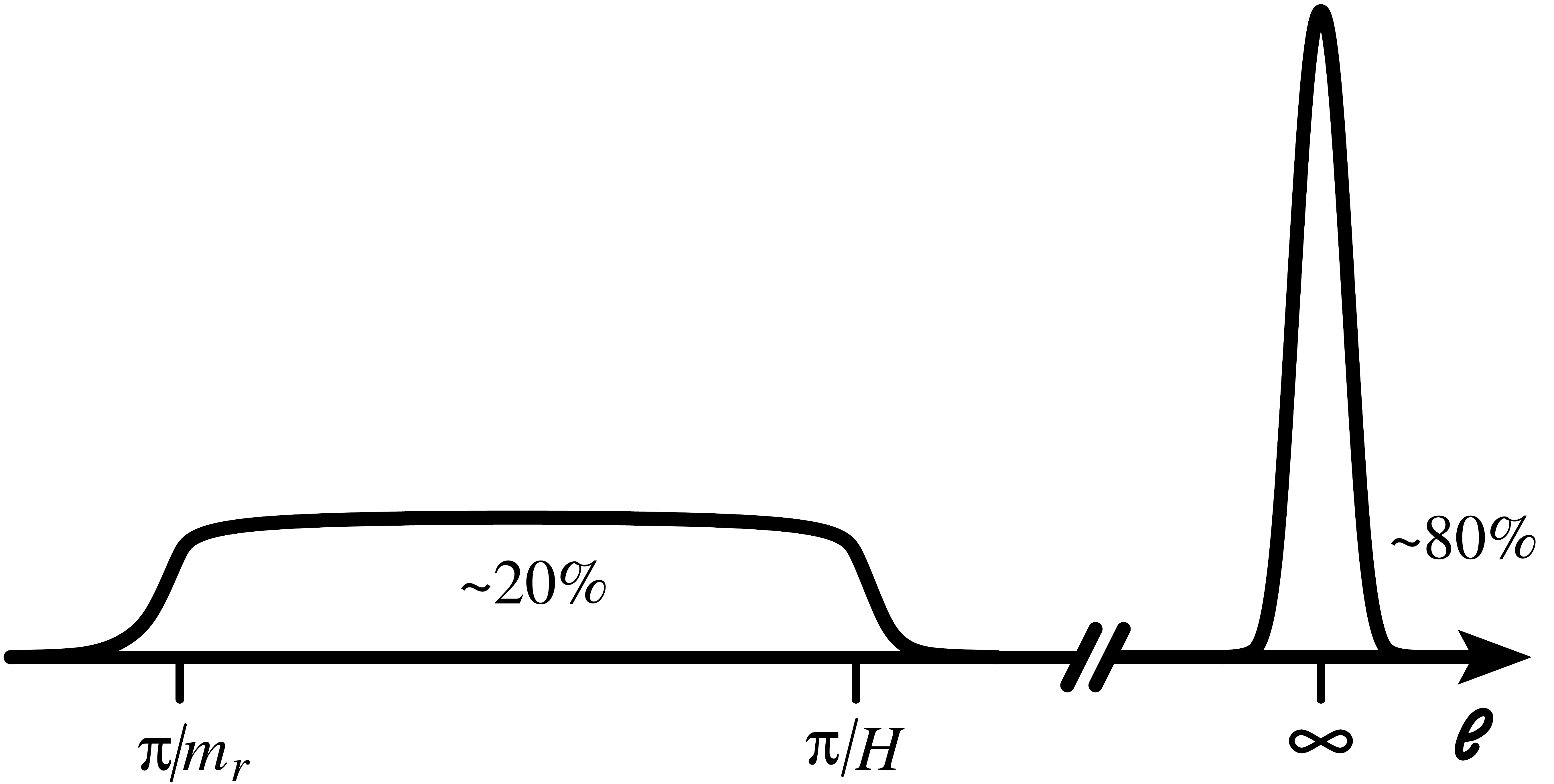}

\caption{ \label{fig:cartoon}
A cartoon of the distribution of string loops, $d\xi_\ell/d\ell$, in the scaling regime. A constant fraction of the total length is in sub-Horizon loops, with equal numbers of loops in each logarithmic interval down to the scale of the string cores.
} \end{center}
\end{figure}

Our analysis suggests that, in the infinite volume limit, the distribution of string length in the attractor solution is of the form depicted in Figure~\ref{fig:cartoon}.
Roughly 80\% of the string length per Hubble patch is contained in long strings (infinite string loops),
while the remaining 20\% is distributed in loops ranging from the core to the Hubble size, with equal numbers of loops in each decade of length.
The total string length grows logarithmically according to eq.~\eqref{eq:xifit} with the relative 4:1 ratio fixed.

In the literature the parameter $\xi$ is sometimes defined restricting to long strings only. 
 However, the loop distribution that we observe implies 
that the two definitions only differ by 20\%, and more importantly the factor of proportionality is
constant in time.

Since we do not directly use the quantitative behaviour of $\xi_\ell$ in our subsequent estimates of the final
axion abundance, we have only performed this analysis in the fat string case. 
The similarity in the behaviour of all of the other properties studied suggests that the picture 
for the physical case should be qualitatively similar.

\section{The Spectrum} \label{sec:spectrum}
We now discuss how the string network radiates energy, and study in detail how the energy in the system is split
among the different components (strings, axions, radial modes), and the way that the energy in axions is distributed in modes of different frequencies.
We then analyse the evolution of the axion number density, and the extrapolations of this to the physically relevant point. However, before turning to the results of our simulations, we first establish the physically relevant quantities and describe how these affect the axion relic abundance.

The conservation of energy implies that, in order to maintain the scaling regime, the string network 
must constantly lose energy into radiation. This is because, in the absence of interactions, the number of strings per Hubble patch would increase fast as more strings re-enter the horizon. To keep $\xi$  approximately constant, the excess string length must be destroyed, emitting energy.

The rate at which the system of interacting strings releases energy can be calculated by comparing the energy density in the scaling regime (parameterised by eq.~\eqref{eq:rhos}), to that of a ``free'' network of strings. By free we mean that long strings remain essentially at fixed comoving coordinates, so that $\xi(t) \propto R^2(t)\propto t$. The energy density of such a system is
\begin{equation} \label{eq:rhosfree}
\rho_s^{\rm free}(t) \propto \frac{\mu_{\rm}(t)}{R^2(t)} \propto \frac{\log\left(m_r\, d(t)\right)}{t}\,,
\end{equation}
where $\mu(t)=\pi f_a^2\log(m_r\, d(t))$ is the tension of free strings  and $d(t)\propto 1/R(t)\propto1/\sqrt{t}$ parametrises the average distance between strings.

We consider a network of free strings that has the same string configuration as the interacting system with energy $\rho_s(t)$ given by eq.~\eqref{eq:rhos} at a time $t_0$. The energy of such a system is
\begin{equation} \label{eq:rhofree}
\rho^{\rm free}_s(t)= \frac{\xi(t_0)\mu(t)}{t_0 t}~,
\end{equation}
with $\mu(t_0)=\mu_{\rm eff}(t_0)$, so that this matches that of the interacting network at $t=t_0$. The string energy density of the interacting system in the scaling regime decreases faster than $\rho_s^{\rm free}$, and the difference corresponds to energy that is released. The rate $\Gamma$ at which the interacting network emits energy into radiation  is therefore given by 
\begin{align} \label{eq:Gammasol}
\Gamma&=\left[ \dot{\rho}^{\rm free}_s(t_0)-\dot{\rho}_s(t_0) \right ]_{t_0=t}
=\rho_s \left[2H-\frac{\dot \xi}{\xi}-\frac{\mu_0}{\mu_{\rm eff}} \left(H 
+\frac{\dot \gamma}{\gamma}-\frac12 \frac{\dot\xi}{\xi} \right)\right] \nonumber \\ 
\Gamma&
 \stackrel {\log\gg1}\longrightarrow
2H\rho_s=\frac{\xi(t)\mu_{\rm eff}(t)}{t^3}\,. 
\end{align}
The last limit holds at late times when the log is sufficiently large, as is the case in the physically relevant 
regime.\footnote{To derive $\Gamma$ we used eq.~(\ref{eq:rhosfree}) although this only applies for long strings. Free sub-Hubble loops are expected to redshift as non-relativistic matter, however since as we saw in the previous section they represent only a small fraction of the total energy density of strings the corresponding correction to $\Gamma$ is small.}

The rate of energy loss from strings can be split into two contributions $\Gamma=\Gamma_a+\Gamma_r$,
corresponding to the rate of energy transfer to axions and radial modes respectively. The corresponding continuity 
equations for the axion and radial energy densities are then
\begin{align} \label{eq:rhodots}
&\dot \rho_a+4 H \rho_a=\Gamma_a+\dots\,, \nonumber \\
&\dot \rho_r+z H \rho_r=\Gamma_r +\dots \,,  
\end{align}
where $z$ is a factor ranging from 3 (for non-relativistic radial modes) to 4 (for relativistic radial modes,
and in the fat string case), and the dots represent subleading contributions from energy transfer via axion-radial mode interactions.

At sufficiently late times in the scaling system's evolution, most of the energy released by the string network is expected to go
into axions, since radial modes are heavy relative to Hubble and harder to excite. The energy density in axions is therefore mostly fixed by conservation of energy. 
In contrast, the number of
axions produced depends on the energy spectrum with which they are emitted by the string network.\footnote{Regardless of their initial energy, cosmic expansion will redshift the momenta of the emitted axions so that most of them are non-relativistic soon after the QCD crossover. Consequently, the spectrum affects the subsequent phenomenology and the final relic abundance only through its impact on the axion number density at that time.} 
Indeed it is the nature of the axion spectrum that is the source of the largest uncertainty 
in the relic abundance of axions produced from strings, and this has been the subject of disagreement
for many years~ \cite{Davis:1985pt,Harari:1987ht,Battye:1993jv,Hagmann:2000ja}. 
Before reporting the arguments underlying the different possibilities, we first review how the axion number density depends on the  properties of the spectrum.

Since strings typically have curvature of order Hubble, the natural expectation --- always assumed in the literature, but never confirmed in simulations --- 
is that the spectrum of axions emitted at each instant is peaked at momenta of order the Hubble parameter at that time. Meanwhile production of modes with momentum below Hubble or above the string core scale is expected to be strongly suppressed. Between these scales an approximate power law is usually assumed, which determines the hardness of the spectrum. 
If the spectrum
is soft, meaning that it is sharply peaked in the IR scale (around Hubble in this case), a relatively large number of axions will be released
to account for the total energy lost by strings. If the spectrum is harder, with a larger UV tail, fewer axions
will be produced, although each will be more energetic. The expectation that the attractor solution is approximately scale invariant corresponds to a prediction that the location of the spectrum's peak relative to Hubble, and the power law fall off, are constant up to possible logarithmic corrections.

From eq.~\eqref{eq:rhodots}, if we neglect the energy emitted in radial modes (and axion-radial interactions), which as we will see is a small fraction, we have that
\begin{align}
\frac{1}{R^{4}(t)}&\frac{\partial}{\partial t}\Big(R^4(t)\rho_a(t)\Big)=\Gamma_a(t)\approx \Gamma(t) \,, 
\end{align}
and therefore the energy density in axions at a time $t$, when they are still massless is
\begin{align} \label{eq:rhoasol}
\rho_a(t)&= \int^t\!\! dt' \left(\frac{R(t')}{R(t)}\right)^4 \Gamma(t') \quad
 \stackrel {\log\gg1}\longrightarrow \quad \frac{\xi(t) \mu_{\rm eff}(t)}{3t^2}\,\log\left 
 (\frac{m_r}{H}\right)
\,,
\end{align}
that is, integral of the energy emitted at each previous instant, appropriately redshifted (and we omitted the initial time in the integral since it is dominated by late times). 

We also introduce the differential energy transfer rate 
\begin{equation} \label{eq:dGammadef}
\frac{\partial \Gamma}{\partial k}\left[k,t\right]:\quad \Gamma(t)=\int dk \ \frac{\partial \Gamma}{\partial k}\left[k,t\right] ~,
\end{equation}
which depends only on the axion momenta $k$, the time, and the core size $m_r$. It is convenient to further split this up as
\begin{equation} \label{eq:Fdef}
\frac{\partial \Gamma}{\partial k}\left[k,t\right]=\frac{\Gamma(t)}{H(t)} F\left[\frac{k}{H},\frac{m_r}{H}\right]\,, \qquad
\int dx\, F\left[x,y\right]=1\,,
\end{equation}
where the function $F(x,y)$ fully characterises the shape of the spectrum (through the variable $x$), and
its time dependence (through the variable $y$).
Combining eqs.~\eqref{eq:rhoasol}, \eqref{eq:dGammadef} and \eqref{eq:Fdef} we get a formula for
the axion spectral energy density
\begin{equation} \label{eq:drhodk}
\frac{\partial \rho_a}{\partial k}\left[k,t\right]=\int^t dt' \frac{\Gamma'}{H'}\left( \frac{R'}{R}\right)^3 F\left [\frac{k'}{H'},\frac{m'_r}{H'}\right ]\,,
\end{equation}
where primed quantities are computed at the time $t'$, the redshifted momentum is defined as $k'=k R/R'$, 
and we have left a possible time dependence in the core mass scale $m_r$ to include the fat string case. 
Eq.~\eqref{eq:drhodk} is just the time integral of the instantaneous spectra appropriately redshifted, and the  change in power of the redshift factors compared to eq.~\eqref{eq:rhoasol} is due to the extra power of $k$ in 
the differential spectrum.

The total number density of axions is therefore given by
\begin{equation} \label{eq:nat}
n_a(t)=\int \frac{dk}{k} \frac{\partial \rho_a}{\partial k}=
\int^t dt' \frac{\Gamma'}{H'}\left( \frac{R'}{R}\right)^3 \int \frac{dx}{x} F\left [x ,\frac{m'_r}{H'}\right ]\,.
\end{equation}

\begin{figure}[t]
\centering
\includegraphics[height=5cm]{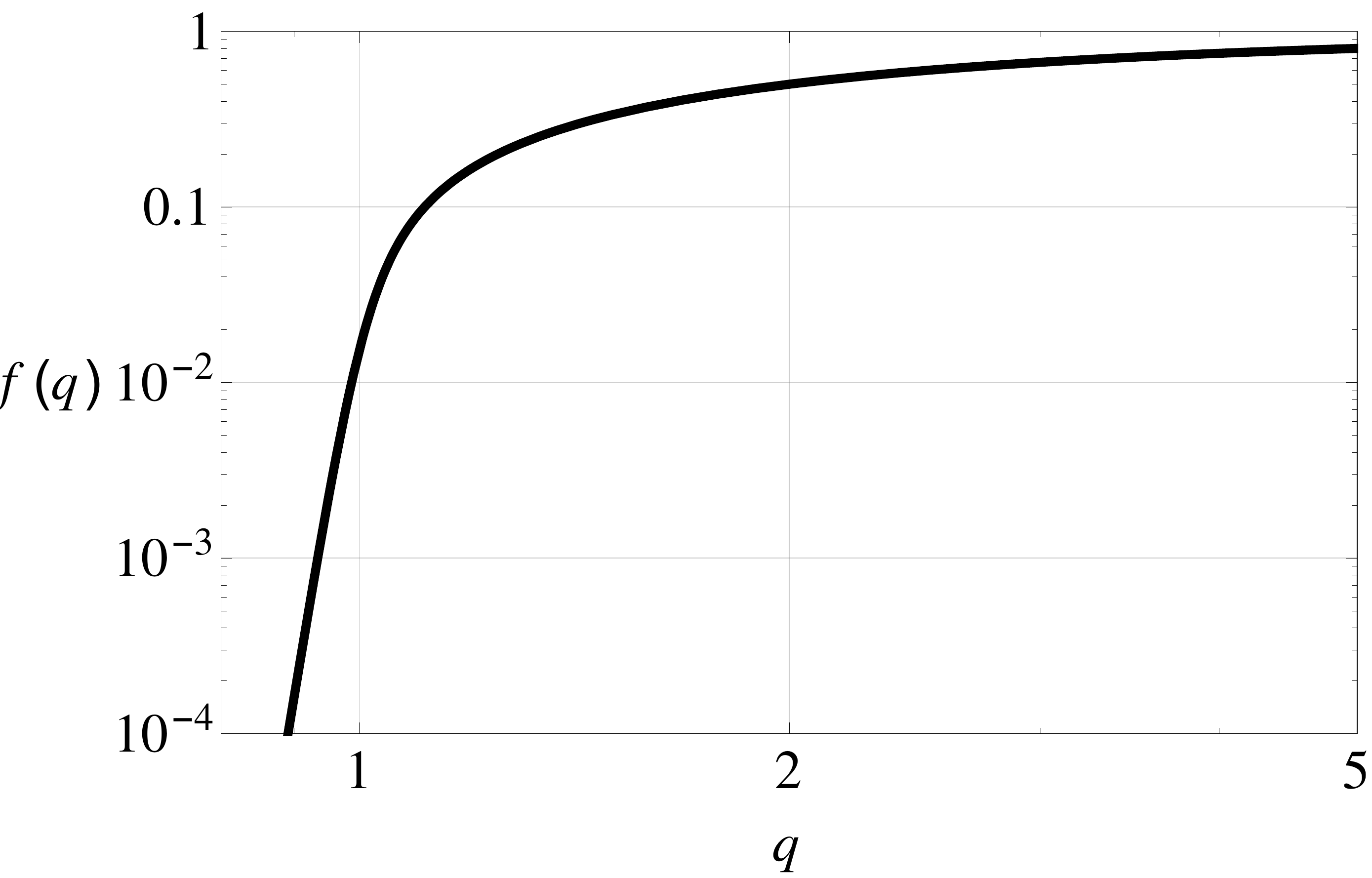}
\caption{\label{fig:nafactor} The dependence of the axion number density, relative to its value in the limit $q\rightarrow \infty$, $f(q)\equiv n_a/(8H\xi\mu_{\rm eff}/x_0)$ on the power $q$ of the spectrum of axions emitted by the string network, at the physically relevant scale separation $\log \left(m_r/H \right) \approx 70$.}
\end{figure}

To  see how the number density depends  on the shape of the spectrum we consider
an analytic form that reproduces the theoretical expectation:
\begin{equation}
F\left[x,y\right]=\left \{
\begin{array}{lc}
\frac{1}{x_0}\left(\frac{x_0}{x}\right)^q \frac{q-1}{1-\left(\frac{x_0}{y}\right)^{q-1}}  & \quad x_0<x<y \\
0 & \quad x<x_0 \vee  x>y  ~,
\end{array} \right.
 \end{equation} 
i.e. a single power law $1/k^q$ with an IR cutoff at $k=x_0 H$ and a UV one at 
$k=y H \approx m_r$ (the extra factors are required to have the right normalisation). 
Substituting this into eq.~\eqref{eq:nat} and taking the large time limit, which corresponds to  
keeping the leading log contributions, we get
\begin{equation} \label{eq:natapprox}
n_a(t)\approx\frac{8 H \xi(t)\mu_{\rm eff}(t) }{x_0} \frac{1-q^{-1}}{1-(2q-1)e^{(1-q)\log(m_r/H x_0)}}\,.
\end{equation}
Given the large size of the log, the last factor strongly depends on whether the power $q$ is larger,
equal, or less than 1 (see Figure~\ref{fig:nafactor}). 
In fact we can rewrite the expression above as
\begin{equation} \label{eq:nathe}
n_a(t)\approx\frac{8 H \xi(t)\mu_{\rm eff}(t) }{x_0} \times
\left \{
\begin{array}{lc}
1-1/q & \quad q>1 \\ & \\
\frac{1}{\log\left(\frac{m_r}{H x_0}\right)} & \quad q=1 \\ &\\ 
\frac{1-q}{q(2q-1)}\left[\frac{Hx_0}{m_r}\right]^{1-q} & \quad  \frac12<q<1 \,,
\end{array}  \right. 
\end{equation}
in which the last factor varies from ${\cal O}(1)$ for $q>1$ to 
${\cal O}(10^{-2})$ for $q=1$, and is exponentially small (in terms of the log) for $q<1$. 

The dependence of the axion number density on $q$ can be easily understood in terms of our previous qualitative discussion. For $q>1$ the spectrum is soft,
most of the energy is emitted with momenta of order Hubble, and the final number density
is of order the total energy density contained in strings, $H^2 \xi \mu_{\rm eff}$, divided by
the average axion momentum ${\cal O}(H)$. 
For $q=1$ energy is equally distributed in logarithmic intervals of momentum.
Therefore, although most of the axions are still emitted with momenta of order $H$, the total number of axions emitted 
is smaller by a factor of the log. For $q<1$ the spectrum is UV dominated, and the majority of the energy is distributed to axions with large momentum
 so that the axion number density is power suppressed by the UV scale. 
 
 The different behaviour of the number density for different choices of $q$ can be linked to the change in the
average momentum of the axions in the spectrum. If we define the inverse average momentum as 
\begin{equation}
\langle k^{-1} \rangle = \frac{1}{\rho_a}\int \frac{d\rho_a}{dk} \frac{dk}{k}\,,
\end{equation}
the number and energy densities are related via $n_a=\langle k^{-1}\rangle \rho_a$. 
Depending on whether $q$ is larger or smaller than unity the average momentum 
is parametrically of order $H$ or $m_r$ respectively. 

The huge ratio of scales $m_r/H$ in the physically relevant part of parameter space results in an enormous range of possible values of $n_a$. 
It is therefore clear that understanding the spectrum is of paramount importance if results obtained at the values of the log accessible in the simulations are to be extrapolated to the physical values.
In particular, even a small change in the behaviour of the spectrum could change the extrapolated value of the relic abundance 
by many orders of magnitude.

We can now identify the main source of disagreement in the literature.  
Refs.~\cite{Davis:1985pt,Davis:1986xc,Battye:1993jv,Battye:1994au} claim
that at late times, when the scale separation is large, the coupling of strings to axions is small and the rate that axions are emitted is suppressed, and as a result the dynamics of axion strings are close to those of local strings.
If this is the case, 
the expectation based on the Nambu--Goto effective 
theory is that loops will oscillate many time before 
emitting their energy, producing a spectrum that is sharply peaked at small frequencies, 
of order their initial size, i.e. Hubble. 
Consequently, they predict that $q > 1$, and that the number density of axions produced by strings in the scaling regime will dominate over the contribution from misalignment,
here taken as a reference value $n_a^{\rm mis}=\theta_0^2 H f_a^2$, with $\theta_0^2\approx 5$.\footnote{In this expression for $n_a^{\rm mis}$, $H$ is the Hubble parameter when the axion mass becomes cosmologically relevant, which is around the time of the QCD crossover although the exact value depends on $f_a$.} 
Setting $\mu_{\rm eff}=\pi f_a^2 \log$ following the theoretical expectation in eq.~\eqref{eq:mueff}, 
$\log\approx 70$ and $x_0\approx 10$ we get
\begin{equation}\label{eq:shell}
\frac{n_a^{q>1}}{n_a^{\rm mis}}\approx \frac{8\pi}{\theta_0^2 x_0}\xi \log \sim 30 \xi\,,
\end{equation}
which ranges from 30 to 300 depending if $\xi$ is taken to be close to 1 or 10.

Conversely, refs.~\cite{Harari:1987ht,Hagmann:1998me} 
 claim that string loops do not oscillate, but instead efficiently shrink emitting all of
their energy at once and producing a spectrum with $q=1$. In this case 
the number density from strings is suppressed
\begin{equation}\label{eq:siki}
\frac{n_a^{q=1}}{n_a^{\rm mis}}\approx \frac{8\pi}{\theta_0^2 x_0}\xi \sim 0.5\, \xi\,,
\end{equation}
 and can even be subleading with respect to that from misalignment if
$\xi$ is taken to be 1.

In the rest of this section we present a detailed analysis of the spectrum obtained from simulations with the aim of understanding which of these possibilities is more likely. We therefore postpone further discussion
 to the end of the section, when we compare our findings with
the existing literature.

To analyse the spectrum emitted by the scaling solution we fixed the initial conditions in simulations
to be as close to the attractor as possible, corresponding to the black curves in Figure~\ref{fig:xit}. 
This isolates the radiation emitted in the scaling regime as much as possible, and reduces contamination from pre-existing radiation. 
In Appendix~\ref{sec:app_init} we show that, starting with different initial conditions, the spectrum and number density converge to those of the scaling solution, so that the results we obtain do not depend on this convenient choice.

\subsection{Energy Budget} \label{sec:energybud}
In analysing the distribution of energy in the scaling solution, and the rate at which axions are produced by strings,
it is useful to organise the total energy density stored in the complex scalar field into three
components, namely
\begin{equation} \label{eq:rho_split}
\rho_{\rm tot}=\rho_s+\rho_a+\rho_r\,.
\end{equation}
Here $\rho_{\rm tot}=\langle T_{00}\rangle$ is the total energy density of the scalar field as given by the average Hamiltonian
density eq.~\eqref{eq:ham1}; $\rho_s$ is the contribution contained in strings; $\rho_a$ is the energy density in axion particles; and $\rho_r$ is that
in radial modes. At early times axions, strings, and radial modes are strongly coupled to each other  
so this separation is ill-defined, but at later times the individual components decouple
 and the separation becomes meaningful. In the scaling regime, the theoretical expectation is that when the energy density in strings is parameterised as in eq.~\eqref{eq:mueff}, $\gamma(t)$ will be of order 1 and vary only slowly with time.

 The way that we actually compute the various components in eq.~\eqref{eq:rho_split} is as follows:  
The axion energy density is calculated from the spatial average of $\dot a^2$
away from the core of the strings (close to the string cores, the motion of strings gives a significant contribution to $\dot a^2$). By using $2 \langle \frac{1}{2} \dot a^2 \rangle$ rather than $\langle \frac12 \dot a^2 + \frac12( \nabla a)^2\rangle$ we avoid the part of the energy that corresponds to the strings' tension, which is mostly contained in
$( \nabla a)^2$. We have checked that our results for $\rho_a$ redshift as expected 
(i.e. as relativistic matter) and that they are robust against different types of string-core masking
(more details about our procedure for screening strings and the consistency tests can be found
in Appendix~\ref{sec:app2}).  The radial energy density is computed by averaging the part of the Hamiltonian density that 
involves only the radial mode, i.e.
$\langle \frac12 \dot r^2+\frac12 ( \nabla r)^2+V(r)\rangle$, again away from the string cores.
Finally $\rho_s$ is simply extracted from the difference $\rho_s=\rho_{\rm tot}-\rho_a-\rho_r$,
which avoids double-counting energy contributions in the other components.
The string energy density defined in this way includes
the energy density stored in the axion-radial interactions, corresponding 
to the terms $(r/f_a+r^2/2f_a^2)(\partial a)^2$ in the Hamiltonian (second line of eq.~\eqref{eq:rhoar}), only part of which (that in regions close to string cores) genuinely contributes to the string energy. The remainder corresponds to interaction energy between axion and radial modes.
Such interactions could in principle trigger parametric resonance and have a substantial effect, however we have checked that they only give negligible oscillating corrections to the energy 
densities in our
simulations.\footnote{The axion-radial interaction terms are small because $|r|/f_a\ll1$ away from string cores, and the amplitude of radial waves is rapidly decreased by redshifting.}

We can compare $\rho_s$ extracted in this way to the prediction obtained using the theoretical expectation for
 the string tension, based on the typical separation between strings, and the measured values of $\xi(t)$. 
In particular, we compute the effective tension $\mu_{\rm eff}=\rho_s(t) t^2/\xi(t)$ from the definition eq.~\eqref{eq:rhos}, using $\xi(t)$ 
and $\rho_s(t)$ from the simulation. This can be compared to the theoretically expected form
\begin{equation} \label{eq:muth}
\mu_{\rm th}=\mu_0 \log\left(\frac{m_r\,\gamma_c}{H \sqrt{\xi}} \right)\,,
\end{equation}
which is obtained by replacing $\gamma(t)$ in eq.~\eqref{eq:mueff}, by a constant $\gamma_c=1/\sqrt{4\pi}$ as a reference (we choose $1/\sqrt{4\pi}$ somewhat arbitrarily based on the average distance between strings if they were all parallel, but any roughly similar value would also be theoretically reasonable).

\begin{figure}[t]\begin{center}
\includegraphics[height=6.cm]{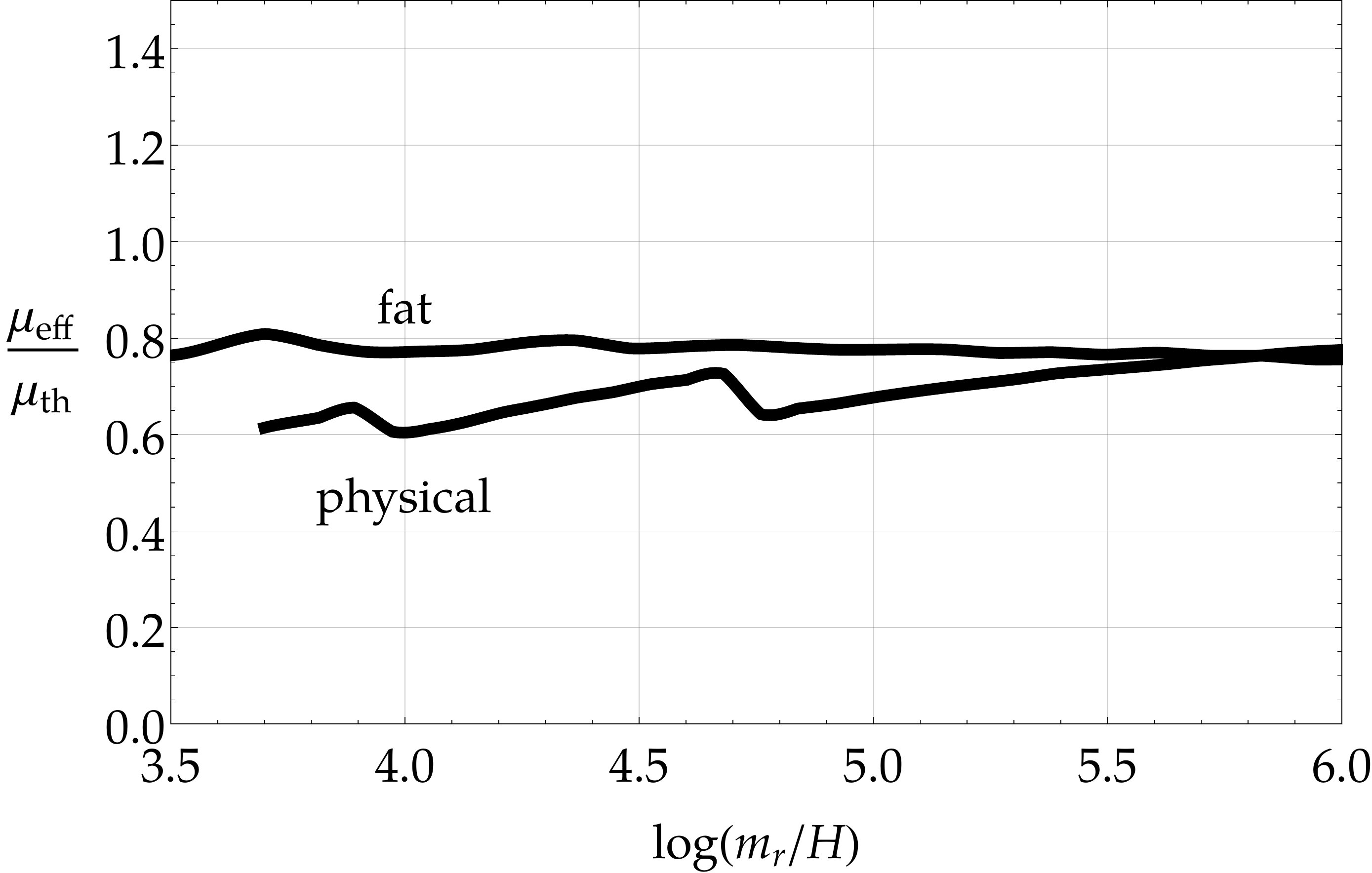} 
\caption{\label{fig:mueff} 
The closeness of $\mu_{\rm eff}=t^2 \rho_s(t)/\left(\xi(t)\right)$ to the theoretical prediction $\mu_{\rm th}$, defined in eq.~\eqref{eq:muth}, (plotted in terms of the ratio of these two quantities)
is a non-trivial check that our procedure to extract the string energy density $\rho_s$ 
is reliable. More importantly, it shows that the relation eq.~\eqref{eq:rhos} can be used to predict the energy in the string network for a given string density and time, by replacing $\mu_{\rm eff}$ by $\mu_{\rm th}$, and that the theoretically predicted logarithmic growth in the string tension is seen in simulations.
}
\end{center}
\end{figure}

In Figure~\ref{fig:mueff}, we plot the ratio $\mu_{\rm eff}/\mu_{\rm th}$ as a function of time, for the fat string and the physical cases. 
The closeness of $\mu_{\rm th}$ and $\mu_{\rm eff}$ over the entire time range that the system is in the scaling regime is highly non-trivial. These quantities could have differed by orders of magnitude, or had different time dependences, but instead the naive theoretical prediction reproduces the results from the simulation to within 20\% throughout. As well as showing that our method of computing $\rho_s$ is meaningful, it is a strong sign that eq.~\eqref{eq:rhos} with $\mu_{\rm eff}$ replaced by the theoretical expectation eq.~\eqref{eq:muth} correctly captures the dynamics of the string system. This includes the logarithmic growth of both $\xi\left(t\right)$ and the string tension due to the increasing scale separation, as well as the variation of $\rho_s$ compared to $\rho_a$ and $\rho_r$ with time.

The small difference between $\mu_{\rm eff}$ and $\mu_{\rm th}$ is not worrisome.
First of all $\mu_{\rm th}$ does not include relativistic effects. These are expected to be small on the basis that the energy density is dominated by long strings, whose motion
is damped by the Hubble expansion; indeed such effects are expected to increase the ratio
$\mu_{\rm eff}/\mu_{\rm th}$ which is instead smaller than (and close to) unity. 
Second, we do not have a reliable way to compute $\gamma$ analytically since its value
is determined by the loop distribution and the shape of the strings, 
which can also depend on time. Further, the parametrisation $\mu=\mu_0\log{(m_r\gamma/H\sqrt{\xi})}$ with an IR  cutoff $\sim m_r/H\sqrt{\xi}$ and constant $\gamma$ applies only to long strings (for loops with radius smaller than Hubble, which make up less than $10\%$ of $\xi$, the tension is expected to be cutoff at smaller values).
A simple modification of $\gamma$ can fix the ratio
$\mu_{\rm eff}/\mu_{\rm th}=1$ at all times, however 
for the moment we are not interested in such a detailed understanding of 
$\mu_{\rm eff}$, and we content ourselves with the degree of agreement obtained in Figure~\ref{fig:mueff}.

\begin{figure}[t
]\begin{center}
\includegraphics[height=5.5cm]{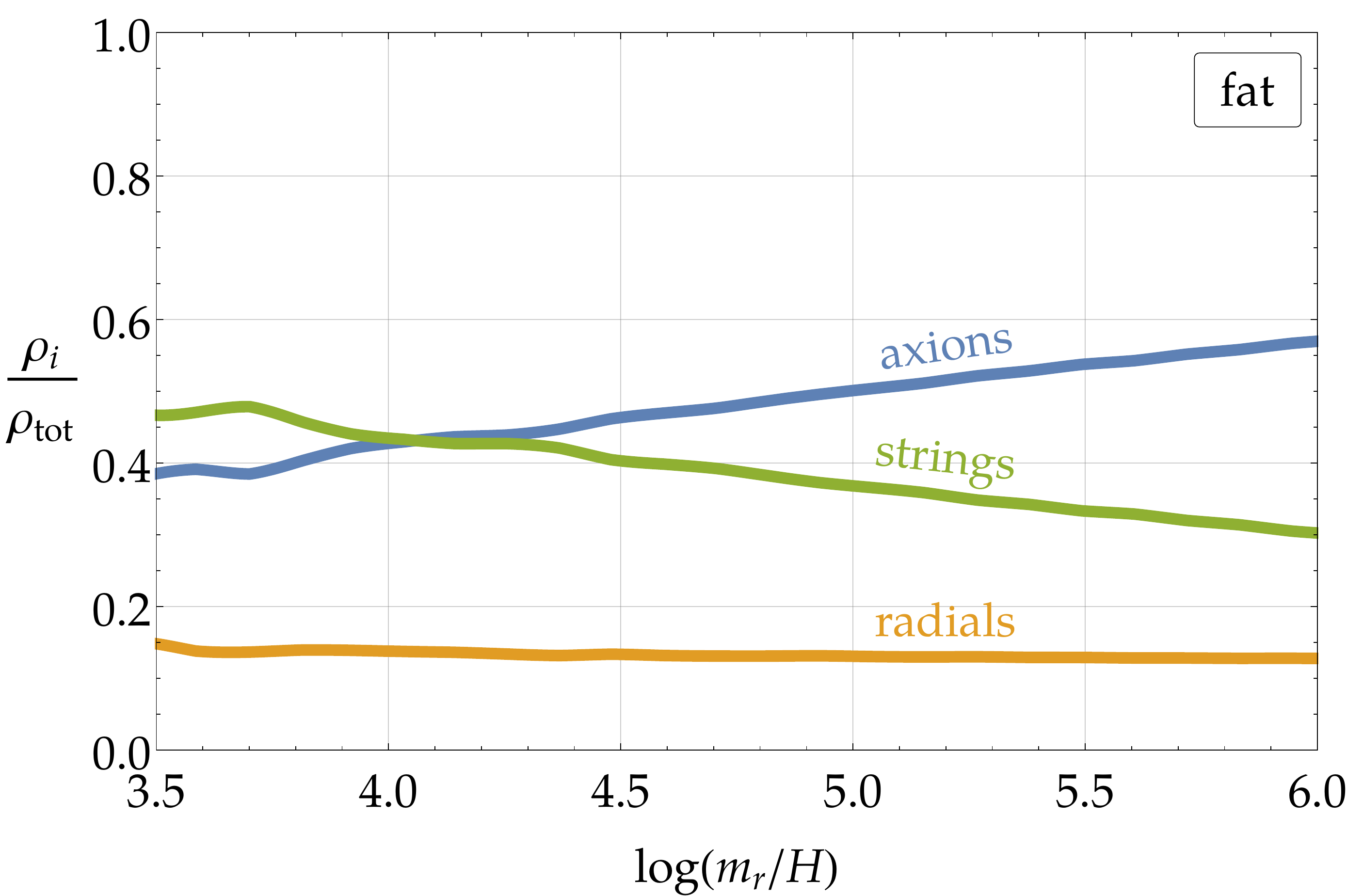}\hspace{0.5cm}
\includegraphics[height=5.5cm]{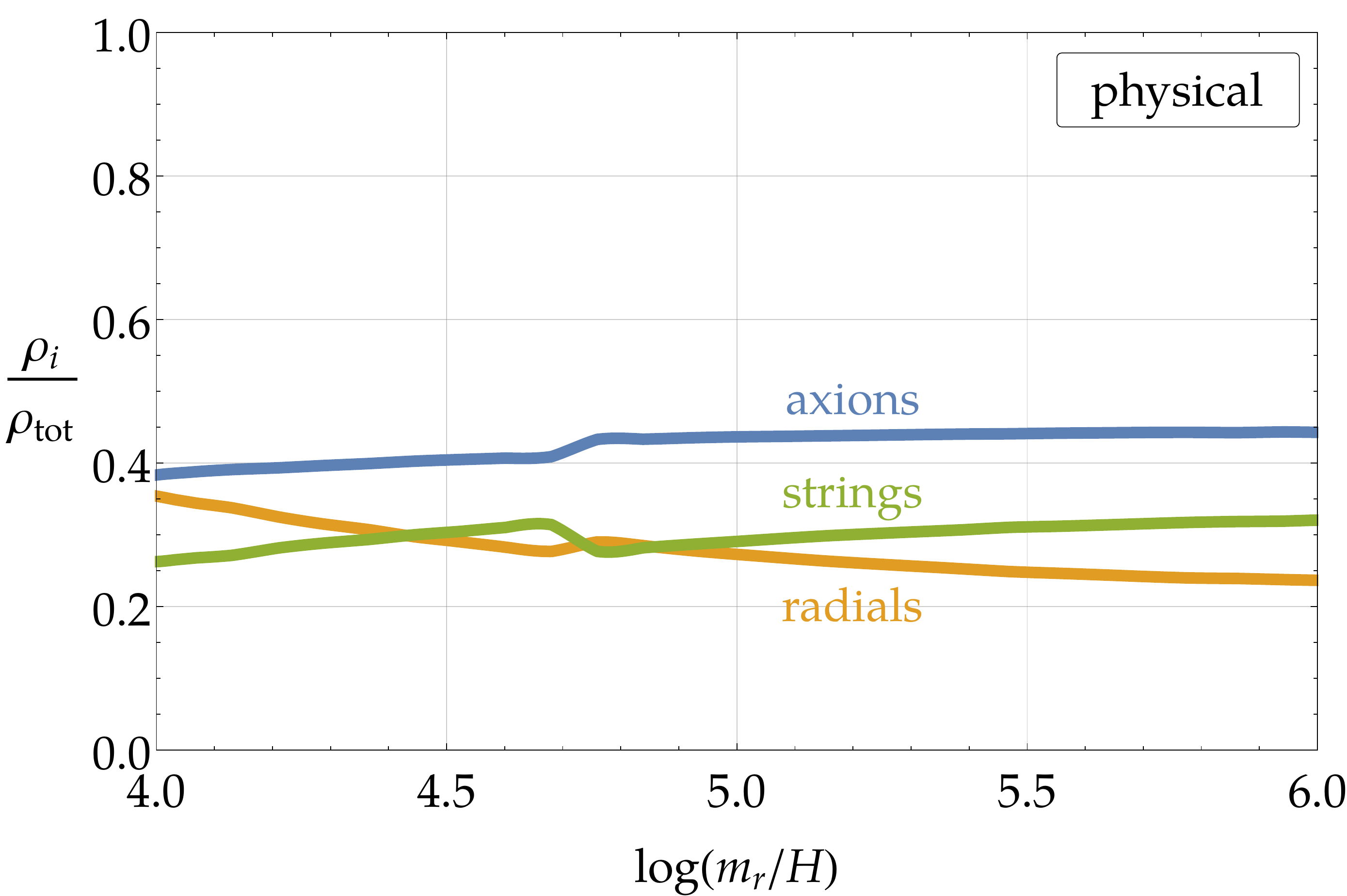}\\ \hspace{0.1cm}
\caption{\label{fig:energybudget} 
The fraction of the total energy density in free axions, radial modes and strings, as a function of time using
the fat string trick (left) and in the physical case (right). The string contribution $\rho_s(t)$ is extracted from
the difference $\rho_s=\rho_{\rm tot}-\rho_a-\rho_r$ as explained in the text.
} \end{center}
\end{figure}

Turning to consider the energy in axions and radial modes, 
in Figure~\ref{fig:energybudget} we show the proportion of the total energy that is in the three components of eq.~\eqref{eq:rho_split} as a function of time, in the fat string and the physical cases. 
We plot this only for times corresponding to the range 
$\log(m_r/H)=3.5\div 6$ for the fat string case, and $\log(m_r/H)=4\div 6$
for the physical one. Data at later time may not be safe from finite volume effects 
(see Appendix~\ref{sec:app2}) while data at earlier times is not representative of the scaling
regime since there is not yet a sufficient hierarchy between Hubble and the core scale to consider
these decoupled (as will be clear when we analyse the axion spectrum, shortly).

First, we note that in the fat string case all three components redshift as $1/t^2$,
up to logarithmic corrections. Therefore the time
dependence in Figure~\ref{fig:energybudget} is only a result of the increase in energy in strings (due to the factor $\xi(t) \mu_{\rm eff}(t)$ growing) and energy transfer from strings to axions and radial modes. Indeed, using eqs.~\eqref{eq:Gammasol} 
and \eqref{eq:rhoasol} these two effects combined predict that $\rho_a/\rho_s \propto \log(m_r/H)$, which  
explains the relative change in $\rho_a$ and $\rho_s$. The proportion of energy in radial modes appears to be approximately constant in time,
accounting for around 13\% of the energy budget.

The situation is less clear for the physical case. The radial mode  now
redshifts as non-relativistic matter $1/t^{3/2}$, slower than the other components. If we use again 
eqs.~\eqref{eq:rhodots} and \eqref{eq:Gammasol}, and assume constant energy transfer rates $\Gamma_i$, then $\rho_r/\rho_a\propto t^{1/2}/(\xi(t)\mu_{\rm eff}(t)\log(m_r/H))$. Therefore, 
 if the rates remain constant until late times, energy in radial modes will eventually dominate that in axions. However, over the time range plotted in  Figure~\ref{fig:energybudget} this expression predicts that $\rho_r/\rho_a$ remains approximately constant, which matches our results from simulations. 
The fact that the string
energy density is not decreasing with respect to that in axions, as happens in the fat string case, may
be an indication that larger scale separations are required for the asymptotic behaviour
$\Gamma\propto \xi(t)\mu_{\rm eff}(t)/t^3$ in eq.~\eqref{eq:Gammasol} to be reached in the physical case, or due to the shorter time that simulations can be run compared to the fat string scenario.

\begin{figure}[t]\begin{center}
\includegraphics[height=6cm]{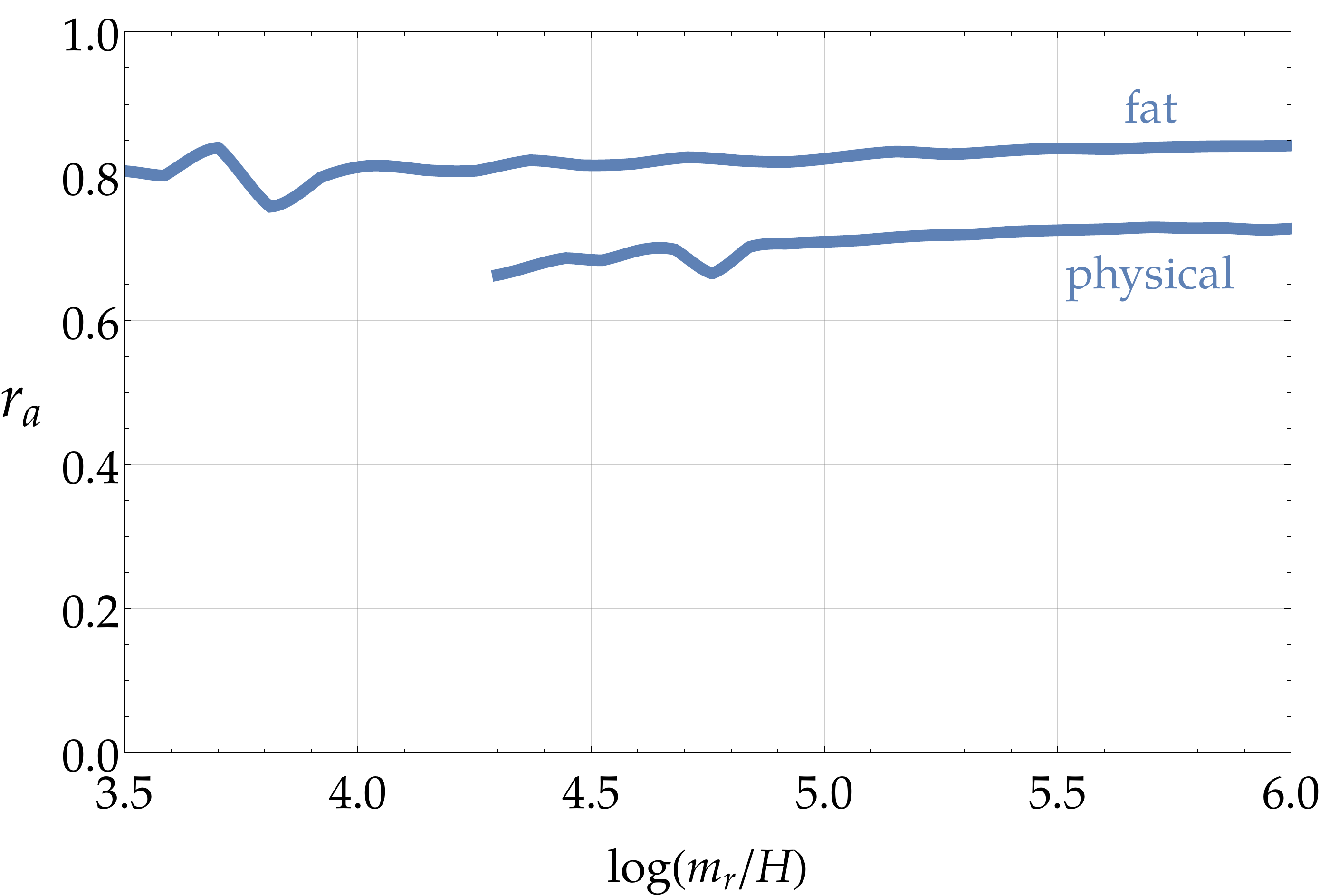}\hspace{0.5cm}
\caption{\label{fig:emissratio}
The fraction of the instantaneous
energy density emitted by strings that is converted into axions (as opposed to radial modes) $r_a=\Gamma_a/\Gamma$ as a function of time, for the
fat string and the physical simulations. At late times more than 70\% (80\%) of the energy released from the string network in the scaling regime goes into axions in the physical (fat string) case. The rest of the energy goes
into the production of massive radial mode states.
}\end{center}
\end{figure}

Perhaps more revealing are the rates at which energy is transferred into axions and radial modes.
In Figure~\ref{fig:emissratio} we show the time dependence of the ratio $r_a=\Gamma_a/\Gamma$ in the simulations, which we compute from eq.~\eqref{eq:rhodots} by taking derivatives of the energy densities calculated as above 
(setting $z=3$ in the physical case, $z=4$ in the fat string one, and neglecting the dots, i.e. including 
possible energy transfer between axions and radial modes in $\Gamma_a$). 
It seems that in both cases the fraction of energy that is transferred to axions is roughly constant, 
and the value in the fat string scenario is compatible with the approximately constant proportion of the total energy that is in radial modes, plotted in Figure~\ref{fig:energybudget}. As discussed, in the physical case a constant energy transfer will eventually lead to radial modes dominating the total energy density, but is compatible with the results in Figure~\ref{fig:energybudget} for the time range that we can simulate.

Similarly to the fraction of the total energy in strings $\rho_s / \rho_{\rm tot}$, the constant value of $r_a$ in the physical case might only be a transient effect. 
In fact we expect that the rate of energy transfer from radial modes to axions will increase if the abundance of radial modes 
grows, so that at sufficiently late times the dominant net effect of energy loss from the string network should be the production of axions. 
Because of this, any interpretation of the fractional rate $r_a$
computed from simulations in the physical case should be taken with a grain of salt.

To summarise, in both the fat string and the physical cases we find that most of
the energy radiated by strings in the scaling regime goes into axions, and these give the largest contribution to the energy density at late times. Meanwhile, the string energy
density is well reproduced by the ansatz from the scaling solution, with the theoretically expected string tension and the values of $\xi(t)$ measured in simulations. There
is a non-negligible component of the energy density in radial modes, 
 and an approximately constant proportion of the energy emitted by strings goes into such states. 
This surprising result clashes with the expectation 
that heavy modes decouple from the evolution of macroscopic soft objects, and that their production is suppressed. 
We will see that
this phenomenon has a close analogue in the rate at which high momentum axions are emitted.

\subsection{Axion Spectrum} \label{sec:axspectr}

We finally arrive at our analysis of the spectrum of axions emitted, the shape of which has a dramatic effect on the axion relic density, as previously discussed.

\begin{figure}[t]\begin{center}
\includegraphics[height=5.3cm]{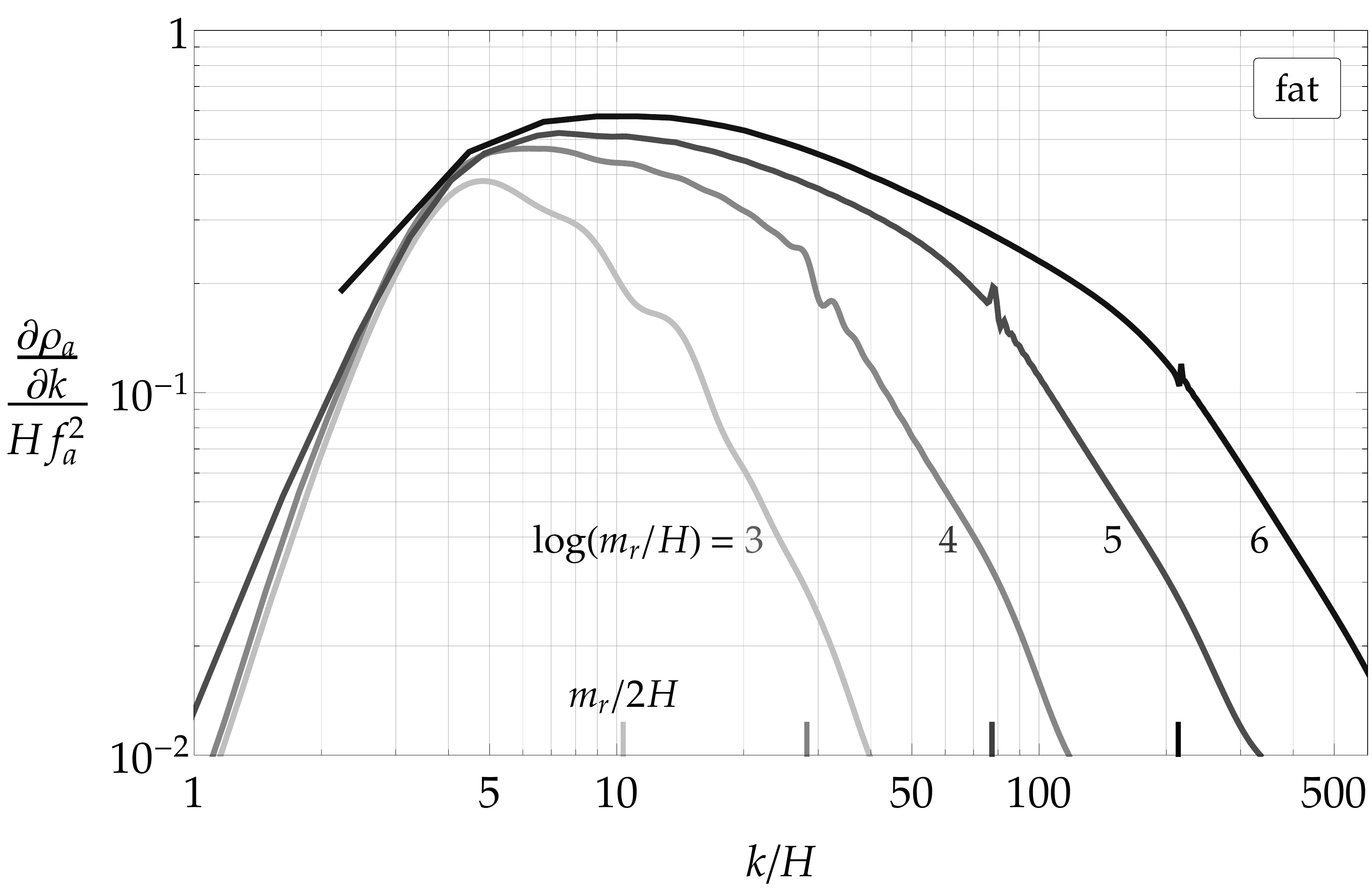}\hspace{0.5cm}
\includegraphics[height=5.3cm]{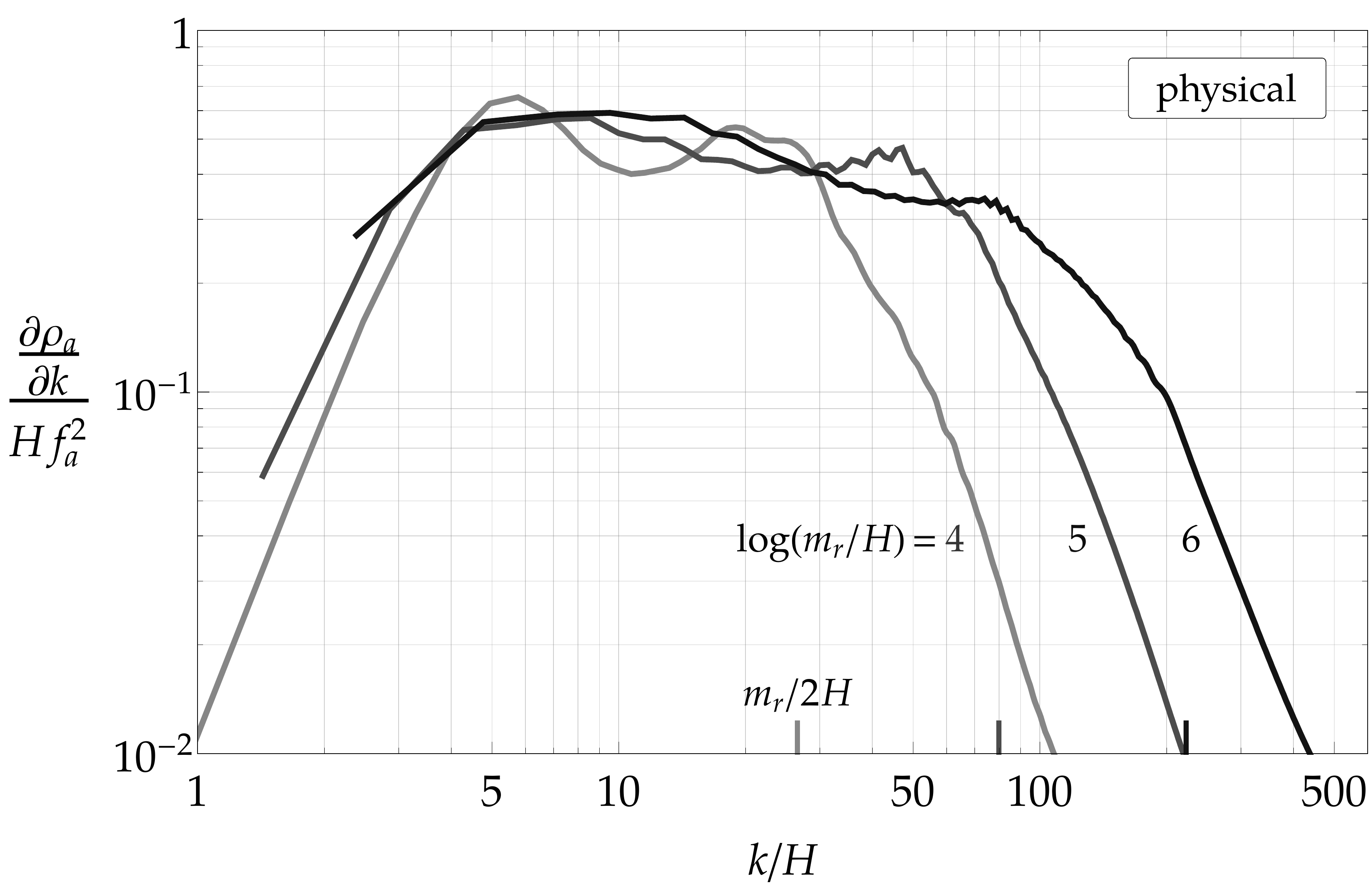} 
\caption{\label{fig:spectra}
The axion energy density spectrum $\partial \rho_a/\partial k$ as a function of the ratio of the physical momentum $k$ to the Hubble parameter. The results are shown for different times (i.e. different values of $\log(m_r/H)$) for a string system in the scaling regime, in the fat string (left) and the physical (right) scenarios. In both cases the spectrum is dominated by a broad peak
at around $k/H=10$, and emission at lower momenta is suppressed. As a reference, for each time shot
we also show the value of $k$ corresponding to the core ($k=m_r/2$), above which there is very little emission, as expected.}\end{center}
\end{figure}

In Figure~\ref{fig:spectra} we show the differential energy density $\partial \rho_a/\partial k$ at different time shots, for both the fat string and the physical cases.
The spectrum is such that $\int dk~ \partial \rho_a/\partial k=\rho_a$,
 and it has been computed from the Fourier transform of $\dot a$, with
the strings appropriately masked out, as explained in Appendix~\ref{sec:app_mask}. 
We plot the spectrum as a function of $k/H$ to 
highlight that it remains peaked around momenta of order Hubble as this decreases,
and we divide it by the factor $H f_a^2$ to remove
the main time dependence (see eq.~(\ref{eq:rhoasol})).  

The spectra have a number of relevant features. 
We start with the fat string case, 
for which the core scale $m_r$ evolves in exactly the same way as physical momenta redshift. 
 As a result, radiation produced by the cores with $k\sim m_r$ remains at $k\sim m_r$ at all later times, and contributes to the same part of the differential spectrum regardless of when it was originally produced. 
Similarly, since the simulations begin with $H\sim m_r$, the typical momentum of pre-existing radiation from the initial conditions is $k\sim m_r$ and states from these early times will subsequently remain at $\sim m_r$. 
Therefore the part of the spectrum sufficiently below $m_r$ is entirely 
from genuine radiation produced by strings, and is independent of the physics of their cores. 

Figure~\ref{fig:spectra} clearly shows that the spectrum is peaked in the IR at around the Hubble scale, 
in particular at $k/H\sim 5\div10$ for all values of the log larger than 3. 
There is another small peak in the UV, which is exactly at $k=m_r/2$, the typical frequency 
of parametric resonance. 
These modes could be produced by the string core itself, or by the 
conversion of non-relativistic radial modes into axions. The position of this peak at a particular time
identifies the part of the spectrum that is sensitive to core scale (i.e. UV) physics. For times corresponding to $\log(m_r/H)\lesssim 3$ the peaks in the UV and the IR cannot be distinguished, and it is only for
$\log(m_r/H)\gtrsim 3.5$ that there starts to be enough of a hierarchy to justify considering the IR
dynamics as decoupled from the core scale physics. For this reason we do not consider smaller values
of the log in our analyses involving the spectrum. 

At late times the size of the simulated
box, $L(t)$, starts cutting off the low momentum part of the spectrum, as can be seen from the interruption of 
 the spectrum at the minimum non-trivial $k=2\pi/L(t)$ in the final shots. Subsequently this would 
start altering the IR peak, so we only analyse the spectrum until
$\log(m_r/H)= 6$ when this effect is still harmless (a more detailed study of the various
systematics can be found in Appendix~\ref{sec:app2}).

As expected, the spectrum is power suppressed 
for momenta smaller than the IR peak or larger than the UV peak --- long wavelength modes are inhibited 
by the horizon, while the high energy ones are suppressed by decoupling. In particular, the spectrum seems to fall as $\sim k^{3}$ in the far IR, and as $\sim 1/k^{2}$ in the far UV. 
The region of physical interest is in between the two peaks. 
The spectrum reaches a stable form towards the end of the simulation, and 
in the last Hubble $e$-folding (i.e. for logs in the interval 5 to 6) its shape remains very similar,
modulo the shift of the UV peak. However, even though $\partial \rho_a/\partial k$
is largest in the IR, its area is dominantly in the UV, since the slope between the two
peaks is less steep than $1/k$. This can be seen more clearly by plotting $\partial \rho_a/\partial \log k$
on a log scale (shown in Figure~\ref{fig:spectralog} in Appendix~\ref{sec:app_scaling}). 
We conclude that although most of the axions produced by the evolution
of the string network  at late times are soft, the majority of the energy density is contained in UV
axions with energy of order the inverse core size.

The situation for the physical case seems similar, although the uncertainties are larger. In fact
it is in studying the spectrum that the advantages of the fat string trick are most pronounced. 
In the physical case, axions produced at early times with momentum of order the core-scale redshift with respect
to $m_r$, which is constant. This means that, although the distance between the IR peak and $m_r/2$ at late times is the same as in the fat string case (for the same value
of $\log(m_r/H)$), the spectrum
is contaminated by core-scale radiation produced earlier and redshifted down to $k\sim \sqrt{H m_r/2}$. This effectively
halves the region of the spectrum that is free from UV dependent contributions. Indeed the 
spectrum shows that the IR and the UV dynamics are not decoupled before logs of order 4.5 or 5,
 and for this reason we do not consider quantities that rely on the spectrum at times corresponding to log~$<4.5$.
Similarly, the UV peak is now replaced by a broader feature, caused by the convolution of core-scale radiation emitted at different times and 
redshifted by different amounts. 

Because of these disadvantages, extracting the shape of the spectrum between 
the peaks is more challenging in the physical case.  The rest of the spectrum shows similar features
to the fat string case, with a strong suppression of modes below $k\sim \left( 5\div10 \right) H$ and above $k=m_r/2$ (for the UV modes, this time with a stronger suppression, $\sim 1/k^{3}$ instead of $\sim 1/k^{2}$). 
As for the fat string case, the IR peak seems to dominate the spectrum, especially at late times, while the
area is still dominated by the UV region.  
Therefore, most of the axions 
produced by the evolution of the string network at late times are again soft, but most of the energy density is in UV axions. However, we stress that 
 the results in 
the physical case are less clear and should be interpreted with caution. 

To the best of our knowledge, the only other serious attempt to extract the axion spectrum in the
scaling regime was in ref.~\cite{Hiramatsu:2010yu}, 
based on results of simulations carried out on a somewhat smaller grid than ours. In that paper the authors
observe an exponential suppression of the spectrum at large momenta, which they interpreted
as indicating a strongly IR peaked distribution. However, the range of momenta that shows
such behaviour seems to lie above the scale of the core, $m_r/2$. 
Unfortunately, the region at smaller momenta has been very poorly binned, 
so that little information on the actual behaviour of the spectrum in this region is available.

\subsubsection{Instantaneous Emission} \label{sec:instant}

In addition to the overall spectrum, the shape of the instantaneous axion spectrum, i.e. the function $F(x,y)$ in eq.~\eqref{eq:Fdef}, is 
crucial for understanding the properties of the emitted axions, and in particular for inferring the evolution of the axion number density at later times.

\begin{figure}[t]
\begin{center}
\includegraphics[height=5.3cm]{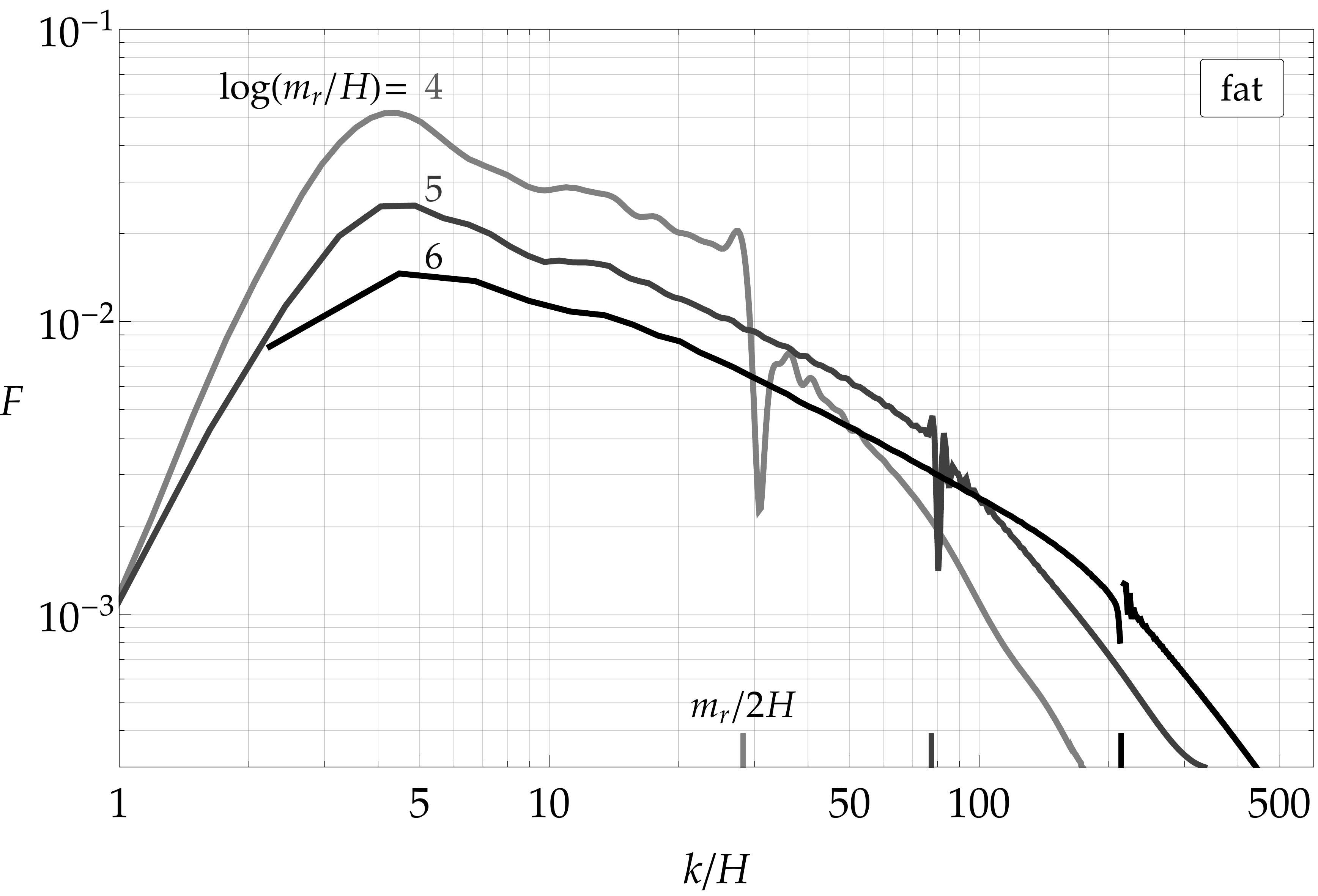}\hspace{0.5cm}
\includegraphics[height=5.3cm]{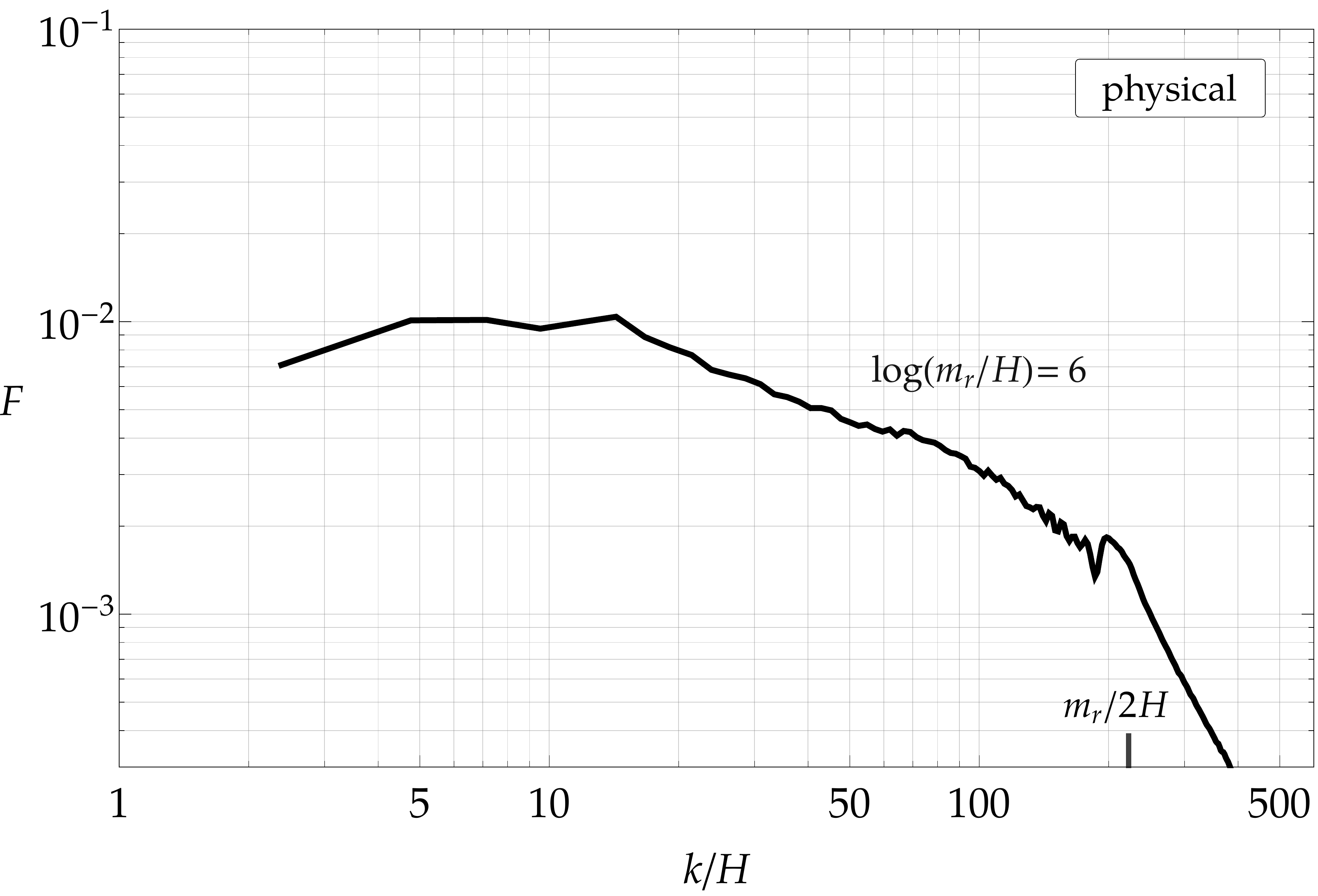} 
\caption{\label{fig:Fs}
The shape of the spectrum of axions emitted instantaneously by the string network $F(k/H,y)$
for the fat string (left) and the physical (right) cases, for different scale separations, i.e. different values of $y=m_r/H$.
}\end{center}
\end{figure}

We compute $F$ from the spectrum by inverting eq.~\eqref{eq:drhodk}, namely
\begin{equation}
F\left [\frac{k}{H},\frac{m_r}{H}\right ] = \frac{A}{R^3}
\frac{\partial}{\partial t} \left ( R^3 \frac{\partial \rho_a}{\partial k}\right) ~,
\end{equation}
where the factor $A=H/\Gamma$ is fixed by requiring that $F$ is normalised to 1. 
To evaluate the derivative numerically we took the difference of $R^3 \partial \rho_a/\partial k$ between two subsequent time shots (separated by $\Delta {\rm log} = 1/4$).
The results are shown in Figure~\ref{fig:Fs}. Since interactions with radial modes
induce small oscillations in time with frequency $\sim m_r$ of the axion energy density 
(see Appendix~\ref{app:compenerg}) 
the procedure to extract $F$ is subject to fluctuations at
frequencies near the core, as evident in the plots (as explained in Appendix~\ref{app:compenerg}
this effect is more pronounced for physical strings than for fat ones).

For the fat string case we plot $F$ at three time shots, i.e. at
three values of the log. 
At all of these times $F$ has an IR peak at momentum around the Hubble scale (and suppressed emission at lower momenta), and a UV peak at the scale of the string cores. 
In particular, the IR peak is located around $k/H\approx 5$ and extends to $k/H\approx 10$, while the UV peak is at $k\sim m_r/2$. As a result, 
the ratio between the maximum and the minimum momenta 
 between which the instantaneous spectrum can show a power law behaviour is bounded by
$k_{\rm max}/k_{\rm min}\lesssim m_r/40H \lesssim N/120$, where $N$ is the number of points
in the spatial grid of the simulations (assuming a lattice spacing $a=m_r^{-1}$ and that the simulations are stopped when the box size $L=3/H$, and also requiring $k/H > 20$ to be safely away from the IR peak). For our simulations this means that $k_{\rm max}/k_{\rm min}\sim 10$. 

Although the separation of the UV and IR peaks is not yet large for $\log=4$, 
at values above $5$ an intermediate momentum range can be clearly identified in which the differential spectrum shows a definite power law behaviour. In this region the instantaneous  spectrum of axion emitted is compatible with a behaviour 
$1/k^q$ with $q\approx 0.7\div 0.8$.
More details about
how we extract $q$, and further plots, can be found in Appendix~\ref{sec:app_scaling}. 
 Consistent with our analysis of the convoluted spectrum, 
this value means
that most of the energy released by the string network goes into high energy axions, although
most of the axions are soft. 

We also note the constant form of $F$ at different times (over a range in which $H$ changes by $2$ orders of magnitude) is another demonstration of the scaling behaviour of the system. In particular, the IR and UV peaks remain their expected positions, and the intermediate power law with $q\approx 0.75(5)$ is constant with time to within the uncertainty. Although there is no noticeable change in $q$ in the interval of logs between 
5 and 6, the uncertainty is relatively large. Consequently, we cannot exclude the possibility that, as for $\xi$, a logarithmic scaling violation occurs. This could lead to, for example, a behaviour $q\sim0.7+ \epsilon \log (m_r/H)$ for some small constant $\epsilon$. In Appendix~\ref{sec:app_init} we also show that $F$ is independent of the string systems initial conditions, confirming that we are indeed analysing the properties of the attractor solution.

As expected, the analysis of the physical case is more difficult. UV modes pollute the convoluted
spectrum well below the scale $m_r$, so useful results only start appearing at
late times. Because of this, in Figure~\ref{fig:Fs} we only show $F$ for the last time shot at $\log=6$. 
Although there are still large fluctuations due to higher frequency modes, an approximate power law
can be recognised at this time, with a value of $q$ that is again smaller than 1 and which appears to be similar to in the fat string scenario.

The form of the instantaneous emission found above clashes with either of the theoretical expectations discussed in eqs.~\eqref{eq:shell} and~\eqref{eq:siki}, which predicted $q\geq1$. Similarly to the radial modes, axions with momentum of order the string core scale have not decoupled from the dynamics of strings at the final times $\log(m_r/H)\lesssim6$, despite the relatively large hierarchy $m_r/H\sim 500$. To investigate this surprising result further, in Appendix~\ref{sec:app_eft} we study the dynamics of the collapse of a single axion string loop. In particular we analyse whether this is converging to the prediction for a Nambu--Goto string in the limit that its initial radius is large compared to its core size. We find that the radius of the axion string loop as a function of time during its initial collapse does indeed get closer to the cosine prediction (and is extremely close to the prediction based on the effective theory of global strings coupled to radiation in the limit of large scale separation \cite{Dabholkar:1989ju}).
At the scale separations that can be simulated ($\log(m_r/R_0)\lesssim 5$ where $R_0$ is the initial loop radius)  the loop does not rebound significantly, as opposed to the 
case where $\log(m_r R_0)\gg1$ in which the loop is expected to rebound many times according to arguments based on the use of the Nambu--Goto string action~\cite{Vilenkin:2000jqa}. On the other hand, whether the loop bounces also depends on its dynamics at small radius, a regime in which the Nambu--Goto approximation is not valid. Since it is at such times that energetic modes are more efficiently produced, this result is consistent with our observation that the spectrum also deviates from the expectation based on a loop oscillating many times.

\subsection{Number density}
\begin{figure}[t]\begin{center}
\includegraphics[height=7.cm]{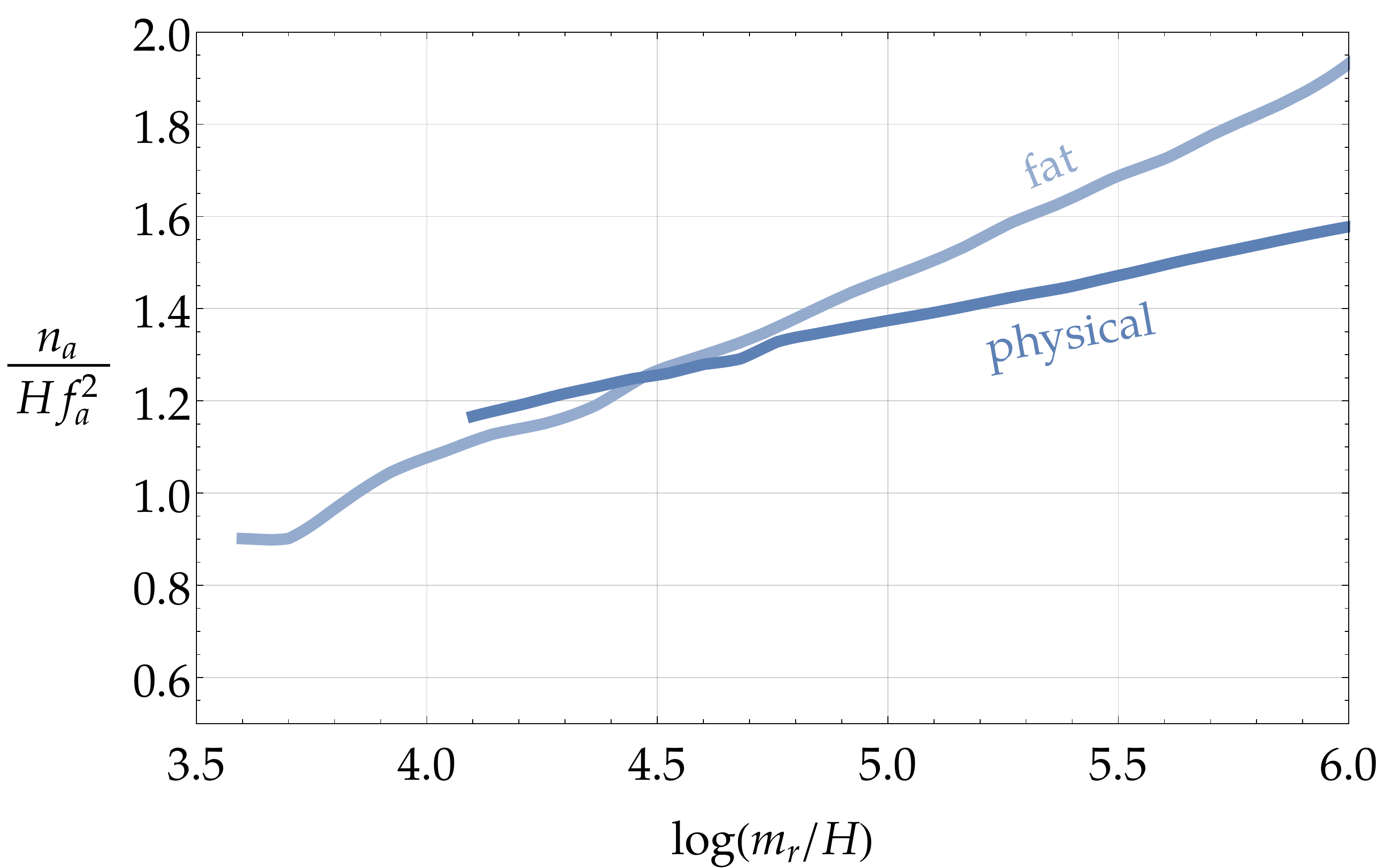} 
\caption{\label{fig:na}
The axion number density, normalised by $H f_a^2$, as a function of time for the fat string
and the physical cases.
}\end{center}
\end{figure}

We are now ready to consider the axion number density. We evaluate this in simulations
 using the differential spectrum $\partial \rho_a/\partial k$ at different time shots, and eq.~\eqref{eq:nat}.
The results for $n_a$, normalised with the factor $H f_a^2$, as a function of time 
 are shown in Figure~\ref{fig:na} for the fat string and the physical scenarios. 
In both cases $n_a/H$ increases logarithmically over the time range plotted,
although the values and the slopes are different.
 Therefore, similarly to the fit of $\xi(t)$, our first conclusion is that at the quantitative level the fat string system
 differs from the physical theory, although they seem to have the
 same qualitative features.

 As we have stressed, the values of the log that can be analysed with numerical simulations
are nowhere near large enough to reach the physically relevant region of parameter space. 
Consequently, $n_a$ must be extrapolated 
 to extract predictions, and the way that this is carried out has a dramatic impact on the results obtained. Such an extrapolation is a viable possibility because we have shown the existence of an attractor solution, and we have found that the energy of the string network is accurately reproduced by eq.~\eqref{eq:rhos} with the theoretical prediction for $\mu_{\rm eff}$, eq.~\eqref{eq:muth}. Additionally, given our results for $\xi(t)$ it is plausible that the fit in eqs.~\eqref{eq:xifit} and \eqref{eq:alphafit} can be extended all the way to the physical scale separation. 
 The remaining component needed in order to predict the axion relic abundance is an understanding of the form of the instantaneous emission spectrum at late times.

As usual
we will discuss the fat string case before moving to the physical one.  
From Figure~\ref{fig:na} it is tempting to extrapolate $n_a/\left(Hf_a^2\right)$ linearly in the log, 
however this procedure is too naive. 
From eq.~\eqref{eq:nathe}, a linear behaviour is expected only for very large 
logs and even then only if the power law of the instantaneous emission spectrum is $q=1$ (and $\xi(t)$ also increases linearly with the log). 
 This
is not the case in simulations since, as we saw in the previous section,  $q\sim 0.75$ 
and the logs are not large enough to neglect $1/\log$ corrections in eqs.~\eqref{eq:Gammasol} and \eqref{eq:nat}. We conclude that the linear 
behaviour observed is most likely a transient effect. Indeed, should the power $q$ remain
constant below 1 the number density would start decreasing exponentially
with the log at late times (from eq.~\eqref{eq:nathe}). Meanwhile, if $q$ grows to be significantly larger than $1$ the number density will increase as the $\log$ squared. 

\begin{figure}[t]\begin{center}
\includegraphics[height=6.4cm]{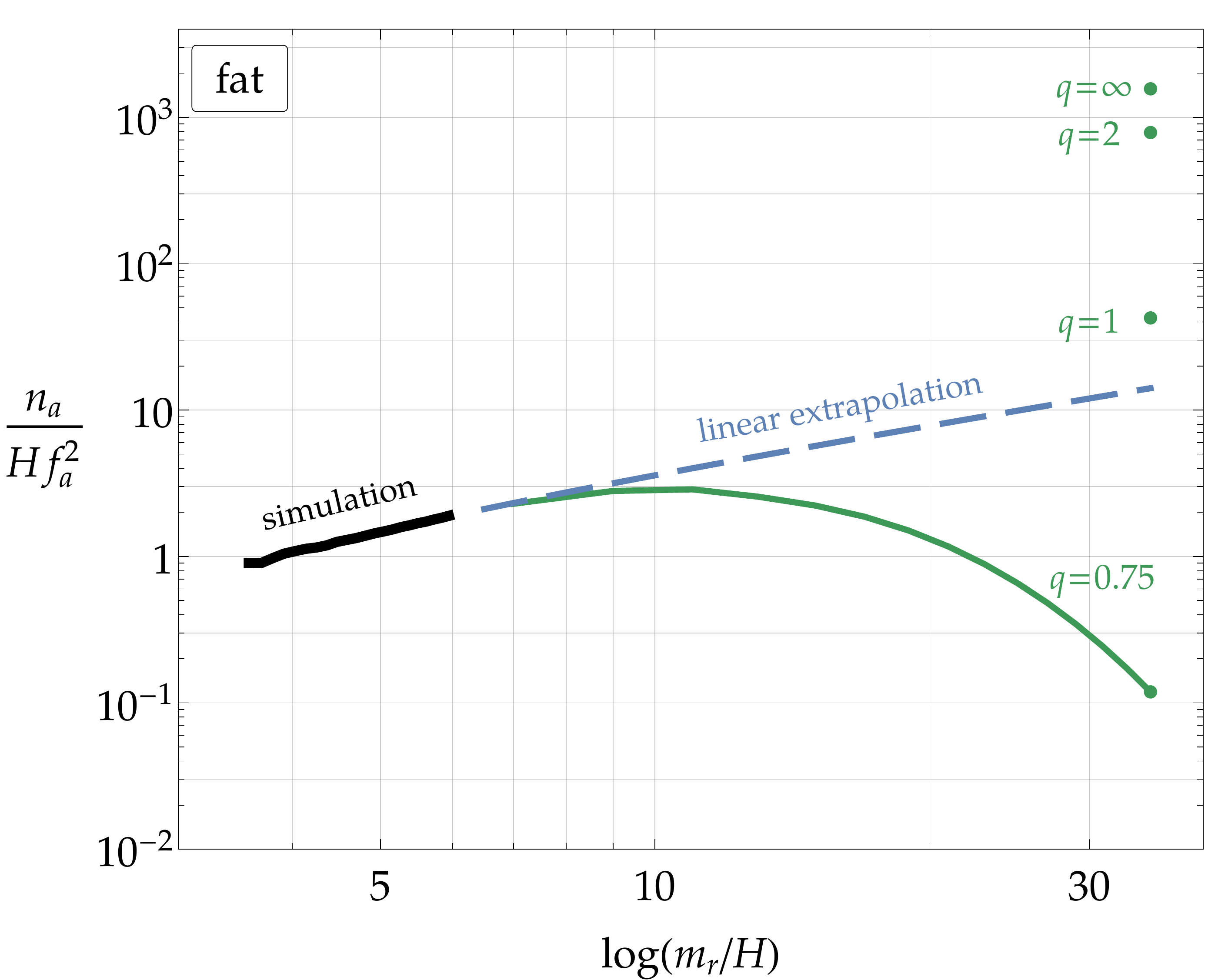}\hspace{0.5cm}
\includegraphics[height=6.4cm]{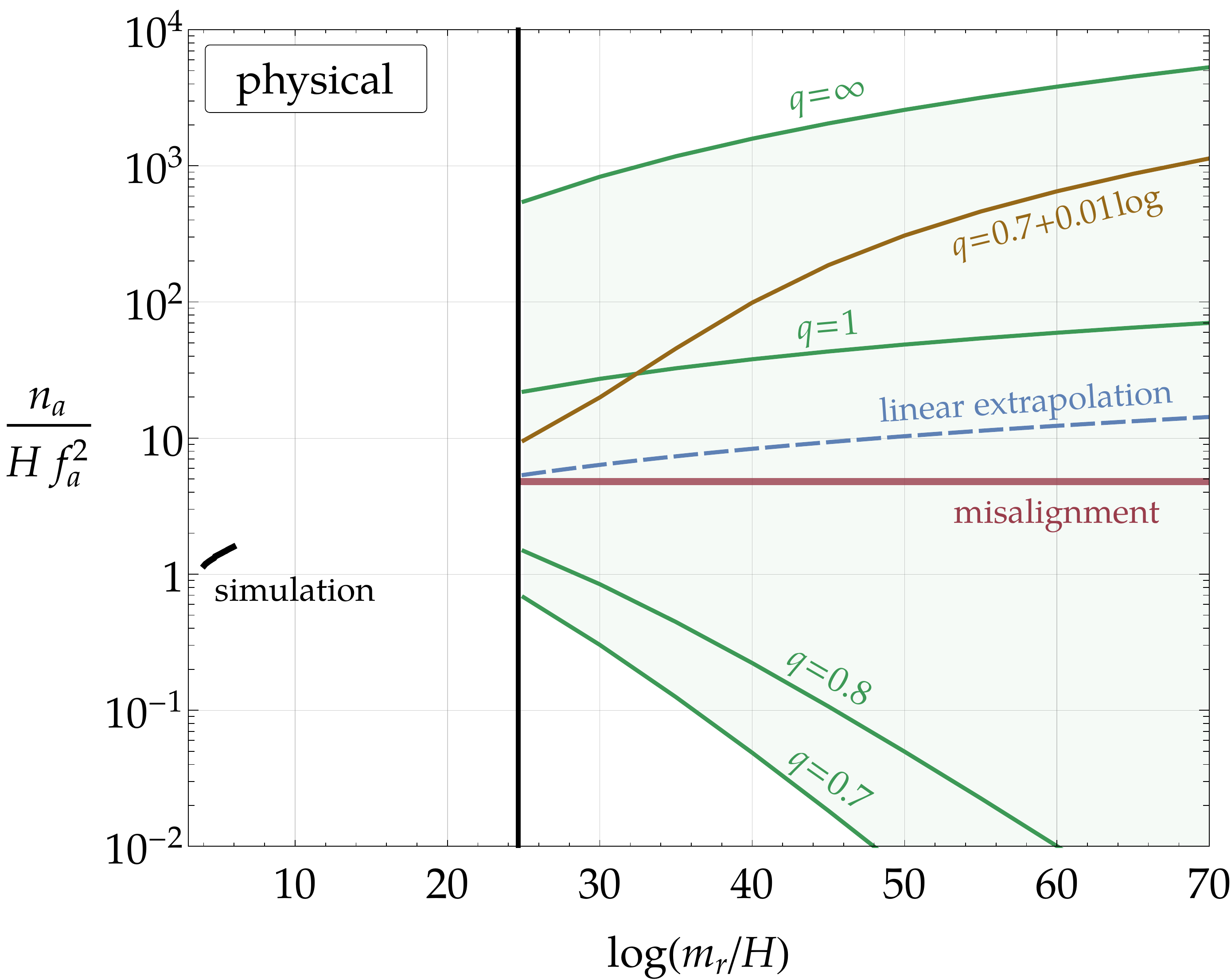} 
\caption{\label{fig:naextrap}
Possible extrapolations of the axion number density to large values of the log with different assumptions
about the late time behaviour of the power $q$ of the instantaneous emission spectrum. For the fat string case 
(left) we compare a naive linear extrapolation of $n_a$ to that assuming $q=0.75$ remains constant.
 We also show the late time number densities if $q$ changes to $1$, $2$, or $\infty$ at an intermediate value of the log.
For the physical case (right)
 we only show extrapolations in the range log~$\gtrsim 23\div 70$, corresponding roughly to the values of  $m_r$ in the range keV$\div 10^{11}$GeV, which are not excluded by fifth force experiments and star cooling bounds. 
 In addition to constant values $q=0.75$, $1$, $2$ and $\infty$, we plot the effect of a small logarithmic scaling violation $q = 0.75+ 0.01 \log\left(m_r/H \right)$.}\end{center}
\end{figure}

The different behaviours at large values of the log
are shown in Figure~\ref{fig:naextrap}, assuming that $\xi$ continues to increase logarithmically. 
 It is clear that a naive linear extrapolation
 leads to significantly different results compared to 
 if $q$ is assumed to remain constant at $0.75$. 
 Since we are not able to reliably study the behaviour of $q$ with time, we cannot 
exclude the possibility that a small scaling violation results in $q$ growing to 1 or even larger values, and we also show the axion number densities at large values of the log in these cases. 
It can be seen that if such a change does occur, the large log behaviour of $n_a$ is completely different to both the linear extrapolation and the $q=0.75$ possibilities, leading to much larger number densities.  The extrapolations in
Figure~\ref{fig:naextrap} have been obtained from eq.~\eqref{eq:nat} using a form for $F$ fitted from the results shown in Figure~\ref{fig:Fs} with $q$ modified between the IR and UV peaks, although the results are not very sensitive to the exact shape of $F$ away from the power law region (more details on the form of $F$ used are given in Appendix~\ref{sec:app_scaling}).

For the physical case there is even more uncertainty in the value of $q$, and consequently on the late time number density. 
 In Figure~\ref{fig:naextrap} (right)
we plot the extrapolated number density for different values of 
$q$, together with the naive linear extrapolation and the available values from the simulation.  As a comparison the reference number density from misalignment ($n_a^{\rm mis}=\theta_0^2 H f_a^2$ with $\theta_0=2.2$)  is also shown.
 Similarly to the fat string case, if $q=1$ then $n_a/\left(H f_a^2 \right)$ increases proportionally to the $\log$,  if $q\gg 1$ it grows with $\log$ squared (again assuming that $\xi$ continues to increase logarithmically), while if $q<1$ it is exponentially suppressed in terms of the $\log$.

It follows that if the spectrum remains UV dominated with small $q$ (i.e. with $q\lesssim 0.85$) 
for large values of the log, the number of axions produced from strings during the scaling regime is negligible.
However, if $q$ increases at larger log, the axions produced by strings can easily match
or dominate those from misalignment. Indeed, even a very mild 
log dependence of the power $q$, for example of the form $q=0.7+0.01\log(m_r/H)$, too small to be excluded by our simulations, would be enough to make the final abundance at $\log=70$ more than 2 orders
of magnitude larger than the misalignment contribution.

As discussed in Section~\ref{sec:sim}, the results of simulations 
only depend on the ratio $m_r/H$, and can be interpreted either in terms of a theory with $m_r \sim f_a$ at an early time when $H \sim f_a$, or a theory with a much smaller $m_r \ll f_a$ at a later time $H \ll f_a$. 
 Fixing the Hubble scale to its value when the axion mass becomes relevant $H \approx m_a \sim \Lambda_{\rm QCD}^2/ M_{\rm Pl}$, the  values of $n_a$ calculated in our simulations can therefore be used to extract the physically relevant axion number density for theories with extremely small values of $m_r$ (as would occur in a UV completion with a complex scalar field that had a tiny quartic coupling). However, constraints from observations of the evolution of stars and fifth force experiments require that $m_r\gtrsim$~keV in viable theories, corresponding to rather large values of the log ($\sim 23$ for the reference value of $f_a=10^{11}~$GeV), which is still beyond the reach of simulations. Due to this phenomenological requirement, we only begin the extrapolations in Figure~\ref{fig:naextrap} at this scale separation. Meanwhile, the typical axion models with $m_r \sim f_a$ correspond to logs $\approx 70$ at the physically relevant time as usual, at the right hand boundary of the plot.

The main message from this section is that an understanding of the late time evolution of the properties of the string network is of paramount importance for a correct extrapolation 
of the axion abundance to the physically relevant parameter range. 
 If only small-log data are available, very precise computations 
of the spectrum and its time dependence will be necessary, or if theoretical arguments that $q\rightarrow \infty$ in the limit of large scale separation are believed, a prediction for the relic abundance can already be made. 
In contrast, a naive extrapolation of the axion number density (such as the linear-log one plotted above) 
could easily give results that are off by several orders of magnitude.

Our final results for the axion number density calculated in simulations are not in disagreement with those obtained by other groups.
 Despite this, the conclusions that have been reached in the previous literature about the abundance at the physically relevant scale separations are incompatible with each other, primarily due to  
 the different ansatz used in extrapolating. In particular:
\begin{itemize}
\item
Refs.~\cite{Davis:1985pt,Davis:1986xc,Battye:1993jv,Battye:1994au} assume that $q>1$ based on arguments related to the evolution and emissivity 
of effective Nambu--Goto strings, which are expected to reproduce the dynamics of global strings
at large logs. Assuming $\xi={\cal O}(10)$ they conclude that strings should produce a large number of axions. Our result would only be compatible with this estimate if the power $q$ of the axion
spectrum increases at large scale separation, 
 however we are not aware of
any reliable evidence that this is indeed the case beyond the Nambu--Goto approximation.
 In fact, as mentioned, energetic modes are expected to be efficiently produced from collapsing loops when their radius is comparable to the string thickness (or by long strings that are similarly close together), and in this regime the Nambu--Goto effective description is not valid.
 
\item
Refs.~\cite{Harari:1987ht,Hagmann:1998me}
 on the contrary assume that $q=1$, supported by computations of the 
axion spectrum produced by collapsing loops at values of the $\log\lesssim 5$. 
Using $\xi={\cal O}(1)$ they conclude that the abundance of axions produced by strings is not
larger than that from misalignment. While the value of $\xi$ assumed is compatible
with those measured in our simulations, the time evolution of $\xi$ that we observe suggests that larger values
need to be used in the physical regime. Additionally, the study of the spectrum from collapsing loops
is not sufficient to infer the spectrum emitted by the full string network in the scaling regime. Instead, this is the result of a combination 
of spectra produced by loops and long strings, and therefore also depends strongly on the distribution
and the evolution of the latter. While the value of $q$ observed in our simulation is not far from 1,
we have no evidence to support an expectation of an asymptotic value of $q=1$ at large logs.

\item
Refs.~\cite{Hiramatsu:2010yu} performed similar simulations of global strings 
to those that we have carried out. They find
a value of $\xi\approx 1$, in the same approximate range as that obtained in our simulations, but they did not observe
any linear increase with the log (probably for the reasons discussed in Section~\ref{sec:scalvio}). 
For the reasons
discussed in Section~\ref{sec:axspectr}, they inferred a very IR dominated spectrum
 and conclude that strings would efficiently produce a large axion number density, even though their $\xi$ is lower than our extrapolated value. This example shows 
how a different interpretation of the spectrum  can lead to a completely different conclusion, 
despite the fact that the spectrum measured in the simulation for small logs might be qualitatively in agreement. At small logs, a very detailed study of the spectrum is mandatory to reliably extrapolate the number density to the physical point.

\item
Finally in refs.~\cite{Klaer:2017qhr,Klaer:2017ond} 
a different UV completion of the string cores was introduced in order 
to simulate fat strings with a large effective tension $\mu_{\rm eff}\sim 70\mu_0$, 
despite the actual core size remaining bounded by the lattice spacing size $\log(m_r/H)\sim 6$. 
While this trick is claimed to remove the need for a large extrapolation connected to the string tension and the effective axion-string coupling, the one associated to the decoupling of the radial modes of the string core remains.
The results in ref.~\cite{Klaer:2017qhr,Klaer:2017ond}  confirm the non-trivial dependence of $\xi$ on the logs, although the actual functional dependence is not clear from the analysis. 
The authors did not present explicit results for the spectrum.
The final number density of axions observed, which also includes the contribution from the domain walls and the destruction of the string network owing to the axion mass,  turns out to be smaller than that from   misalignment. 
This result seems to be compatible with an underproduction of axions
from strings, associated to a spectrum with $q<1$, in agreement with our results if we assume that 
$q$ remains below 1 at larger logs. 

However, note that in this case the spectrum (and the way that strings sustain the scaling regime) is
dominated by core scale physics. This makes the extrapolation of the string core size from $\log(m_r/H)\sim 6$ to $\log(m_r/H) \sim 70$, as subtle as in the normal case. In fact we cannot exclude the possibility
that at larger logs (i.e. relatively thinner strings) the production of energetic modes gets suppressed and $q$ increases. 
 The extrapolation of $\log(m_r/H)$ to the
physically relevant values made in ref.~\cite{Klaer:2017ond} 
is therefore just as sensitive to the uncertainties 
discussed as it is in conventional simulations. A high precision study of the spectrum is therefore required to
identify the correct extrapolation. 
\end{itemize}

\section{Conclusions}  \label{sec:concl}

In this paper we have studied the dynamics of the global strings that form in QCD axion models when the $U(1)$ PQ symmetry is broken after inflation. Using numerical simulations, we have shown that the string network approaches an attractor solution that is independent of its initial conditions, and which is approximately scale invariant. We have also seen that this solution has a number of interesting properties:
\begin{itemize}
	\item The string length per Hubble patch $\xi(t)$, defined by eq.~\eqref{eq:xidef}, increases logarithmically with the ratio of the Hubble parameter and the string core size, with best fit parameters given in eqs.~\eqref{eq:xifit} and \eqref{eq:alphafit}. 

	\item  At any time more than $80\%$ of the string length is contained in long strings and the rest is in loops of size Hubble and smaller, which follow a scale invariant distribution. While such loops  shrink and disappear they are replaced, at the same rate, by loops produced from longer strings.
	
    \item The energy density of the string network is determined by eq.~\eqref{eq:rhos} with --- nontrivially --- an effective string tension $\mu_{\rm eff}$ that is close to the theoretical expectation eq.~\eqref{eq:muth}. 
     This means that to a good approximation, the energy density of the string network in the scaling regime at a particular time can be determined solely from the density of strings $\xi(t)$.
    \item  Over the scale separations that can be simulated, a substantial, and approximately constant, proportion of the energy emitted by the string network goes into heavy radial modes. Contrary to expectations, the radial modes have therefore not decoupled from the string dynamics by the end of the simulations, despite the relatively large scale separations $m_r/H \sim 500$ reached at these times.
	\item  The instantaneous spectrum of energy emitted into axions, plotted in Figure~\ref{fig:Fs} in terms of $F$ defined in eq.~\eqref{eq:Fdef}, has the theoretically expected shape with a peak at momentum around the Hubble scale. At higher momenta it follows a power law until the string core scale above which emission is suppressed. The slope of the power is such that the majority of energy emitted by strings goes into high momentum axions for the scale separations that we can study (although a greater number of low momentum axions are produced).
	\item In simulations with physical strings, the instantaneous emission spectrum of the scaling solution can only be evaluated close to the end of the simulation because	 the residual energy from the initial conditions has less time to redshift. The result obtained matches the conclusions drawn from our analysis of the overall spectrum. The fat string model ameliorates these problems, and for this system we see that $F$ remains similar at different times. The slope of the power law shows no sign of changing as the scale separation increases, although the uncertainty is substantial. 
\end{itemize}

Since the scale separations that can be simulated are far from the physically relevant values, extrapolation over a vast distance is necessary if predictions relevant to the relic abundance of the QCD axion are to be made. 
Given our results, it is plausible that $\xi$ continues to grow logarithmically (although we cannot exclude the possibility that it eventually saturates). Such an increase would enhance the number of axions emitted by a factor $\sim 10$. 
Having tested that the energy density in strings is well matched by eq.~\eqref{eq:mueff} with the theoretical expectation for the string tension, the remaining ingredient necessary for a precise prediction of the relic abundance is the instantaneous axion emission spectrum at large scale separations, in particular the power $q$ in $F$. While $q<1$ at the scale separations that we can study, there is space within the current uncertainty for a small logarithmic correction that would result in  it increasing to $\geq 1$ in the physically relevant part of parameter space. On the other hand, if $q<1$ persists at large scale separations then the number of axions produced by strings is presumably sensitive to the UV completion of the theory, as a significant proportion of the energy is going into exciting heavy modes, although in this case the number density is suppressed relative to that from misalignment.

To determine whether $q$ has a scale dependence 
it is desirable to carry out simulations at larger scale separations. Indeed, even with a relatively modest increase in range, evidence for a change in $q$ might be seen,  which would allow an extrapolation.  Improvement could come from different directions. One possibility is to simply perform simulations on larger high performance computing clusters, after implementing more efficient parallelisation than we have in our present work.  A more involved possibility would be to develop an adaptive mesh algorithm, in which the  scalar field is simulated on a grid with finer mesh spacing in regions of space where this is required, close to the cores of strings. The mesh must be updated as the strings move, which requires a more sophisticated algorithm, however this approach has proved beneficial in simulations of, for example, astrophysical black hole mergers \cite{Clough:2015sqa}. It may also be interesting to further study the dynamics of individual strings, or small numbers of strings, which could give an indication of the behaviour of the full network. A qualitatively different approach would be to develop a numerical simulation in which the strings themselves were treated as the fundamental degrees of freedom, with parameterised dynamics and interactions. Such an effective theory style simulation could allow much larger scale separations to be studied. In \cite{Fleury:2016xrz} this idea was used to numerically simulate ``strings'' in 2 dimensions, although the extension to three dimensions is challenging, and a careful analysis would be necessary to ensure that the full dynamics of the underlying theory is capture.

A full calculation of the relic abundance must also include the axions produced when axion mass turns on, at which time domain walls form and the string network is destroyed. After the axion mass first becomes cosmologically relevant, when $m_a\left(T\right) \sim H$, it continues to increase fast, and quickly $H\left(T\right) \ll m_a\left(T\right) \ll f_a$. As a result, this system has three widely separated scales, and is even harder to numerically simulate than the string network alone. Additionally, if $N_W > 1$ it is currently unknown if the explicit breaking permitted such that the axion still solves the strong CP problem allows the domain walls to decay fast enough to avoid axions overclosing the Universe. We leave a study of the dynamics at these times to future work.

\section*{Acknowledgements}

We are grateful to Aleksandr Azatov for collaboration at an early stage of this project, and very useful discussions. We acknowledge SISSA and ICTP for granting access
at the Ulysses HPC Linux Cluster, and the HPC Collaboration Agreement between ICTP and CINECA for granting access to the A3 partition of the Marconi Lenovo system. 

\appendix

\section{Details of the Simulation} \label{sec:app1}

\subsection{Evolution of the Field Equations}\label{sec:app_evo}

For the purpose of implementing numerical simulations, it is convenient to rewrite the equations of motion, given in eqs.~\eqref{eq:eom} and \eqref{eq:fatstringdef}, in terms of the rescaled field $\psi=R(t) \phi/f_a$, and the conformal time $\tau$, which is defined as 
\begin{equation}
\tau(t)=\int_0^t\frac{dt'}{R(t')}\propto t^{1/2}~.
\end{equation}
The time-dependent mass in the fat string scenario, eq.~\eqref{eq:fatstringdef}, is $m_r(\tau)=(\tau_i/\tau)m_i$, where $\tau_i\equiv\tau(t_i)$ is the time at which $m_r=m_i$. In this way,
 the equations of motion simplify to
\begin{equation} \label{eq:eom1}
\psi'' - \nabla_x^2\psi+ u(\tau) \psi\left(|\psi|^2 -\frac{R^2}{2}\right)=0  \ ,
\end{equation}
where $\psi'$ and $\nabla_x\psi$ are derivatives with respect to the dimensionless time and distance variables $m_r\tau$ and $m_rx$ respectively (or $m_i\tau$ and $m_ix$ for the fat string case). In the physical scenario $u(\tau)=1$, while in the fat string case $u(\tau)=\tau_i/\tau$.

We then solve eq.~\eqref{eq:eom1} numerically on a cubic lattice with periodic boundary conditions.\footnote{We parallelise the algorithm to step forward in time, and run on a cluster with $2 \times 24$-cores.} Space is discretised in a box of comoving side length $L_c$ containing $N^3=1250^3$ uniformly distributed grid points, where the upper value of $N$ is limited by our memory budget. Consequently the space-step between grid points in comoving coordinates is $a_c = L_c/N$, which is constant in time. It is again convenient to work in terms of the dimensionless comoving space-step $m_r a_c$ in the physical case, and $m_i a_c$ for the fat string system.

The physical length of the box is $L(t)=L_c R(t)$ and the physical space-step between grid points $a(t) =L/N = a_c R(t)$ grows $\sim t^{1/2}$. In the physical string scenario, the string core size $m_r^{-1}$ is constant, and therefore the number of grid points in a core $\left(m_r a(t) \right)^{-1} \sim 1/t^{1/2}$ decreases. Meanwhile, in the fat string scenario the string core size increases $\sim t^{1/2}$, and as a result the number of grid points in a string core $\left( m_r (t) a(t) \right)^{-1} =  \left(m_i a_c\right)^{-1}$ is constant. When considering systematic errors from the space steps in the fat string system we use the notation $m_r (t) a(t)$ for the size of the space step due to its direct physical interpretation (although this is entirely equivalent to $m_i a_c$).

The equations of motion are discretised following a standard central-difference Leapfrog algorithm for wave-like PDEs (see e.g. \cite{LeVeque}). The system is evolved in fixed steps of conformal time $a_\tau$, and we work in terms of the dimensionless time-step $m_r a_{\tau}$ in the physical case, and $m_i a_{\tau}$ in the fat string case. The derivatives are expanded to fourth order in the space-step and second order in the time-step.\footnote{The fourth order discretisation of space is probably not required, since additional tests show using a (less precise) second order discretisation does not lead to significant differences in the results, at least for the string length.} In Appendix~\ref{sec:app2} we extensively study the systematics from the discretisation of space and time, as well as from finite volume effects.

As discussed in Section~\ref{sec:scaling}, we set the initial conditions in two ways, the second of which is used to produce a cleaner initial configuration with a fixed number of strings per Hubble volume.
\begin{enumerate}
\item[(a)] Random initial conditions: $\phi$ and $\dot{\phi}$ are both generated through the anti-Fourier transform of Fourier coefficients $\tilde{\phi}(\vec{k})$ and $\tilde{\dot{\phi}}(\vec{k})$, randomly chosen in the interval $[-\frac{f_a}{\sqrt{2}},\frac{f_a}{\sqrt{2}}]$ for $|\vec{k}|\leq k_{max}$ and zero for $|\vec{k}|> k_{max}$ for a fixed choice of $k_{max}\in[0,m_r]$ in each set of simulations. Larger values of $k_{max}$ lead to initial fluctuations with smaller wavelength, of order $2\pi/k_{max}$, and more initial strings. As a result, the initial string density is controlled by the parameter $k_{max}/m_r$. 
Even though  this method produces configurations in which the axion field winds the fundamental domain $[0,2\pi f_a]$ nontrivially, it does not lead to a clean string configuration at the initial time, because $|\phi|$ does not typically resemble the profile function of string-like solutions. Instead, the system takes some time to relax to a string network solution and in doing so releases a large quantity of energy, which produces extra contamination to the axion spectrum measured at later times. To solve this issue, we employ the method below that also allows us to construct axion field configurations with a predetermined string length, or equivalently with predetermined initial $\xi$. 

\item[(b)] Fixed string number: This approach simply consists of evolving a field configuration produced by method (a) with $k_{max}=m_r$ until the required string length inside the box is obtained, and then using that configuration as initial condition for the actual simulation, with a different initial value of the Hubble parameter. Since this involves resetting the Hubble parameter, the strings produced do not in general have the right core-size, but we have checked that they quickly relax to their expected thickness long before the scaling regime is reached. 

\end{enumerate}

\subsection{String Identification and String Length}

To identify strings and measure their length, we first identify grid points that are likely to be close to a string core. We have carried this out in two ways, and have verified that the results obtained are extremely similar. In our main approach, we flag points such that, as a loop that surrounds it is travelled, there is at least one  change in the axion field $\Delta a$ between consecutive lattice points encountered that satisfy $\left| \Delta a \right|/f_a > \pi/2$. In particular, the loop is taken to be a square of side length 2 grid points.
In order to capture strings with all possible orientations, at each point in the grid we consider loops in three orthogonal planes, and we flag a point if a loop in any of these satisfies the condition. We have checked that the results obtained are extremely similar for any reasonable choice of the threshold value for the change between adjacent grid points.

Having identified points close to the string core, we then combine these into strings. In particular, we cluster together flagged points that are adjacent in the $x$-$y$ plane into an individual string point located at the mean of the flagged points. If such a cluster has a non-zero overlap with a cluster at the next level up or down in the $z$ direction, these are connected into a string segment. As expected, the reconstructed strings form loops, and the length of each individual loop (which is required to analyse the loop distribution) as well as the total string length is recorded.

We have validated our string identification algorithm by comparing to the results obtained following the procedure adopted in \cite{Hiramatsu:2010yu}. In this, a lattice plaquette is identified as containing a string if the minimum axion field range that includes the field values on the four surrounding vertices satisfies $|\Delta a|/f_a > \pi$. It can be seen that this leads to the correct results for the prototype string solution, and indeed for any string solution for which the $2\pi$ field change incurred in traversing an enclosing path is distributed sufficiently homogeneously around the loop.

\section{Analysis of Systematic Errors} \label{sec:app2}

To check that our results are free from numerical artifacts, we study how the key observables depend on the unphysical parameters in our simulations. Of particular importance are: (1) the lattice spacing $a$, (2) the time spacing $a_\tau$, (3) the number of Hubble patches $HL$ contained in the box, and (4) the way that the string cores are screened when evaluating the axion spectrum and energy. In this section we study the first three of these, postponing (4) until Section~\ref{sec:app_mask}. The results that we present here are all obtained from the fat string scenario, however we have tested that the conclusions we reach are also valid for the  physical case.

\subsection{Lattice Spacing}
\begin{figure}[t]\begin{center}
\includegraphics[height=5.4cm]{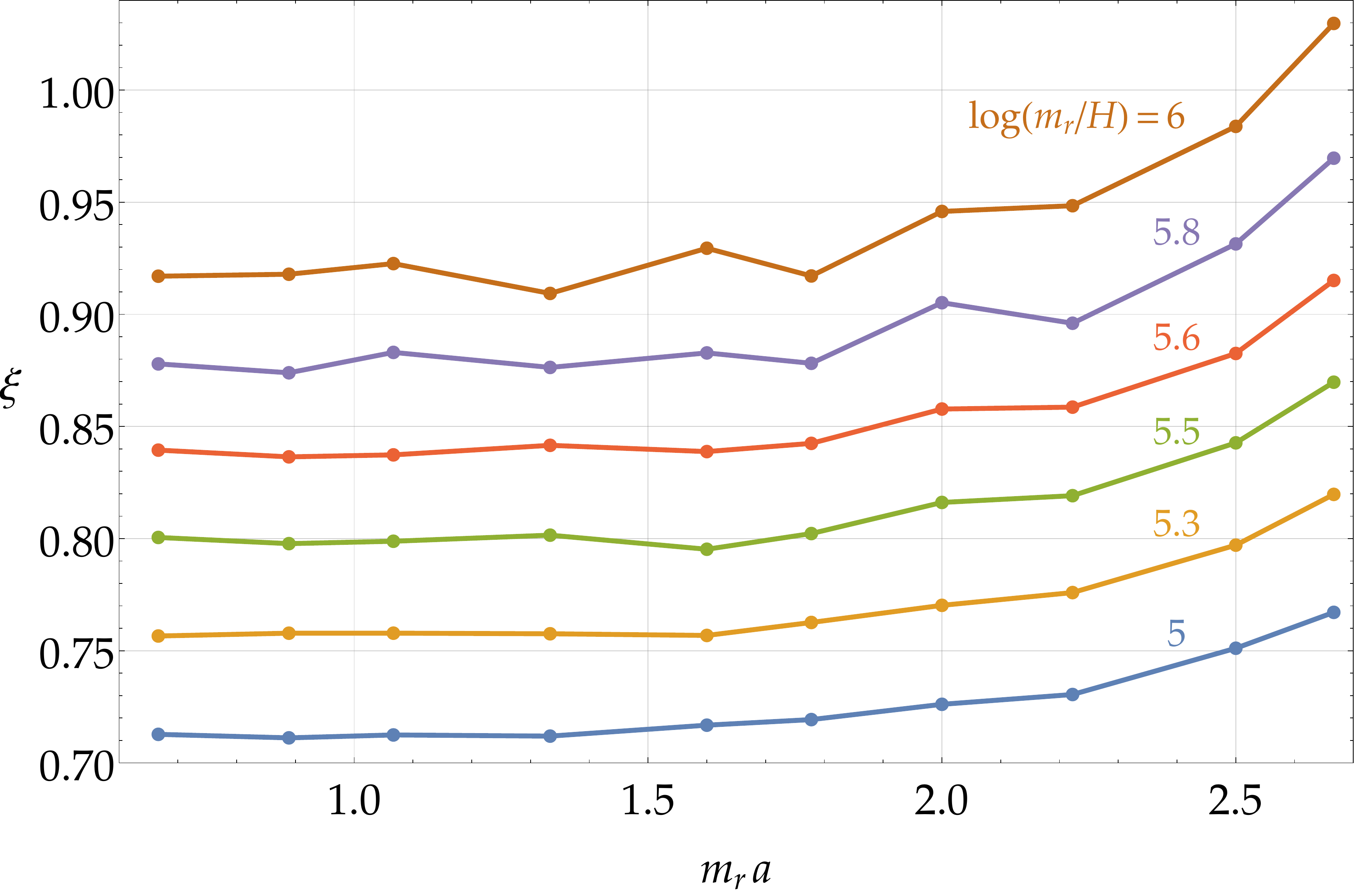}\hspace{.5cm}
\includegraphics[height=5.4cm]{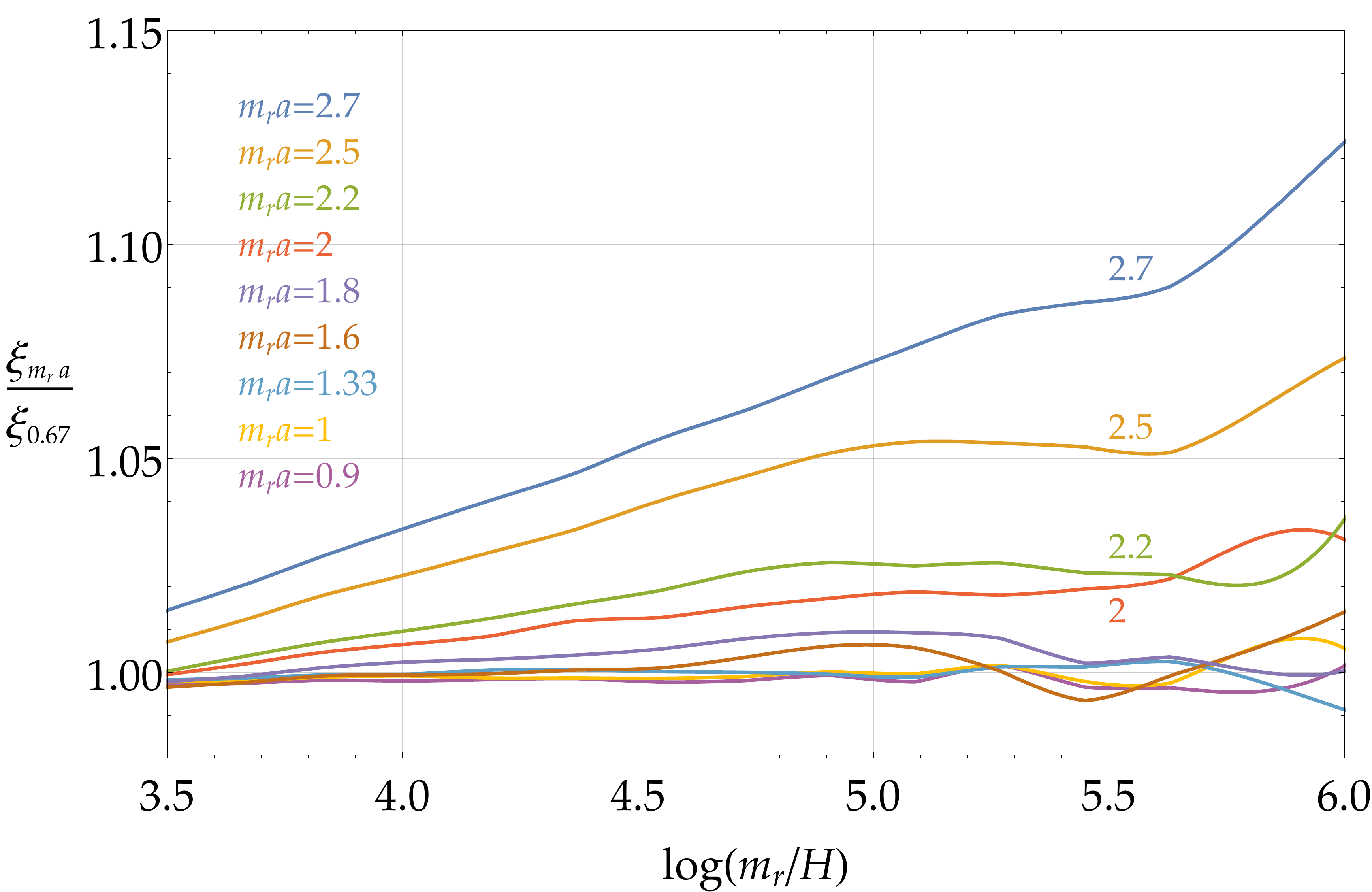}
\caption{Left: The continuum extrapolation of $\xi(t)$ in the step-size, $m_r a\to0$. The values obtained with $m_r a=1.33$ appear to have already converged to the continuum limit, typically with better than percent level accuracy. Right:
Results for $\xi(t)$  for different choices of the step-size $m_r a$, normalised to the same quantities calculated with the finest lattice spacing $m_r a=0.67$. The other unphysical parameters are fixed to $a_\tau/a_c=1/3$ and $H_fL_f=2$.}
\label{fig:spaceTest} \end{center}
\end{figure}

\begin{figure}[t]\begin{center}
\includegraphics[height=6cm]{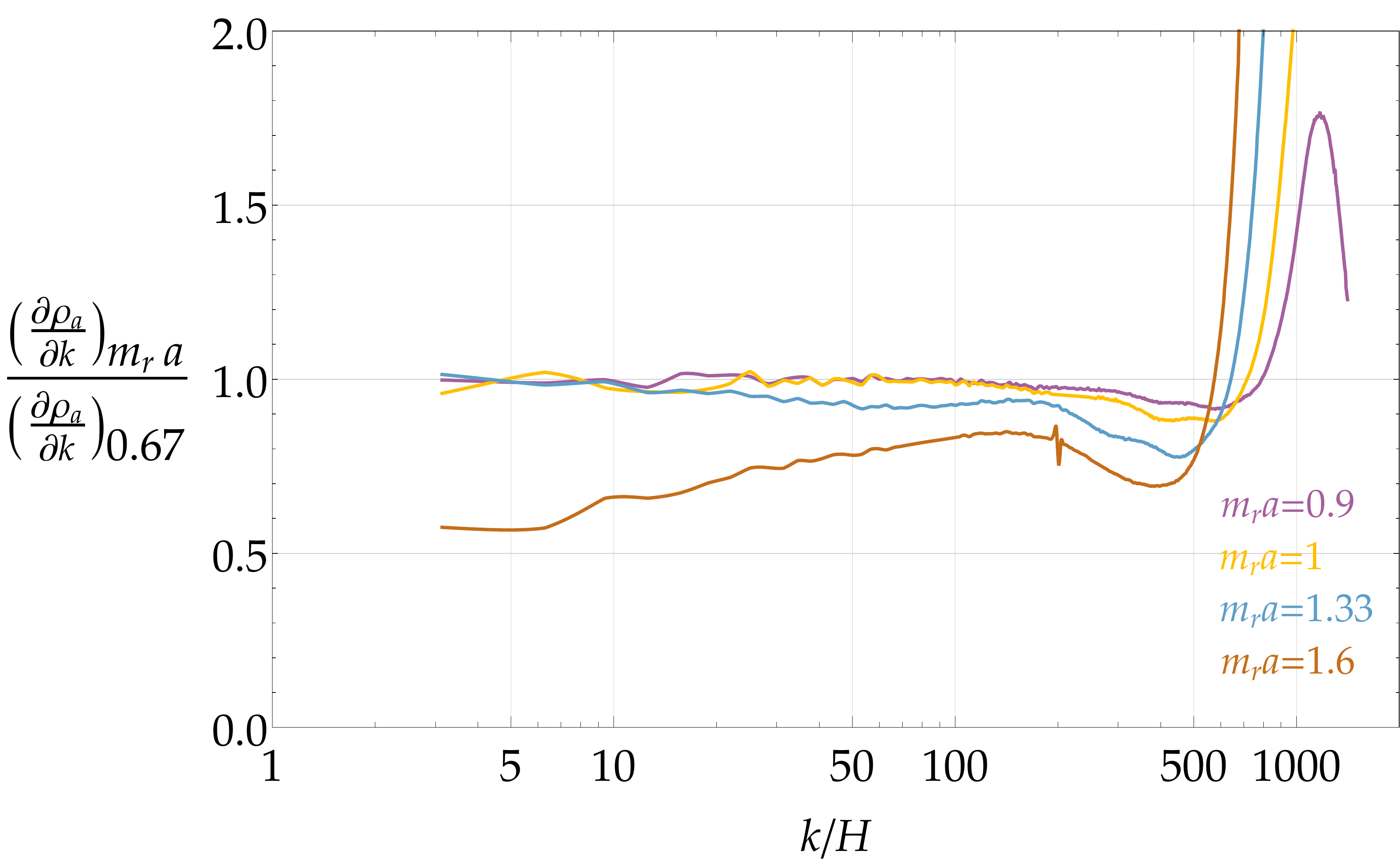}
\caption{Axion spectrum at $\log(m_r/H_f)=6$ for different choices of the step-size $m_r a$, normalised to the same quantities calculated with the finest lattice spacing $m_r a=0.67$. The other unphysical parameters are fixed to $a_\tau/a_c=1/3$ and $H_fL_f=2$. }
\label{fig:spectrumspace} \end{center}
\end{figure}

We study the dependence of $\xi(t)$ and of the axion spectrum on the discretisation parameter $m_r a$ to find the largest value compatible with the continuum limit $m_r a=0$. As mentioned, $m_r a$ is constant and is equal to the number of grid points per string core at all times in fat string simulations (in contrast, for the physical system the number of points per string core decreases with time, and this source of systematic errors only becomes relevant towards the end of simulations). If $m_r a\gg1$, the number of gridpoints per string core is much smaller than $1$ and the evolution of strings is not correctly captured. Conversely, making the space-step unnecessarily small would restrict the maximum scale separations that could be analysed.

We have performed sets of simulations  averaging over $20$ samples, keeping all parameters fixed except $m_r a$, with identical initial conditions. 

In Figure~\ref{fig:spaceTest} we plot the continuum extrapolation for $\xi$ and its dependence on $\log(m_r/H)$ for the different values of the space-step, normalised to those from the finest lattice spacing tested ($m_r a=0.67$). For all $m_r a\leq1.33$, $\xi(t)$ differs from the most precise result by less than $1\%$. However, for larger values of $m_r a$, it is systematically larger, especially at later times. 
In the main text, for our results of $\xi(t)$ and the loop distribution in the fat string scenario,  we used the rather conservative choice $m_r a =1.33$. In the physical scenario the number of grid points inside a string core decreases as the simulation progresses. We use parameters such that, at the final time, we match the resolution in the fat string case
 $m_r a =1.33$ 
 (which means that at early times there are many more grid points inside a string core).

In Figure~\ref{fig:spectrumspace}  we plot the axion spectrum at the time $\log(m_r/H)=6$, for different space steps, again normalised to the results with $m_r a=0.67$. This is slightly more sensitive to the value of $m_r a$ than $\xi(t)$ is, and we see that discretisation effects increase the production of UV states, reducing the energy emitted in the form of low momentum axions. In particular, only simulations with $m_r a\leq1$ seem to have converged to within few percent of the continuum limit, while $m_r a =1.33$ and $1.6$ result in a spectrum that is systematically smaller in the IR. In the main text we have chosen $m_r a=1$ in our analysis of the spectrum in the fat string system; this value is sufficiently small that we are confident that the UV dominated spectrum obtained is not an artifact of the finite space-step. Similarly, in the physical case we use parameters such that $m_r a = 1$ when the final time shot used in calculating the spectrum is taken.

\subsection{Time Spacing}

\begin{figure}[t]\begin{center}
\includegraphics[height=5.1cm]{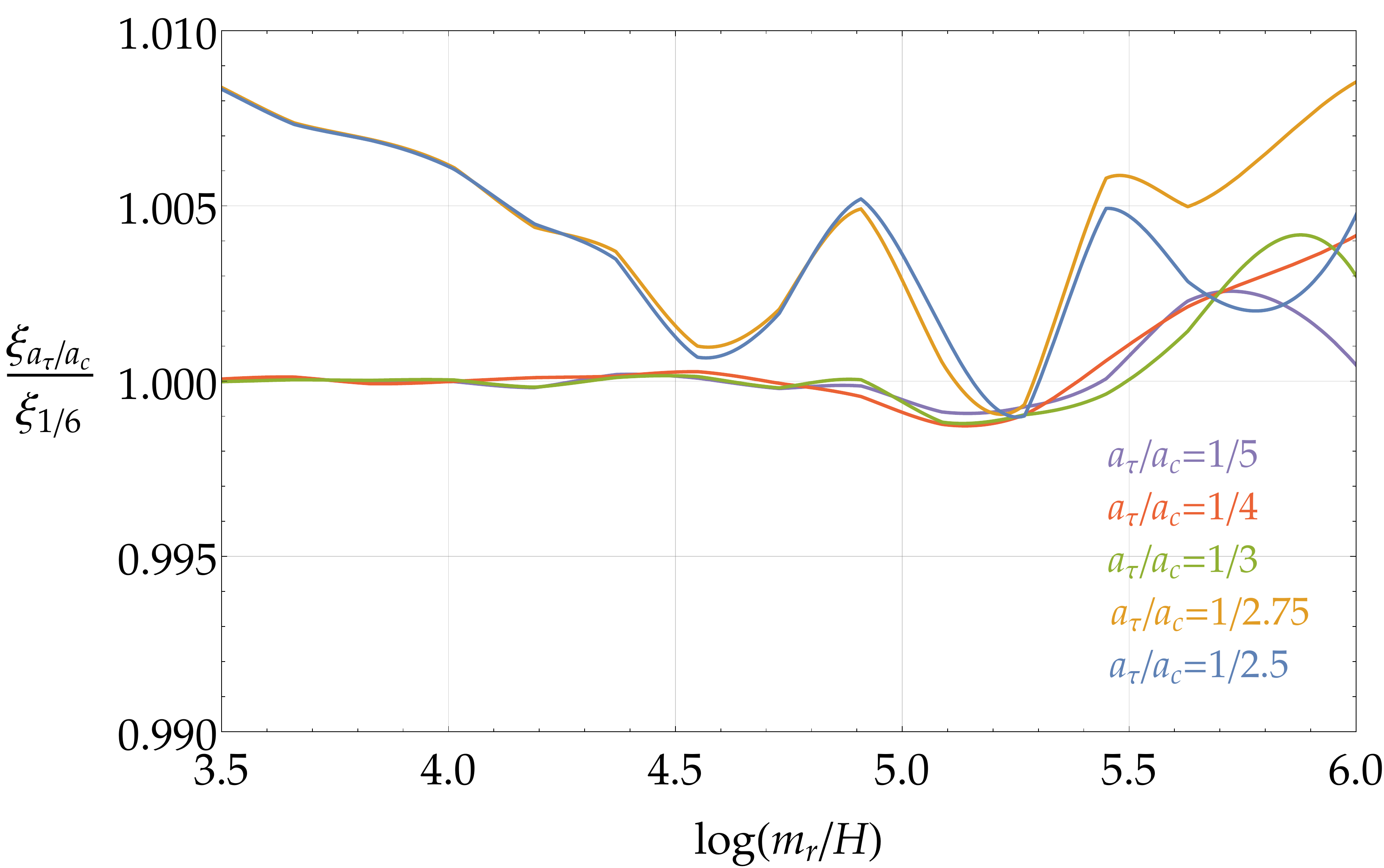}\hspace{0.2cm}
\includegraphics[height=5.1cm]{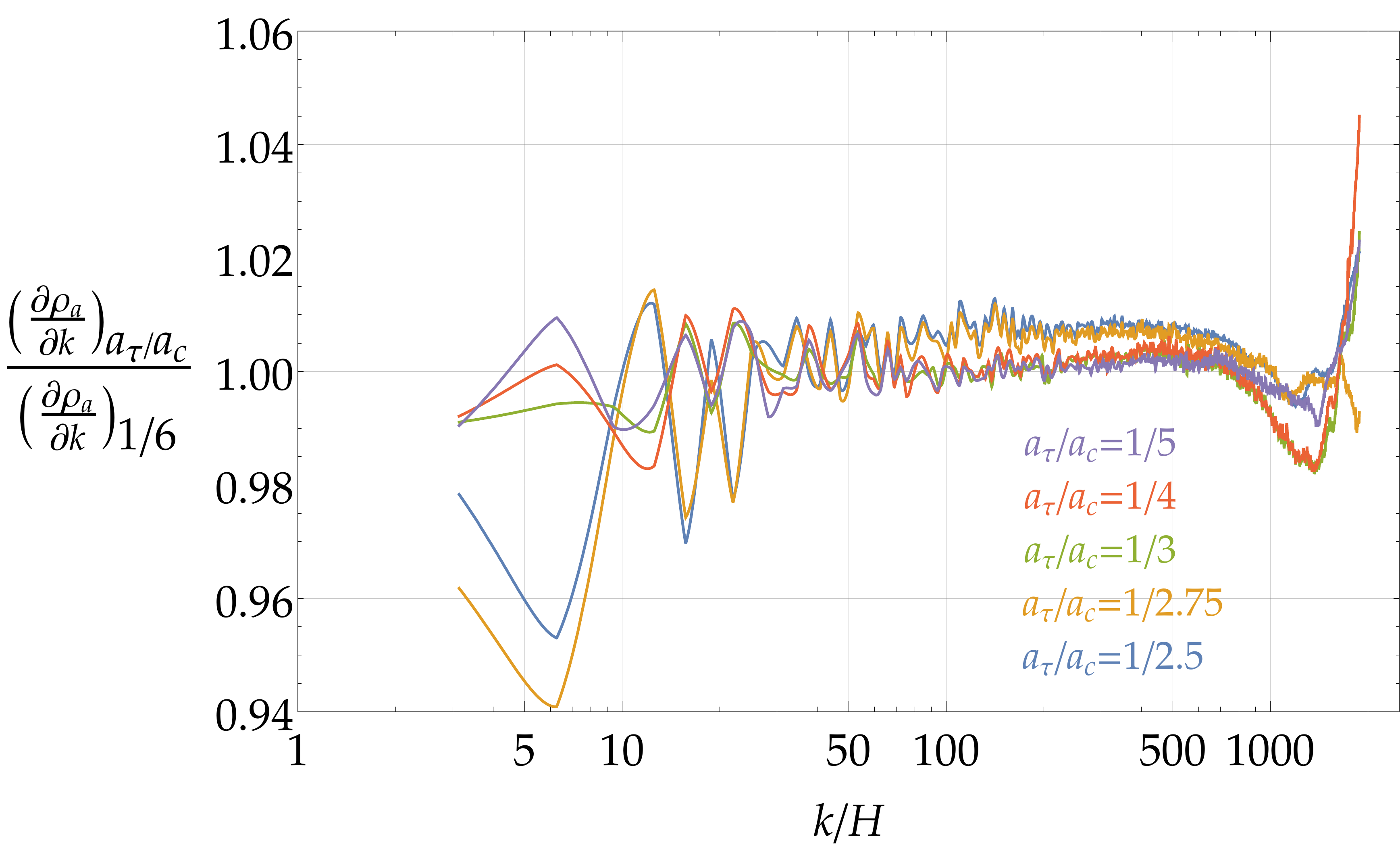} 
\caption{Results for $\xi(t)$ (left) and the axion spectrum at $\log(m_r/H_f)=6$ (right), for different choices of the time-step $a_\tau$. This is measured relative to the comoving space-step $a_c$, which is fixed to $m_i a_c= m_r a=0.67$, and the final box size is fixed to $H_fL_f=2$. The results are normalized to the shortest time step tested: $a_\tau/a_c=1/6$.}\end{center}
\label{fig:timeTest}
\end{figure}

We study the dependence of our results on the time-step similarly. Given the form of eq.~\eqref{eq:eom1}, the general theory of numerical solutions of PDEs tells us that the relevant quantity to which $m_i a_{\tau}$ should be compared is the comoving space-step $m_i a_c$ \cite{LeVeque}.\footnote{The Courant condition for the stability of the finite-difference algorithm sets an upper bound on the ratio $a_\tau/a_c < 1$ for the algorithm to converge, but further analysis is still required to quantify the systematic uncertainty.} Consequently, we perform sets of simulations that differ only in the value of $a_\tau/a_c$, averaging over $10$ samples. In Figure~\ref{fig:timeTest} we plot the results for $\xi(t)$ and the spectrum. For all values of $a_\tau/a_c$, $\xi(t)$ is within $0.5\%$ of the continuum limit, but for $a_\tau /a_c \leq1/3$ the difference is less that $0.1\%$. The spectrum is also affected at less than percent level for $a_\tau /a_c <1/3$, and we use $a_\tau/a_c =1/3$ for our simulations in the main text in both the fat string and the physical scenarios.

\subsection{Finite Volume}
\begin{figure}[t]\begin{center} 
\includegraphics[height=5.5cm]{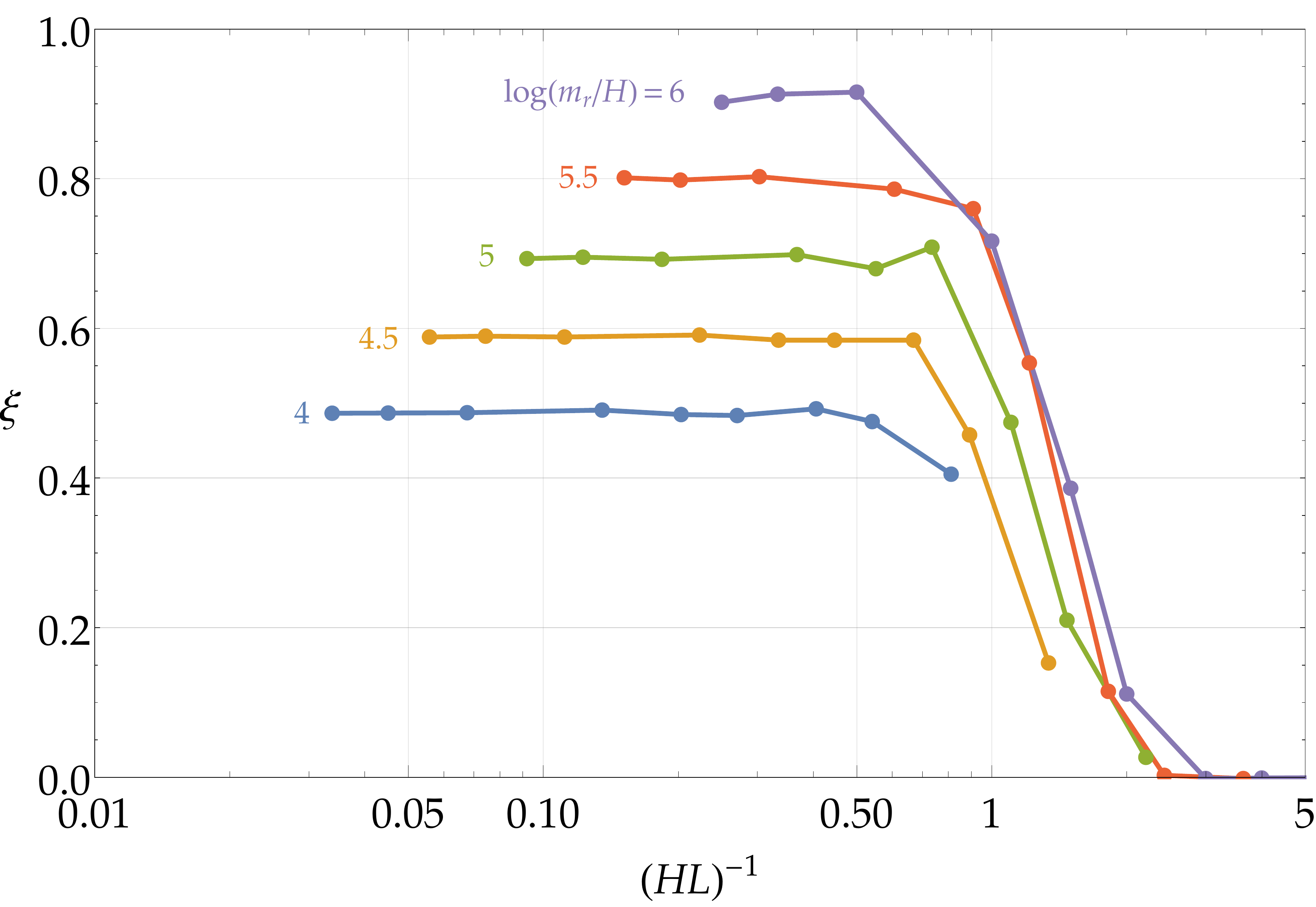} \hspace{.2cm}
\includegraphics[height=5.5cm]{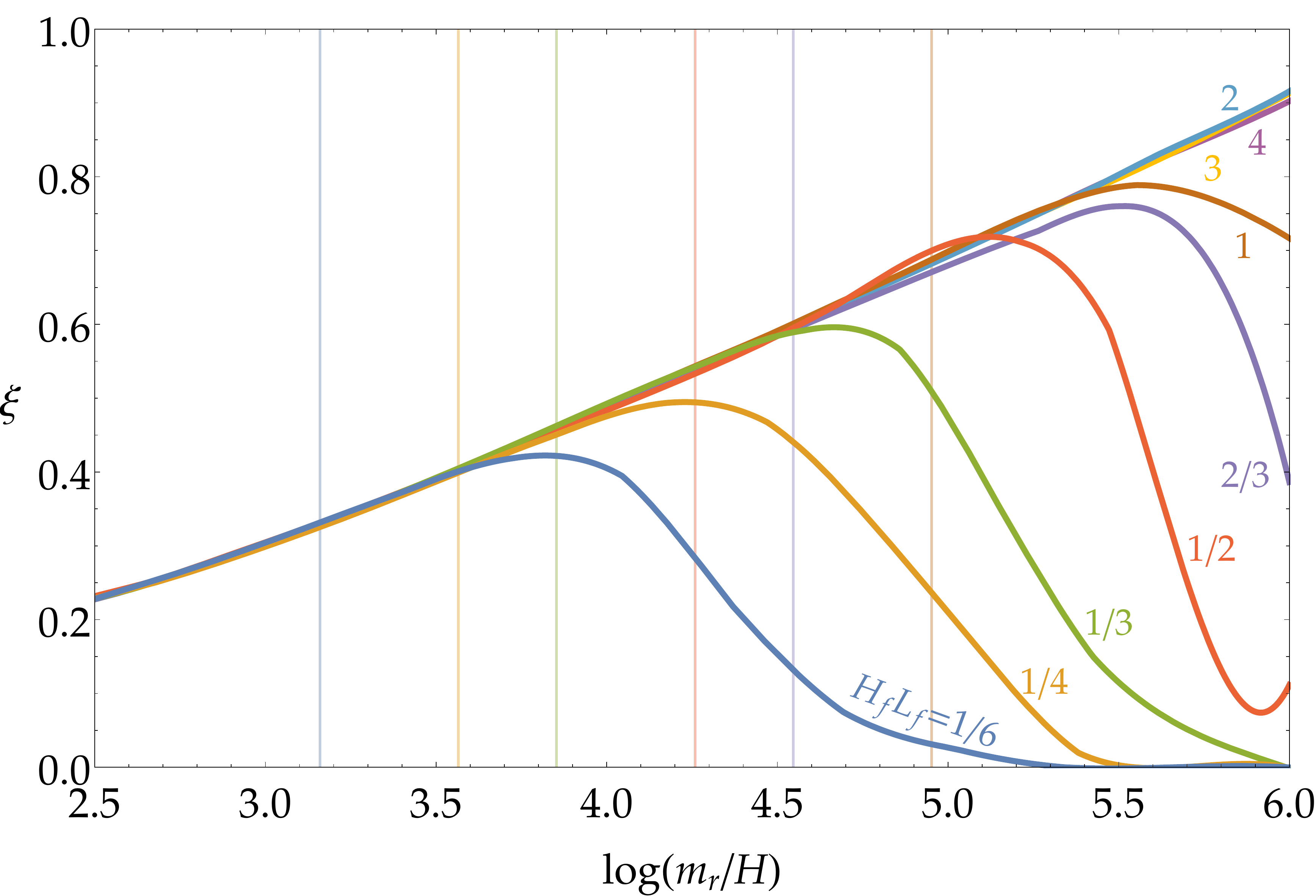}
\caption{Left: The extrapolation of $\xi(t)$ to the infinite box size limit, $(H L)^{-1}\to0$. The value $(HL)^{-1}=1/2$ seems to have already converged, with better than percent level accuracy at most times.
Right: $\xi(t)$ calculated in simulations with different box sizes (measured by $H_fL_f$); vertical lines indicate the times at which $HL=2$ in each simulation.  }
\label{fig:hubbleTest} \end{center}
\end{figure}

\begin{figure}[t]\begin{center}
\includegraphics[height=6cm]{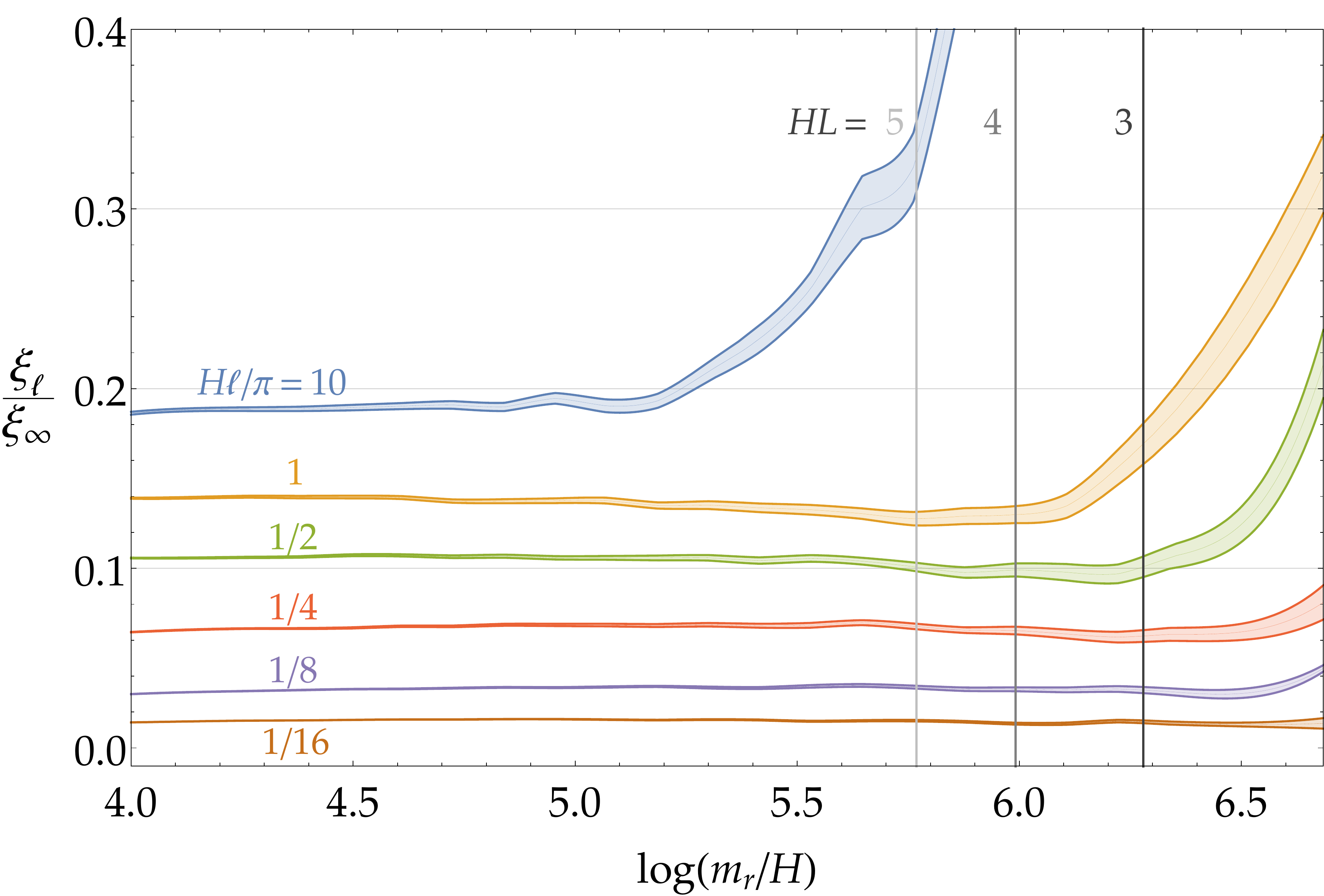}
\caption{$\xi_\ell/\xi_{\infty}$ for the set of simulations with $H_fL_f=2$, for different values of $\ell$. The shaded bands represent statistical errors in the average over different simulations. The other parameters are fixed as $m_r a=1.33$ and $a_\tau/a_c=1/3$.}
\label{fig:xilxinf} \end{center}
\end{figure}

Finite volume effects are a particularly delicate issue. These are controlled by the parameter $HL$, which counts the number of Hubble lengths per box side length, with the physical limit for a spatially flat Universe corresponding to $(HL)^{-1}\to 0$.

Since $HL$ decreases over the course of a simulation, we want to determine the smallest value that it can take at the final time, $H_fL_f$, such that the results obtained match those in the physical limit. Although we use periodic boundary conditions, if $H_fL_f>2$ every point is causally disconnected from itself from the beginning to the end of the simulations. Nevertheless, even choosing $H_fL_f>2$, finite volume effects might still affect some observables such as the loop distribution (see Section~\ref{sec:longvshort}).

We  performed a set of simulations keeping the values of all the parameters except for $H_fL_f$ fixed (all ending at a final scale separation $\log(m_r/H_f)=6$). 
In Figure~\ref{fig:hubbleTest}  we show the convergence of $\xi$ to the infinite volume limit for $HL>2$  for different values of $\log(m_r/H)$ and as a function of $\log(m_r/H)$ (where we indicate with vertical lines the times at which $HL=2$ for each simulation). Finite volume effects do not affect $\xi$ while $HL>2$, and they remain small even slightly later. However, for $HL\lesssim1$ they result in a dramatic change, since the the whole network is in causal contact and starts to be destroyed.

We also study how the loop distribution is affected by finite volume effects. In Figure~\ref{fig:xilxinf} we show the ratio $\xi_{\ell}/\xi_\infty$, defined in Section~\ref{sec:longvshort}, for different values of $\ell$, for a set of simulations with $H_fL_f=2$ and $\log(m_r/H_f)=6.7$. Vertical lines correspond to the times at which $HL=5,4,3$. As mentioned in Section~\ref{sec:longvshort}, the constant value of $\xi_{\ell}/\xi_\infty$ in time is a strong indication that the system is in the scaling regime. However, the finite box size limits the maximum loop radius that can be contained. As a result, finite volume effects lead to $\xi_{\ell}/\xi_\infty$ growing at late times once larger loops are no longer possible, and the larger $\ell$ is the earlier this occurs. For our analysis of $\xi$ and the loop distribution we have chosen $H_fL_f=2$ for both the fat string and the physical systems. This is rather conservative for $\xi$, but slightly sub-optimal for the study of the distribution of relatively long loops with $lH/\pi\gtrsim1/2$ (although earlier time shots are also plotted in Figure~\ref{fig:loopdistr}, so the effect of the finite volume is clear).

\begin{figure}[t]\begin{center}
\includegraphics[height=5.4cm]{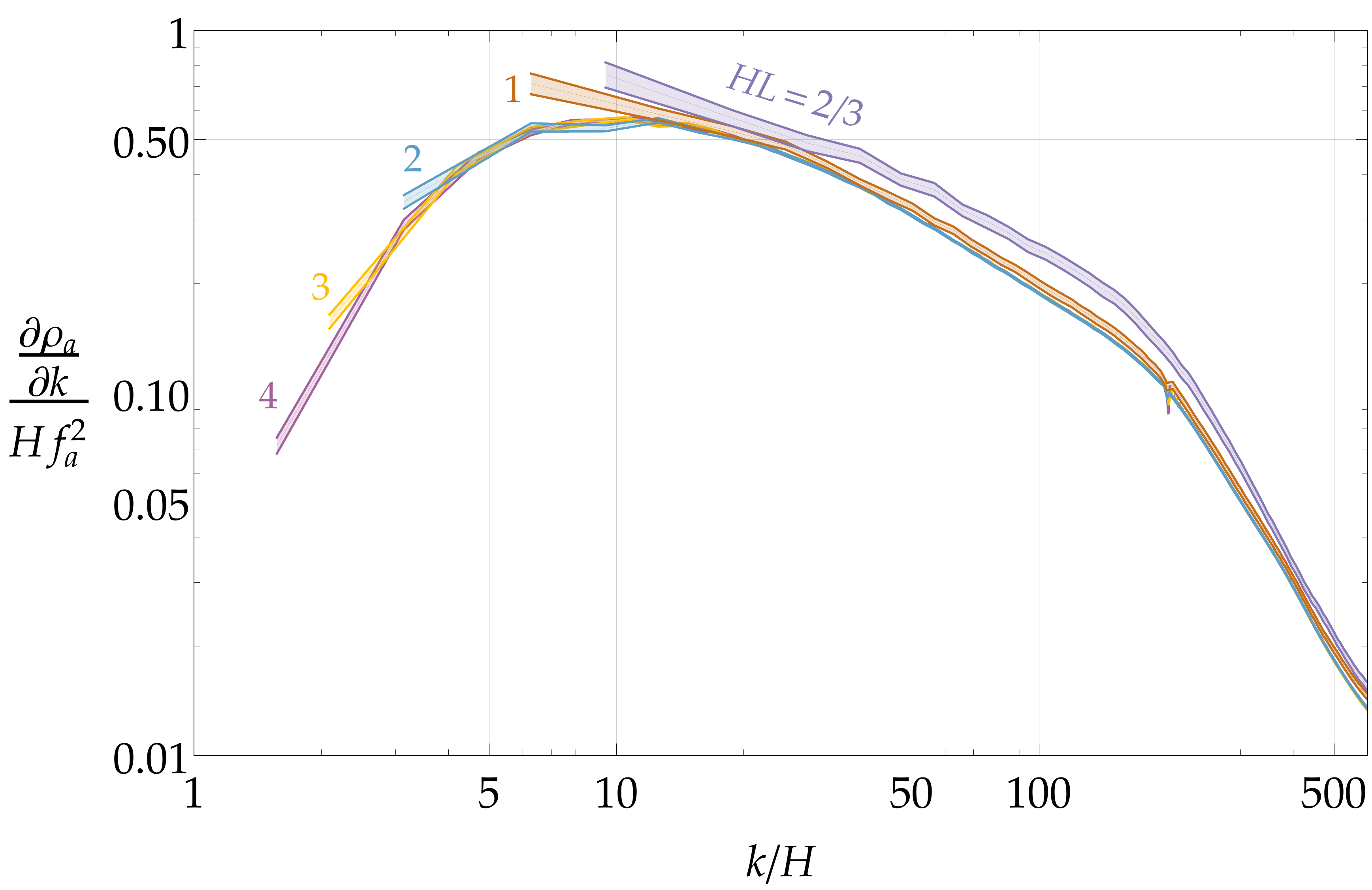}\hspace{0.5cm}
\includegraphics[height=5.4cm]{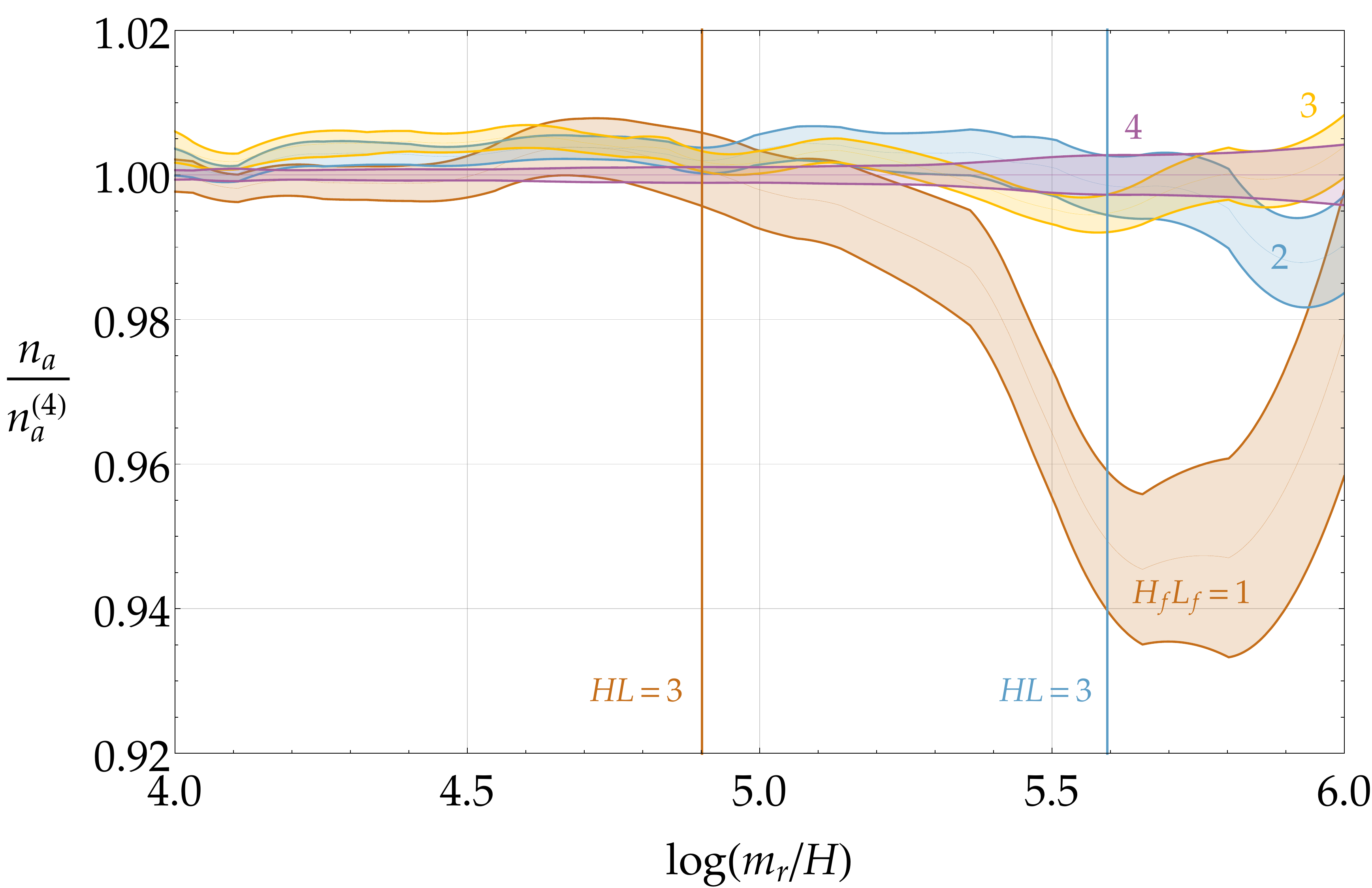} \end{center}
\caption{The spectrum at $\log(m_r/H)=6$  (left) and axion number density (normalised to the number density calculated in a simulation with the largest box size tested: $H_fL_f=4$) as a function of $\log(m_r/H)$ (right) for different final numbers of Hubble patches at the end of the simulation. Error bars represent statistical errors in the average over different samples.}
\label{fig:hubbleSpectrum}
\end{figure}

In Figure~\ref{fig:hubbleSpectrum} (left) we plot the axion spectrum at $\log(m_r/H_f)=6$ for simulations with different values of $H_f L_f$. The peak at the Hubble scale is well reproduced for $H_fL_f\geq2$, even if its position is dangerously close to the IR cut-off (indeed, both are of order Hubble). On the other hand, for $H_fL_f\lesssim1$, not only is the Hubble peak not present, but there is also a significant overproduction of UV modes, related to the shrinking of a significant fraction of the string network, which towards the end of the simulation is made up of loops with radius smaller than the Hubble distance.

In Figure~\ref{fig:hubbleSpectrum} (right) we plot the axion number density $n_a$ as a function of $\log(m_r/H)$ for simulations with different values of $H_f L_f$; with vertical lines corresponding to times at which $HL=3$ for each simulation. $n_a$ has been normalised to the values obtained in the simulation with $H_fL_f=4$, which is the least affected by finite volume effects. The number density seems to be slightly more sensitive to the finite box size than $\xi$. While $HL>3$ the effect on $n_a$ is less than one percent,  for $HL<3$ finite volume effects result in a systematic underestimation of a few percent. This is reasonable, since $n_a$ is strongly dependent on the IR of the axion spectrum, which is the part most sensitive to finite volume effects. For this reason in the main text we chose $H_fL_f=3$ when studying both the spectrum and the number density, for the fat string and also the physical system.

We have also studied the dependence on $H_fL_f$ of the effective string tension $\mu_{\rm eff}$. This depends on the distribution of string length in loops of different shapes and sizes, since e.g. the logarithmic divergence in the string tension is expected to be cut off at a smaller scale for small loops than long strings. Consequently, $\mu_{\rm eff}$ could in principle be sensitive to finite volume effects. However, we do not observe any change in the string tension for $H_fL_f\geq2$.

\section{String Screening, Energies and the Axion Spectrum} \label{sec:app_mask}

\subsection{Components of the energy}
\label{app:compenerg}

In this section we discuss in more detail the way that the total energy density of the system $\rho_{\rm tot}$ splits into the contributions defined in eq.~\eqref{eq:rhoar}, and demonstrate that our method of screening the strings when extracting the energy in free axions and radial modes is consistent and does not introduce systematic errors. 
We also study the energy contributions in the absence of strings, and the effect of the interaction term between axion and radial modes. Finally, we describe the algorithm that we use to calculate the axion spectrum. 

As summarised in Section~\ref{sec:energybud}, the total energy density is written as the sum $\rho_{\rm tot}=\rho_s+\rho_a+\rho_r$. The energy density in free axions $\rho_a$ has been calculated as the spatial average $\rho_a=\langle\dot{a}^2\rangle$ far away from string cores, and in these regions axion and radial modes are to a good approximation decoupled at sufficiently late times. Similarly, the radial energy density is computed from $\rho_r=\langle \frac12 \dot r^2+\frac12 (\vec \nabla r)^2+V(r)\rangle$, again away from the string cores. 
We define the parameter $d_s$ such that the minimum distance from the strings' centre of the regions that are included in the average is  $d_{s}m_r^{-1}$, so that, at all times, $d_{s}$ is the screening distance in units of the core size for both the fat string and physical scenarios.

There is actually an alternative approach to calculating the energy density of axions, which does not require screening. This consists of including the interaction terms in the axion energy density, i.e. evaluating $\langle(1+r/f_a)^2\dot{a}^2\rangle$ over the whole space. 
On the string centre the factor $(1+r/f_a)^2$ vanishes, and the contribution from the core regions is thus suppressed. 
Calculated in this way, the axion energy density will include the contribution from the axion-radial interaction energy. When we discuss the energy contributions to $\rho_{\rm tot}$, we will show that energy densities in axions calculated in both ways, i.e. with $\langle\dot{a}^2\rangle$ screened and with $\langle(1+r/f_a)^2\dot{a}^2\rangle$, are in close agreement.

\begin{figure}[t]\begin{center}
\includegraphics[height=9cm]{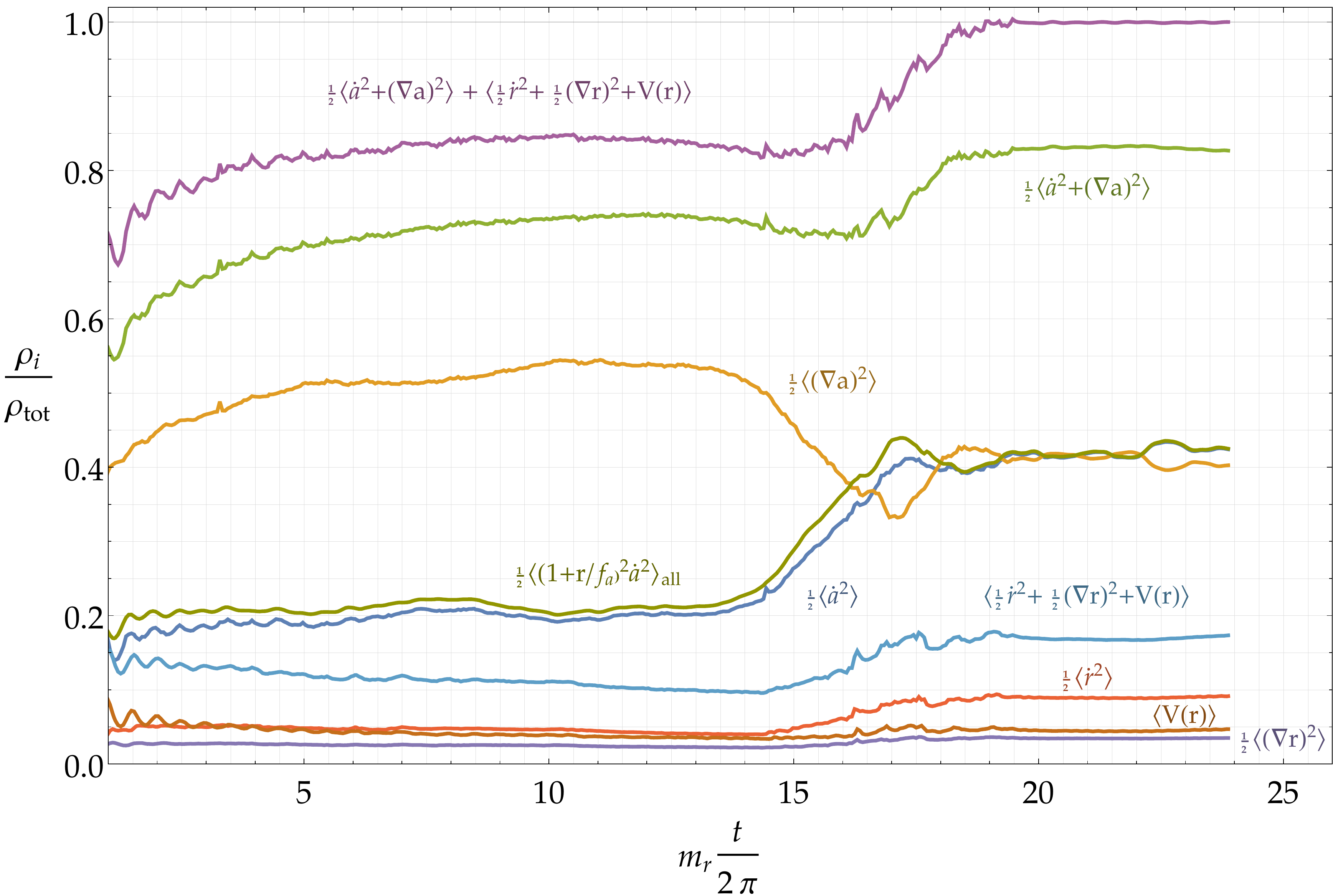}\hspace{0.5cm}\end{center}
\caption{Different contributions to the total energy density $\rho_{\rm tot}$, each one normalised to $\rho_{\rm tot}$, as a function of time for fat string system. All contributions are calculated as the spatial average at a distance $d_s=1$ from the strings' center, except for $\frac{1}{2}\langle(1+r/f_a)^2\dot{a}^2\rangle$, which is calculated over the whole space. The string network begins to be destroyed by the finite box size at $m_rt/2\pi\sim19$, corresponding to the purple line going to $1$.}
\label{fig:energy_total}
\end{figure}

In Figure~\ref{fig:energy_total} we plot the ratio of different components of the energy density to $\rho_{\rm tot}$ as a function of $m_rt/2\pi$ for a single simulation of the fat string system carried out on a small grid, with a space-step $m_r a=1$. The goal here is not to understand how energies behave in the scaling regime (which has not been reached in the simulation analysed here), but rather to check that our approach to calculating the energy in axion and radial modes is consistent. All of the energy contributions plotted are calculated as the spatial average over regions a distance of at least $d_s=1$ away from string centres, except for $\frac{1}{2}\langle(1+r/f_a)^2\dot{a}^2\rangle$, which is averaged over the whole space.  The simulation has been run until the box only contains $1/8$ of a Hubble volume, i.e. $H_fL_f=1/2$. As a result the entire system is in causal contact, and all the strings are destroyed at about $m_rt/2\pi\approx 19$, before the end of the simulation.

Many key features concerning the way that the total energy is split up can be understood from Figure~\ref{fig:energy_total}. First, the kinetic part $\frac12\langle\dot{a}^2\rangle$ of the axion energy outside the  string cores is systematically smaller than the gradient part $\frac12\langle(\vec{\nabla}{a})^2\rangle$ outside the string cores while strings are present. This is expected since the former only gets contributions from free axions (in the approximation that the motions of strings does not have a significant effect on $\dot{a}$ far from string cores), while the latter contains energy from both axion waves and a large fraction of the tension of strings. However, once the string network starts shrinking $\frac12\langle(\vec{\nabla}{a})^2\rangle$ decreases, and both $\frac12\langle\dot{a}^2\rangle$ and the radial energy $\langle \frac12 \dot r^2+\frac12 (\vec \nabla r)^2+V(r)\rangle$ increase. This indicates that as the strings are destroyed the energy stored in their tension is transferred to free axion and radial modes.

As expected for free classical waves, after the network shrinks the axion energy $\langle\frac12\dot{a}^2+\frac12(\vec{\nabla}{a})^2\rangle$ is repeatedly interchanged between its kinetic and gradient parts, which on average are equal. Similarly, the total energy in radial modes $\langle \frac12 \dot r^2+\frac12 (\vec \nabla r)^2+V(r)\rangle$ behaves as expected (for free states with a mass decreasing adiabatically in time, due to the fat string trick), with $\langle \frac12 \dot r^2\rangle$ and the sum $\langle\frac12 (\vec \nabla r)^2+V(r)\rangle$ giving equal contributions. After the string network disappears, the ratios of axion and radial energy to the total energy stay approximately constant. This is a sign that axions and radial models (as well as the total energy) redshift at the same rate, i.e. as massless radiation, or equivalently as massive radiation with a mass decreasing with time as $\sim 1/R(t)$.

The two ways of computing the axion energy density: evaluating $\langle\dot{a}^2\rangle$ outside strings or $\langle(1+r/f_a)^2\dot{a}^2\rangle$ over the whole space, are compatible within few percent while strings are present. The small discrepancy arises because the factor $(1+r/f_a)$ is not exactly a step function at the edge of the string cores. As we will see in the calculation of the spectrum, the difference is stored in UV axion modes (as expected) and therefore do not affect the calculated axion number density. Once the strings disappear the two measurements of the energy density give almost identical results since the difference is due to the non-vanishing interaction energy $\frac12(r^2/f^2_a+2r/f_a)\dot{a}^2$, which goes to zero as the Universe expands.

The sum $\langle \frac12\dot{a}^2+\frac12(\vec{\nabla}{a})^2+\frac12 \dot r^2+\frac12 (\vec \nabla r)^2+V(r)\rangle$ converges to the total energy $\rho_{\rm tot}$ only once all the strings have been destroyed. Prior to this, a significant proportion of the energy is stored in string cores, and this is largest at small values of the log. The difference corresponds to part of the string energy, with the remainder coming from the difference $\langle\frac12(\vec{\nabla}{a})^2-\frac12\dot{a}^2\rangle$ away from the string cores. 
The small mismatch between $\langle \frac12\dot{a}^2+\frac12(\vec{\nabla}{a})^2+\frac12 \dot r^2+\frac12 (\vec \nabla r)^2+V(r)\rangle$ and $\rho_{\rm tot}$ in the absence of strings is due to the non-vanishing axion-radial interaction energy, which however decreases as the Universe expands.

\begin{figure}[t]\begin{center}
\includegraphics[height=8cm]{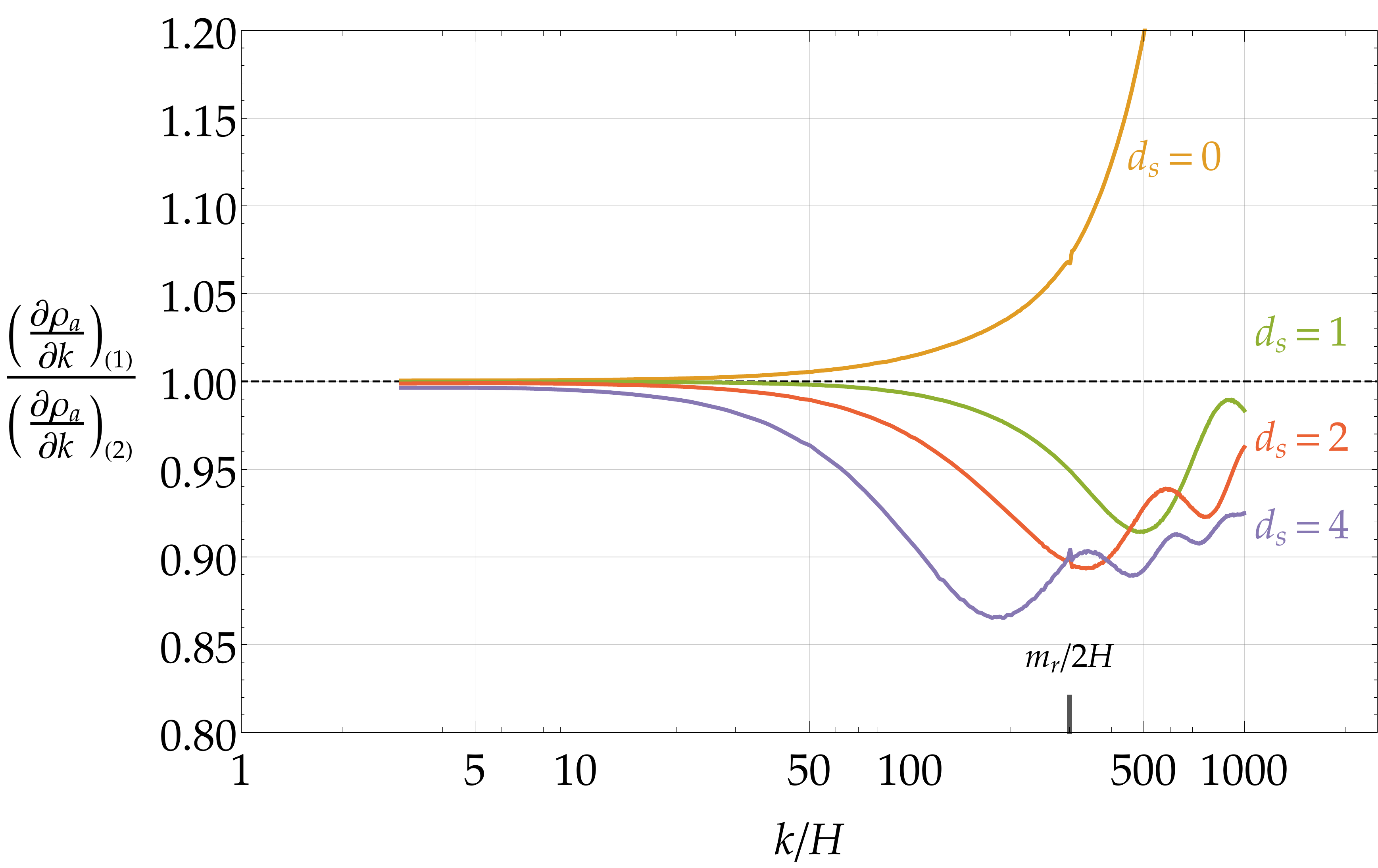}\hspace{0.5cm}\end{center}
\caption{The spectrum calculated with method (1), described in the text, for different screening distances $d_s$, normalised to the spectrum calculated with method (2).}
\label{fig:coresSpectrum}
\end{figure}

We also note that the energy densities  in axions and radial modes have small oscillations with frequency ${\cal O}(m_r)$. 
The amplitude of the oscillations decreases with time due to redshifting, and as a result for a fixed value of the log they are larger in the physical case compared to the fat string scenario. 
Calculating the instantaneous axion emission spectrum involves taking the difference between the axion spectra (appropriately redshifted) at two times, which are separated by more than $m_r^{-1}$. Consequently, the result obtained for momenta around $m_r$ is sensitive to the phase of the oscillations at the two times. This is the origin of the fluctuations of the instantaneous spectrum in Figure~\ref{fig:Fs}, which as expected are more significant in the physical case. To reduce this effect in the physical case we averaged over simulations starting at slightly different times, with relative differences of order $2\pi/m_r$.

\subsection{The Axion Spectrum}

We now describe the details of the calculation of the differential axion spectrum $\partial\rho_a/\partial k$. In the absence of strings, the axion energy density $\rho_a=\langle\dot{a}^2\rangle$ is
\begin{equation}\label{eq:rhoa}
\begin{split}
 \rho_a&=\frac{1}{L^3}\int{d^3x_p \ \dot{a}^2(x_p)} 
       =\frac{1}{L^3}\int{\frac{d^3k}{(2\pi)^3} |\tilde{\dot{a}}(k)|^2 } \ ,
\end{split}
\end{equation}
where $x_p=R(t) x$ are physical coordinates, and $\tilde{\dot{a}}(k)$ is the Fourier transform of $\dot{a}(x_p)$. The axion spectrum $\partial\rho_a/\partial |k|$ is defined by $\int d|k|\ \partial\rho_a/\partial |k|=\rho_a$, and is therefore given by
\begin{equation}\label{eq:drhodkb}
\frac{\partial\rho_a}{\partial |k|}=\frac{|k|^2}{(2\pi L)^3}\int d\Omega_k|\tilde{\dot{a}}(k)|^2 \ .
\end{equation}
In the following, as in the main text, we use the notation $\partial\rho_a/\partial k\equiv\partial\rho_a/\partial |k|$.

In performing the integral $\int d\Omega_k$ in eq.~\eqref{eq:drhodkb}, which converts the three-dimensional spectrum to the one-dimensional spectrum, the three-dimensional momenta $\vec{k}\equiv\frac{2\pi \vec{n}}{L}$ with $|\vec{n}|\leq\frac{N}{2}$ and $|\vec{n}|\in\left]m-\frac{1}{2},m+\frac{1}{2}\right]$, $m\in\mathbb{N}$, are grouped into the same one-dimensional momentum bin labeled $\frac{2\pi m}{L}$. Momenta with $\frac{N}{2}<|\vec{n}|<\sqrt{3}\frac{N}{2}$ have not been included since they lie far above the string core scale and do not matter for either the spectrum or the number density.

In the presence of strings, eqs.~\eqref{eq:rhoa} and \eqref{eq:drhodkb} are no longer valid because of the contribution of the string cores to $\dot{a}(x_p)$ and so to $\tilde{\dot{a}}(k)$. We adopted two methods to mask strings out of the calculation of $\partial\rho_a/\partial k$:
\begin{enumerate}
	\item The first is the Pseudo Power Spectrum Estimator (PPSE) introduced in the analysis of cosmic microwave background data \cite{Hinshaw:2003ex}, and first used in the context of cosmic strings in ref.~\cite{Hiramatsu:2010yu}. This method involves calculating the spectrum $\partial\rho_a/\partial k$ after removing the regions of space that are less than $d_sm_r^{-1}$ from the strings' center. Briefly, it is based on eq.~\eqref{eq:drhodkb} with $\tilde{\dot{a}}(k)$ taken to be the Fourier transform of $\theta(x_p)\dot{a}(x_p)$, where $\theta(x_p)=1$ for $x_p$ a distance of more than $d_sm_r^{-1}$ from the strings' center, and $\theta(x_p)=0$ otherwise. $\partial\rho_a/\partial k$ is then appropriately corrected to account for the bias introduced by the window function $\theta(x_p)$ (more details may be found in \cite{Hinshaw:2003ex}).

	\item The second approach takes advantage of the automatic screening of string cores provided by the factor $(1+r/f_a)$, and consists of using eq~\eqref{eq:drhodkb} with $\tilde{\dot{a}}(k)$ taken to be the Fourier transform of $(1+r(x_p)/f_a)\dot{a}(x_p)$. This is similar to the previous method, except that the string cores do not have to be identified, and the degree of masking varies smoothly as the core is approached. 
\end{enumerate}

We have tested that the computation of the axion spectrum $\partial\rho_a/\partial k$ is independent of the method used, and in particular we studied the dependence of the first method on the screening distance $d_s$. In Figure~\ref{fig:coresSpectrum} we plot the ratio of spectra at $\log=6$ computed using the two methods for different values of the screening parameter $d_s$, including the case $d_s=0$, where the strings are not screened. We see that the difference between the spectrum from method $1$ with $d_s=1$ and the one from method $2$ is $5\%$ at $k=m_r/2$ and rapidly decreases for smaller momenta. This is expected since, given the form of the string profile function for the radial mode, method $2$ is roughly equivalent to  masking distances up to $m_r^{-1}$. Meanwhile, increasing $d_s$ suppresses the spectrum at momenta of order $2\pi/(d_sm_r^{-1})$, but leaves IR momenta unchanged. Again this is not surprising, since masking distances of order $d_sm_r^{-1}$ only loses information about momenta larger than $2\pi/(d_sm_r^{-1})$. The spectrum for $d_s=0$, which is affected by the presence of strings, is much more UV dominated than the others, but at momenta lower than $m_r/2$ still does not differ by more than $10\%$.

In the main text we calculated $\rho_a$ and $\partial\rho_a/\partial k$ using method $1$ with $d_s=1$. The resulting systematic uncertainty from the masking algorithm at $k\sim m_r/4$ is less than $1\%$, while for $k\sim m_r/2$ it can be estimated to be of order $5\div10 \%$. As explained in Appendix~\ref{sec:app_init}, we extract the power law of the spectrum between the IR and the UV peaks, $q$, by considering momenta smaller than $m_r/6$. As a result the systematic uncertainty on $q$ introduced by the masking procedure is negligible. The fact that $\rho_a$ and $\partial\rho_a/\partial k$ are stable under the change of algorithm, and that $\rho_a$ redshifts as radiation, gives us further confidence that, even in presence of strings, our results only include free axion modes, as required.

\section{Convergence from Different Initial Conditions} \label{sec:app_init}

\begin{figure}[t]\begin{center}
\includegraphics[height=6cm]{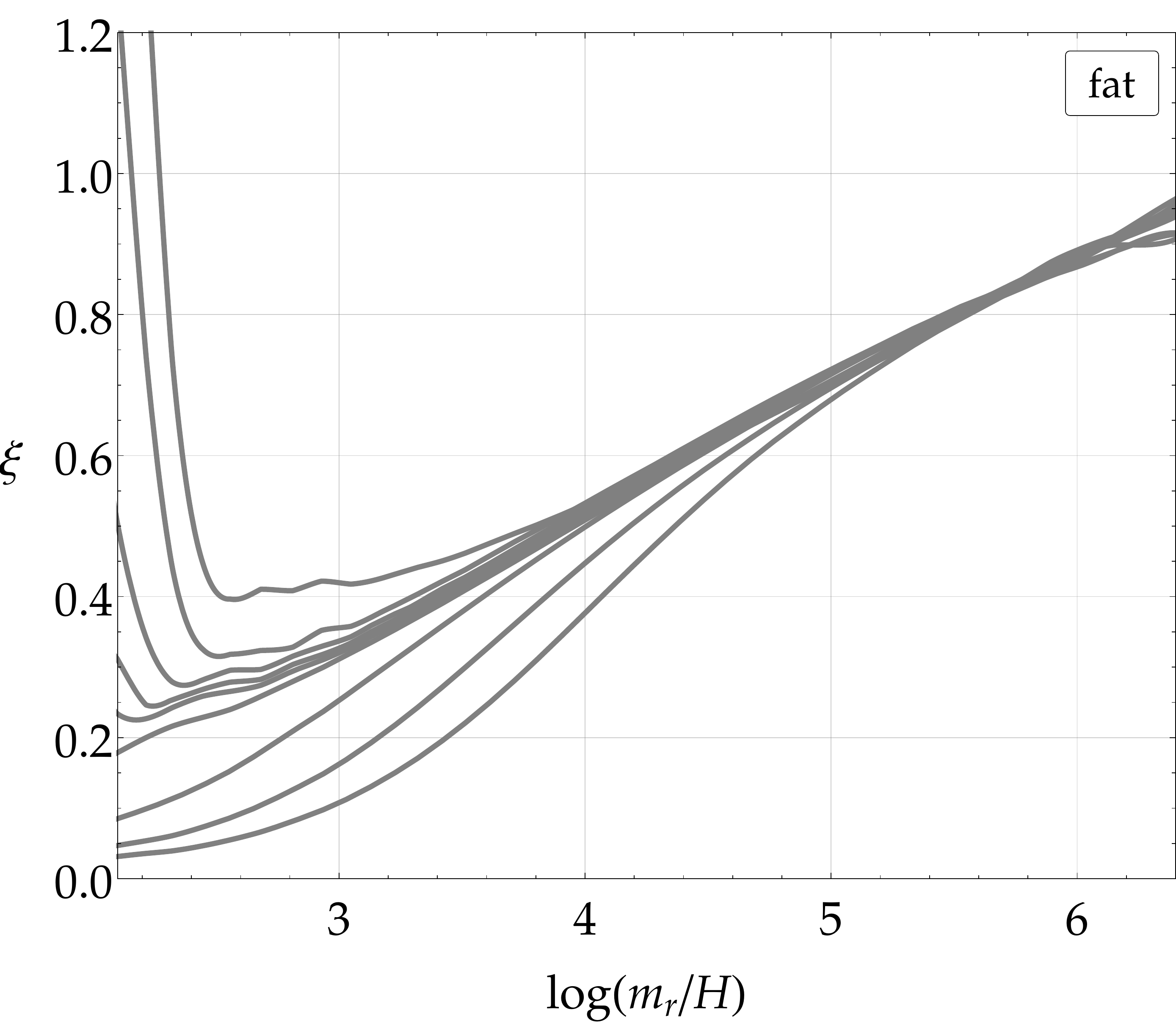}\hspace{1cm}
\includegraphics[height=6cm]{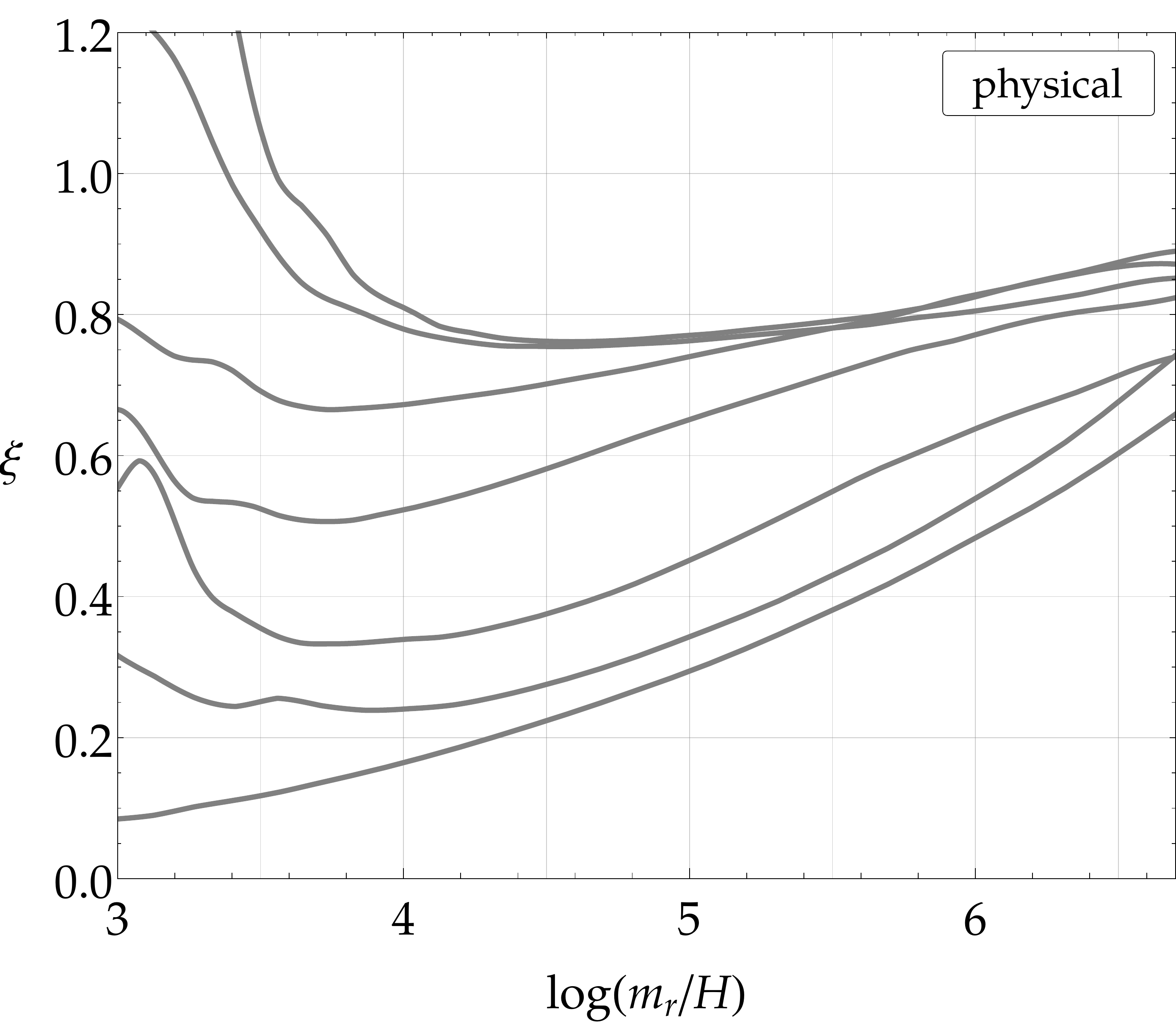} \end{center}
\caption{Convergence of $\xi(t)$ to the attractor solution starting from random initial conditions with different values of $k_{max}/m_r$, between $1/2$ and $1/50$, for the fat string (left) and physical (right) scenarios.}
\label{fig:randomInitial}
\end{figure}

In Section~\ref{sec:attrac} we demonstrated the existence of the attractor solution by considering the convergence of the string density $\xi(t)$, starting from initial conditions with fixed values of $\xi$. Here we present further evidence for the attractor solution. First, rather than fixing initial conditions with a predetermined initial string density by the method (b) described in Section~\ref{sec:attrac} and \ref{sec:app_evo}, we use random initial conditions given by the method (a) varying the maximum coefficients $k_{max}/m_r$. Second, we show how other key properties of the string network, in particular the axion spectrum and axion number density, also converge to the same late time values starting with different string densities.

In Figure~\ref{fig:randomInitial} we plot $\xi(t)$ starting from random field initial conditions for different choices of $k_{max}/m_r$ at the initial time $H_i=m_r$. Although the results obtained are slightly less regular than those starting with fixed numbers of strings, shown in Figure~\ref{fig:xit}, the convergence to the attractor, and the logarithmic increase, is clear.

\begin{figure}[t]\begin{center}
\includegraphics[height=5.5cm]{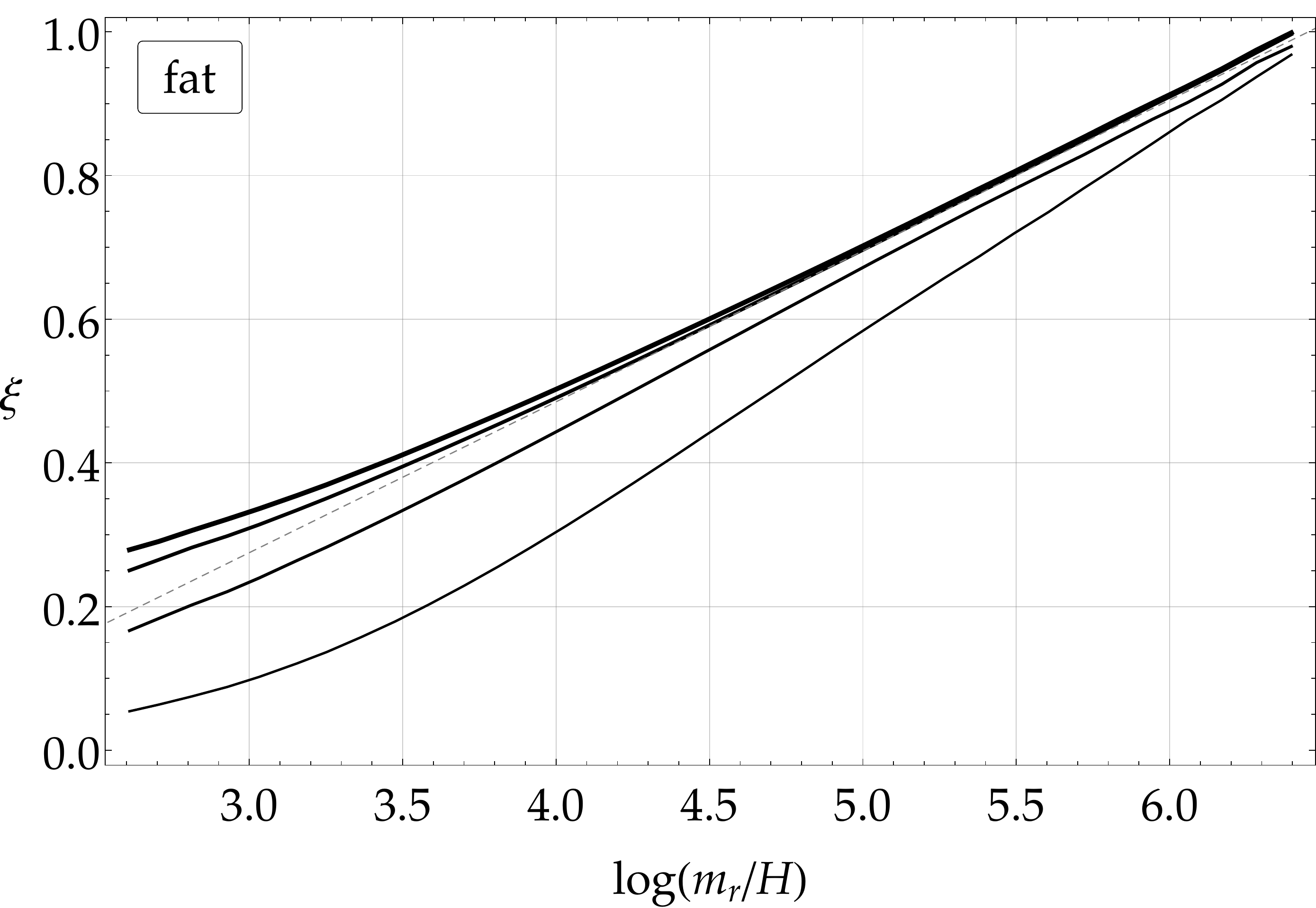}\hspace{0.5cm}
\includegraphics[height=5.5cm]{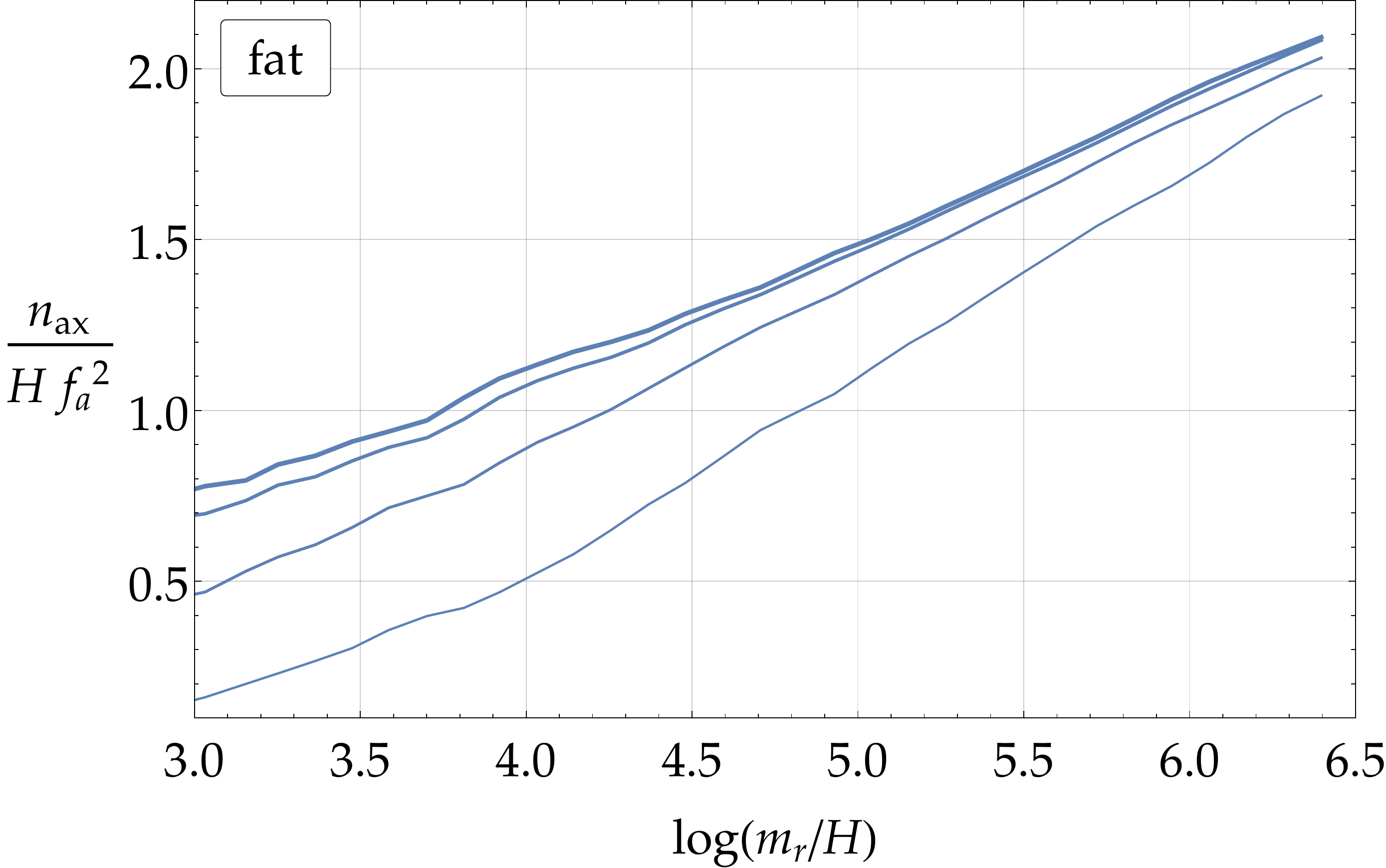} \end{center}
\caption{Convergence to the attractor for $\xi(t)$ (left) and $n_a$ (right) for fat strings. Denser initial conditions are represented by thicker lines, and the results in both cases are obtained from the same sets of initial conditions, with fixed string densities.}
\label{fig:xinaConvergence}
\end{figure}

To analyse the convergence of other properties of the string network, we present results starting from fixed string number densities, which lead to the values of $\xi(t)$ plotted in  Figure~\ref{fig:xinaConvergence} (left). We have also confirmed that the quantities that we study converge to the same attractor solution starting from random initial conditions.
In Figure~\ref{fig:xinaConvergence} (right) we show the results for the axion number density obtained from the same set of simulations as $\xi(t)$ in the left panel. 

Notice that while at $\log=3$ the values for $\xi$ and $n_a$ are spread respectively 
by a factor of 3 and 15, at $\log=6$ the spread reduces to around $10\%$.

\begin{figure}[t]\begin{center}
\includegraphics[height=5.5cm]{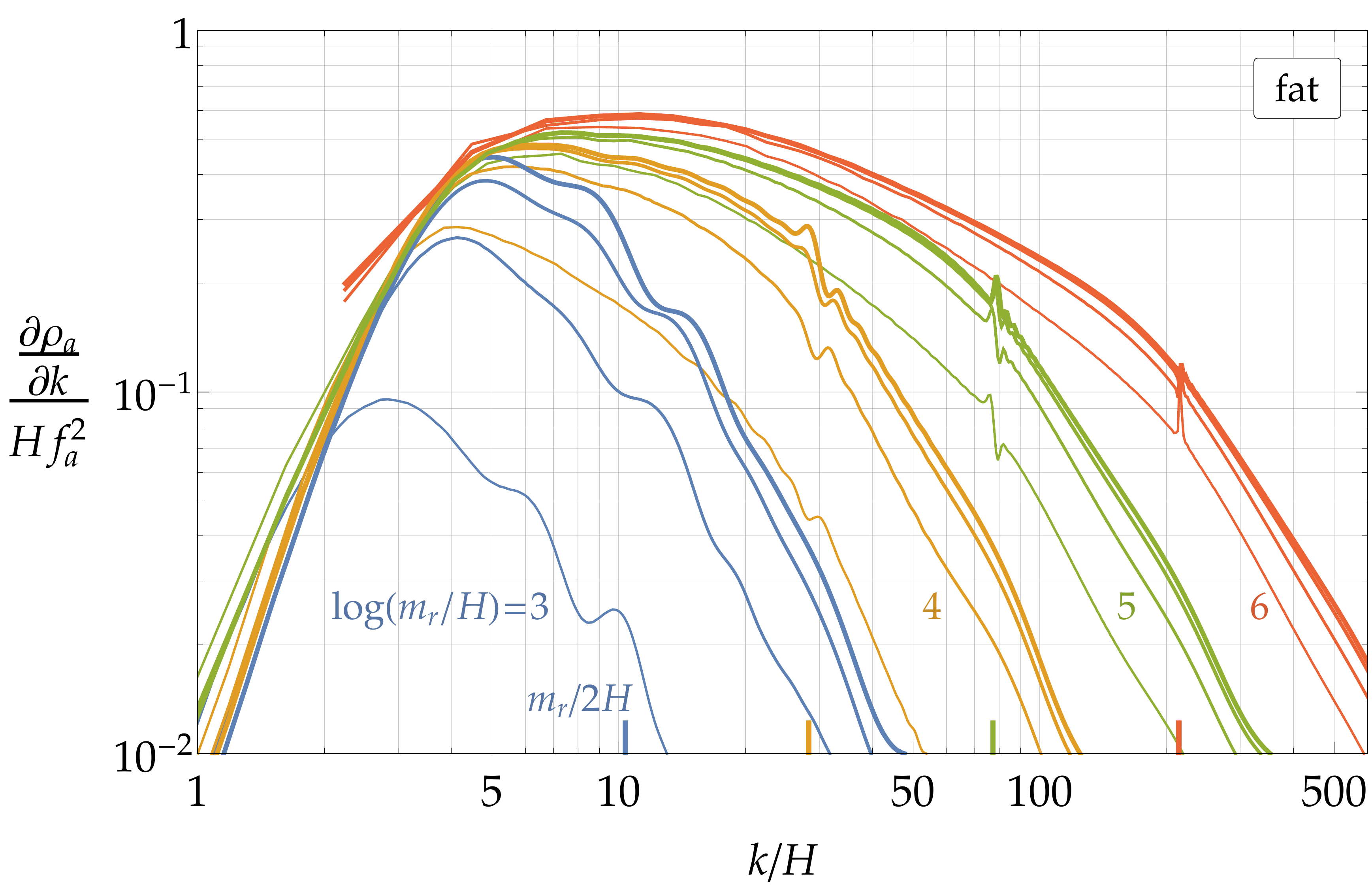}\hspace{0.5cm}
\includegraphics[height=5.5cm]{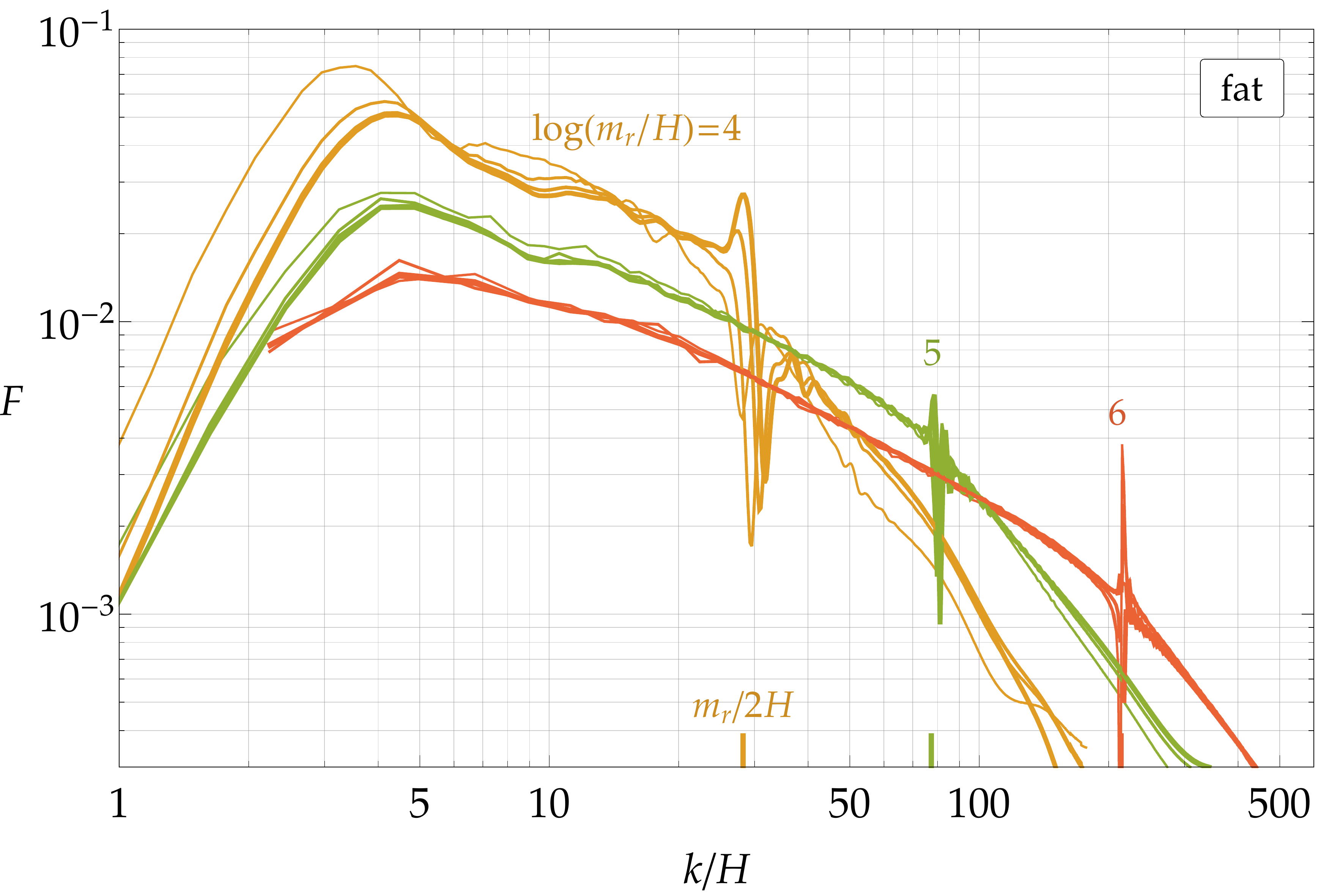} \end{center}
\caption{The convergence of the total axion spectrum (left) and the instantaneous emission spectrum (right), starting from the same set of initial conditions as Figure~\ref{fig:xinaConvergence}, in the fat string system. Simulations with more strings initially are represented by thicker lines, and spectra at the same time instant are represented by the same colour.}
\label{fig:spectrumConvergence}
\end{figure}

The convergence of the axion spectrum $\partial\rho_a/\partial k$ to the attractor, shown in Figure~\ref{fig:spectrumConvergence} (left), is also revealing. At $\log(m_r/H)=3$, the axion spectrum is highly  suppressed in simulations with a lower initial string density than in the others. However, by $\log(m_r/H)=6$ the spectra are very similar. This is especially the case for the modes around Hubble, which have nearly the same amplitude for all initial conditions. More dramatically, the instantaneous emission shown in Figure~\ref{fig:spectrumConvergence} (right) converges extremely fast at late times. This is because, unlike the overall spectrum, it only depends on the properties of the network at a fixed time instant, which are practically the same for all initial conditions by the end of the simulations. We see that underdense networks tend to have more instantaneous emission in the IR at early times, which is expected since at these times the typical distance between strings is larger in this case.

Taken all together, our results provide convincing evidence that the properties of the string scaling solution at late enough times  are sufficiently insensitive to the initial conditions chosen.

\section{Extraction of the Scaling Parameters} \label{sec:app_scaling}

In this appendix we describe how we extract the slope $\alpha$ in the fit of $\xi(t)$, eq.~\eqref{eq:alphafit}, and the power law $q$ of the instantaneous spectrum, and estimate the associated uncertainties. We also describe how we obtain the extrapolations of the number density in Figure~\ref{fig:naextrap}.

Given the convergence shown in Figure~\ref{fig:xit}, initial conditions that result in a constant value of $d\xi/d\log(m_r/H)\equiv d\xi / d\log$ for the longest time correspond to string networks that are the closest to the scaling regime. 
Therefore, we estimate $\alpha$ by calculating $d\xi/d\log$ for initial conditions with different fixed initial string densities, and then restrict to those such that $d\xi/d\log$ is approximately constant at late times.\footnote{Since the derivative $d\xi/d\log$
is more sensitive to local fluctuation we smooth it by convoluting with a Gaussian $g(x)$ with $\sigma=1/4$, i.e. $\int dy \ g(x-y) d\xi(y)/d\log$.} The constant common value that $d\xi / d\log$ reaches in such simulations is a good estimate of~$\alpha$, and the spread indicates the uncertainty. 

\begin{figure}[t]\begin{center}
\includegraphics[height=6.5cm]{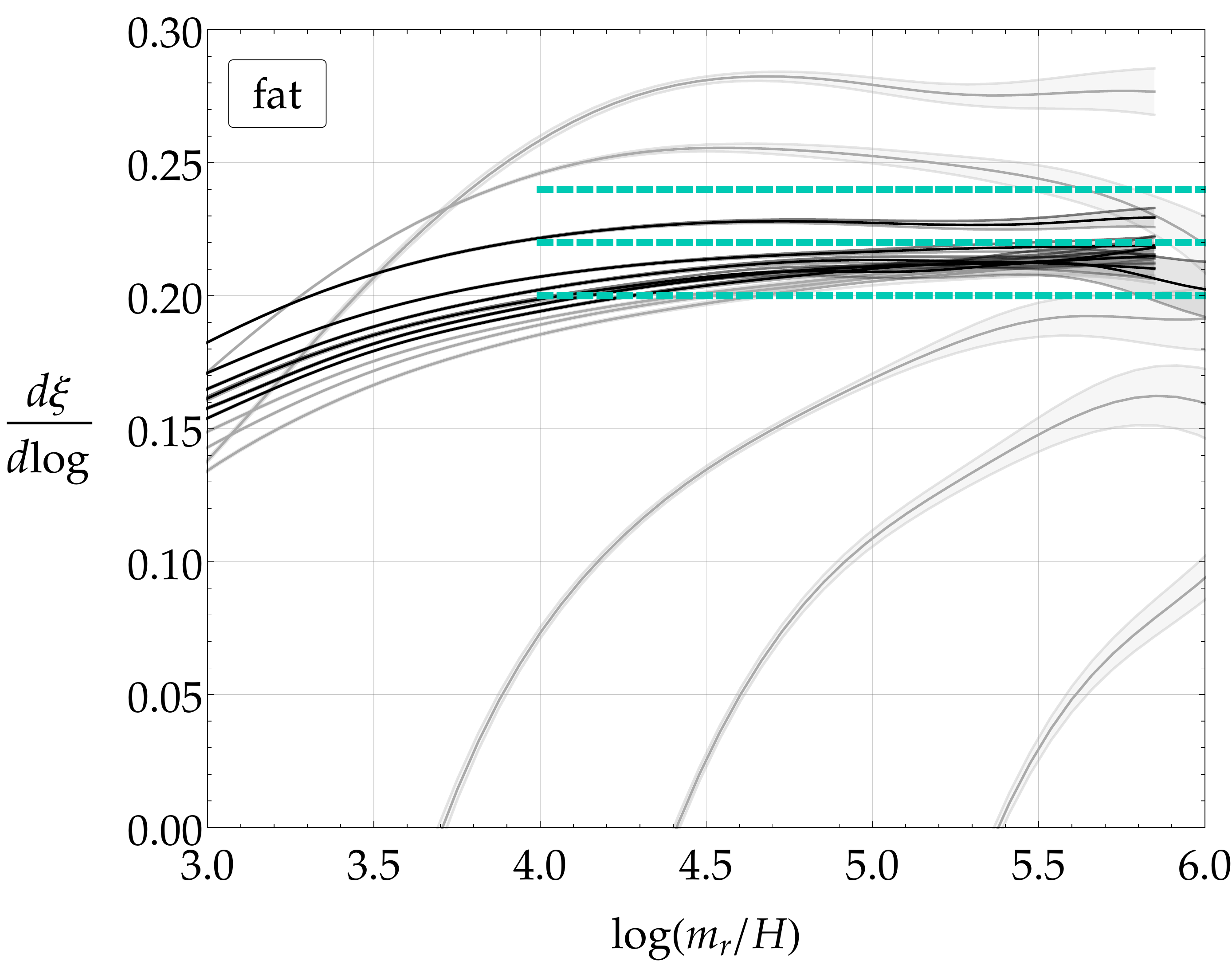}\hspace{0.5cm}
\includegraphics[height=6.5cm]{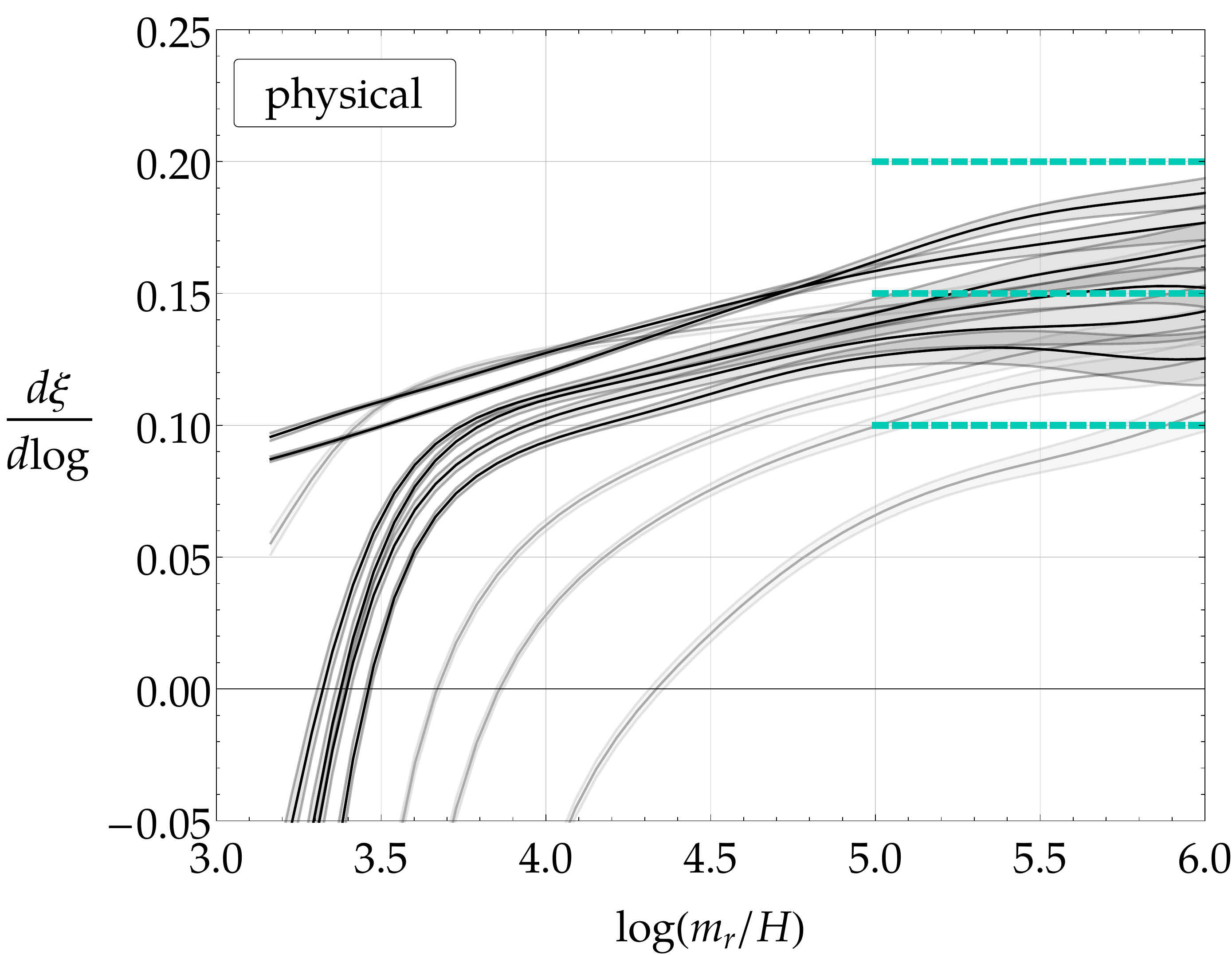} \end{center}
\caption{The derivative $d\xi/d\log(m_r/H)$ for different initial conditions. The black curves are approximately constant for the largest amount of time and therefore the closest to scaling, and their spread is used to estimate the error on $\alpha$. Blue dashed lines represent the estimate of $\alpha$, and are plotted at times when the system is reasonably close to scaling. Error bands represent statistical errors in the average over different samples.}
\label{fig:dxidlog}
\end{figure}

In Figure~\ref{fig:dxidlog} we plot $d\xi/d\log$ for different initial string densities, for both the fat string and the physical case. Slopes of different simulations tend to converge asymptotically at to a common constant value. This is evident in the fat case, for which the constant approached is $d\xi/d\log=0.22\pm0.02$, where the error has been estimated from the spread of the results from simulations that have constant $d\xi/d\log$ for $\log(m_r/H)\gtrsim4$ (plotted in black). In the physical case, the scaling regime is reached at larger values of the log and $d\xi/d\log$ is changing over most of the simulated range. As a result, the uncertainty on $d\xi/d\log=0.15\pm0.05$, estimated as the spread of $d\xi/d\log$ over the simulations for which $d\xi/d\log$ is constant for $\log(m_r/H)\gtrsim5$, is larger.

We now turn to the extraction of the power law of the spectrum $q$. For this purpose, we consider simulations starting from the initial conditions that lead to string networks that are the closest to scaling (corresponding to results for $\xi(t)$ plotted in black in Figure~\ref{fig:xit}).\footnote{As shown in Appendix~\ref{sec:app_init}, the results obtained starting from different initial conditions are very similar, well within the uncertainties on $q$ that we quote.} As shown in Figure~\ref{fig:Fs}, for $\log(m_r/H)\gtrsim5$ the instantaneous emission spectrum, parametrised by $F(k/H,m_r/H)$, has an approximate power law behaviour $1/k^q$ for momenta that are large enough with respect to the peak at around the Hubble scale (at $k/H\sim 5\div10$) and are small compared to the UV cutoff at $k=m_r/2 $.

\begin{figure}[t]
\begin{center}
\includegraphics[height=5.5cm]{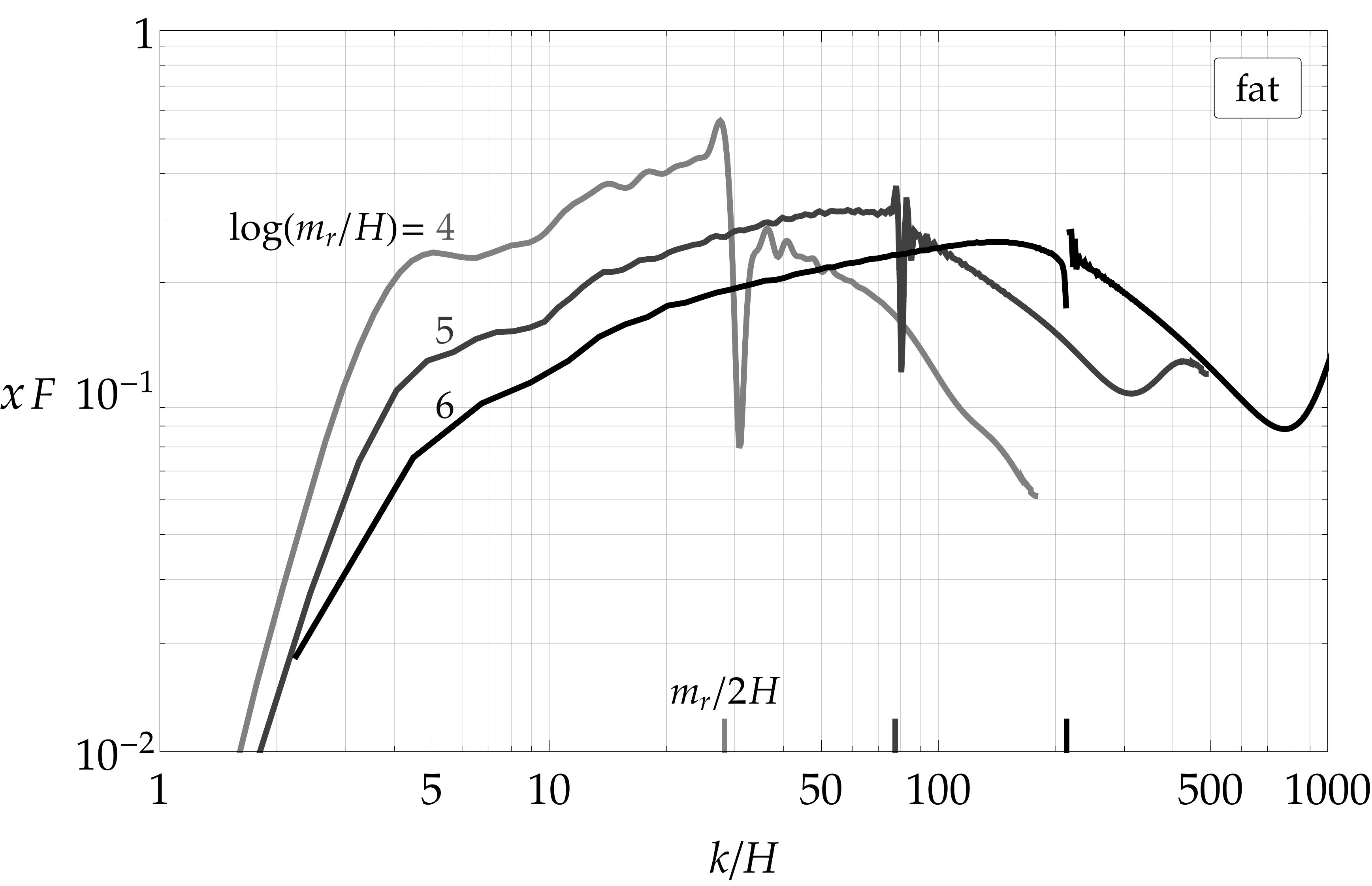}\hspace{0.2cm}
\includegraphics[height=5.5cm]{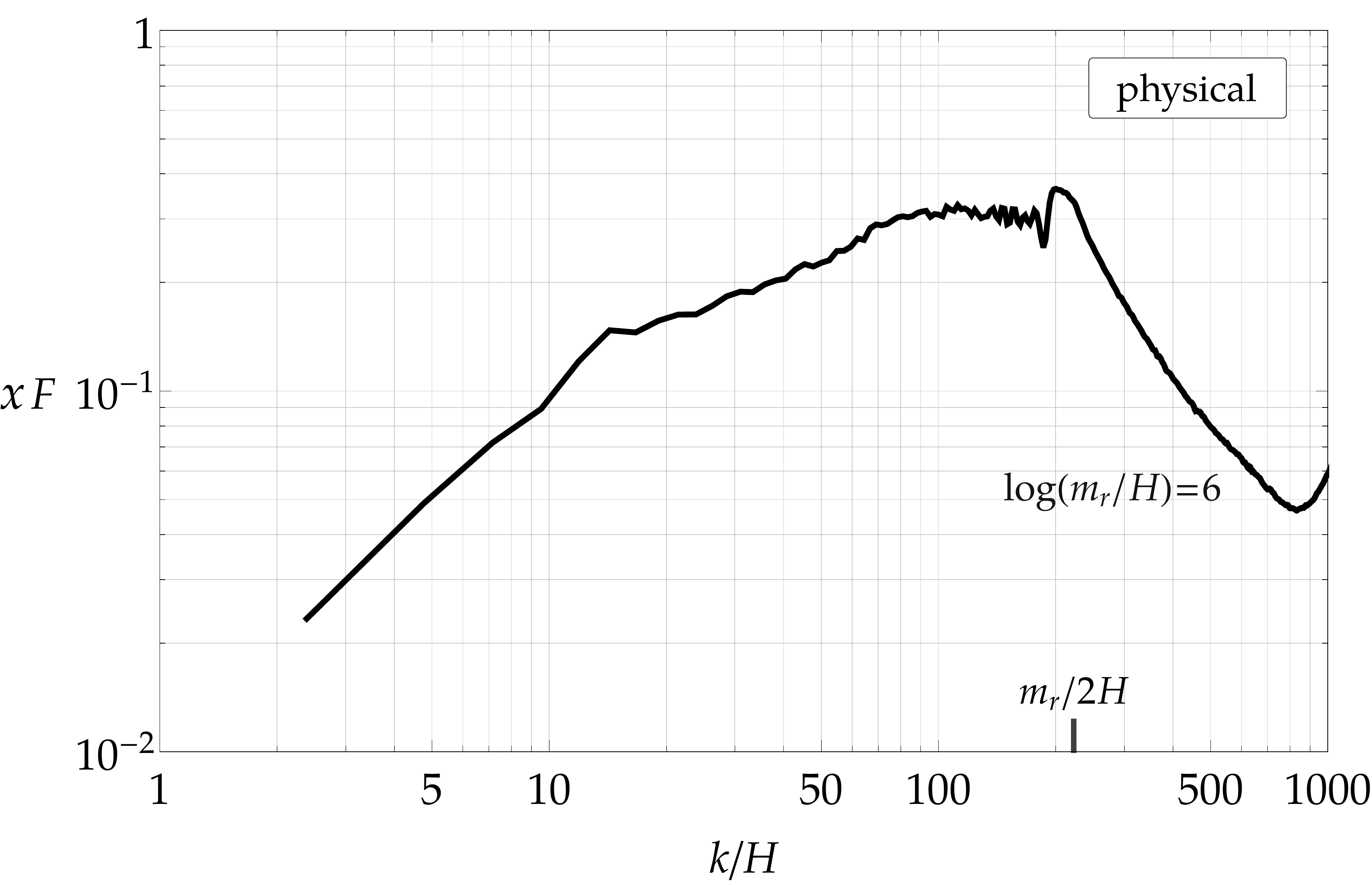} 
\caption{
Plot of $x F(x,y)$ with $x=k/H$ for the fat string (left) and the physical (right) cases, for different values of $y=m_r/H$. The growth of $x F(x,y)$ for $x$ between the IR and UV cutoffs corresponds to a UV dominated spectrum, i.e. $q<1$.
}
\label{fig:xF}
\end{center}

\begin{center}
\includegraphics[height=5.3cm]{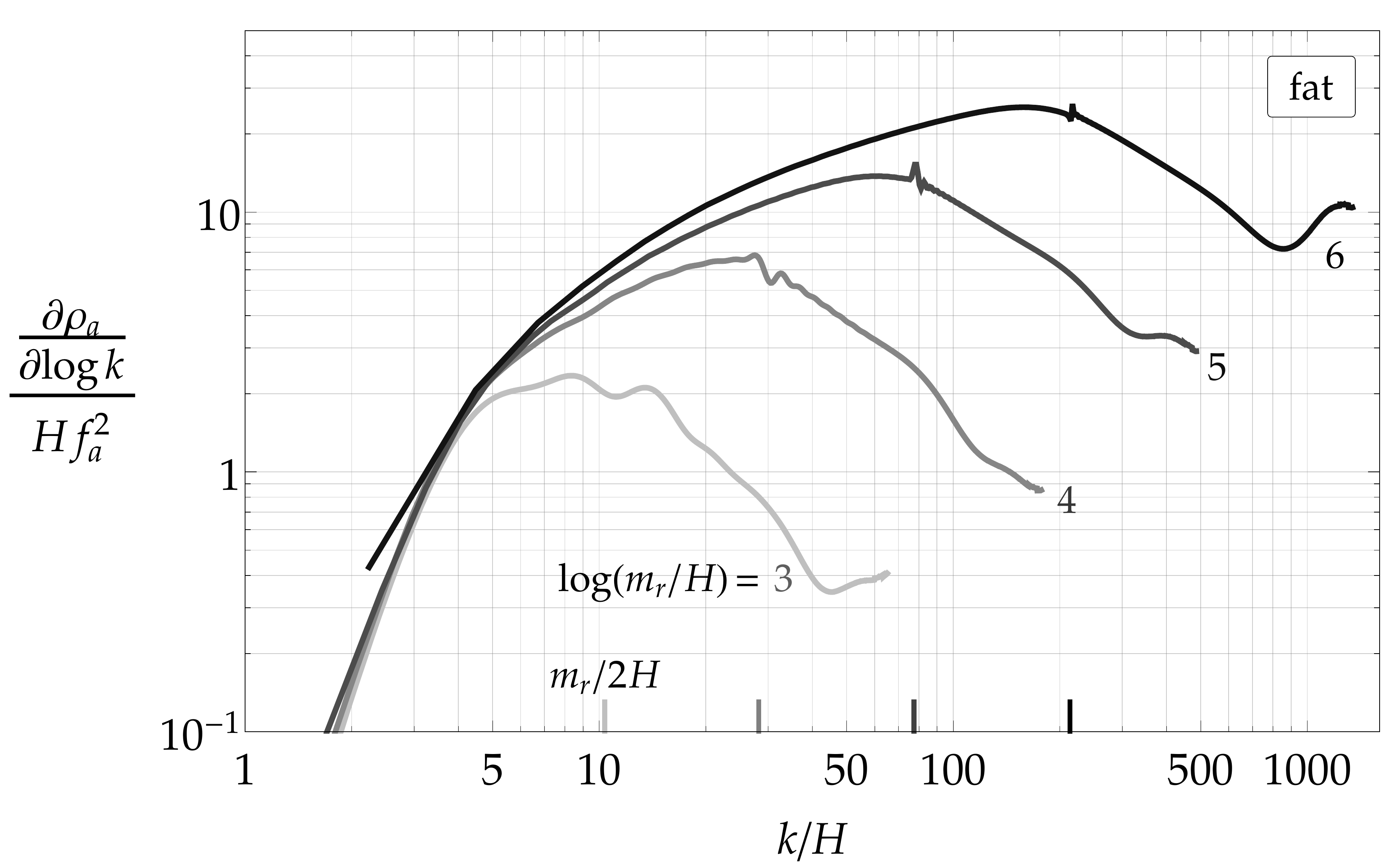}\hspace{0.2cm}
\includegraphics[height=5.3cm]{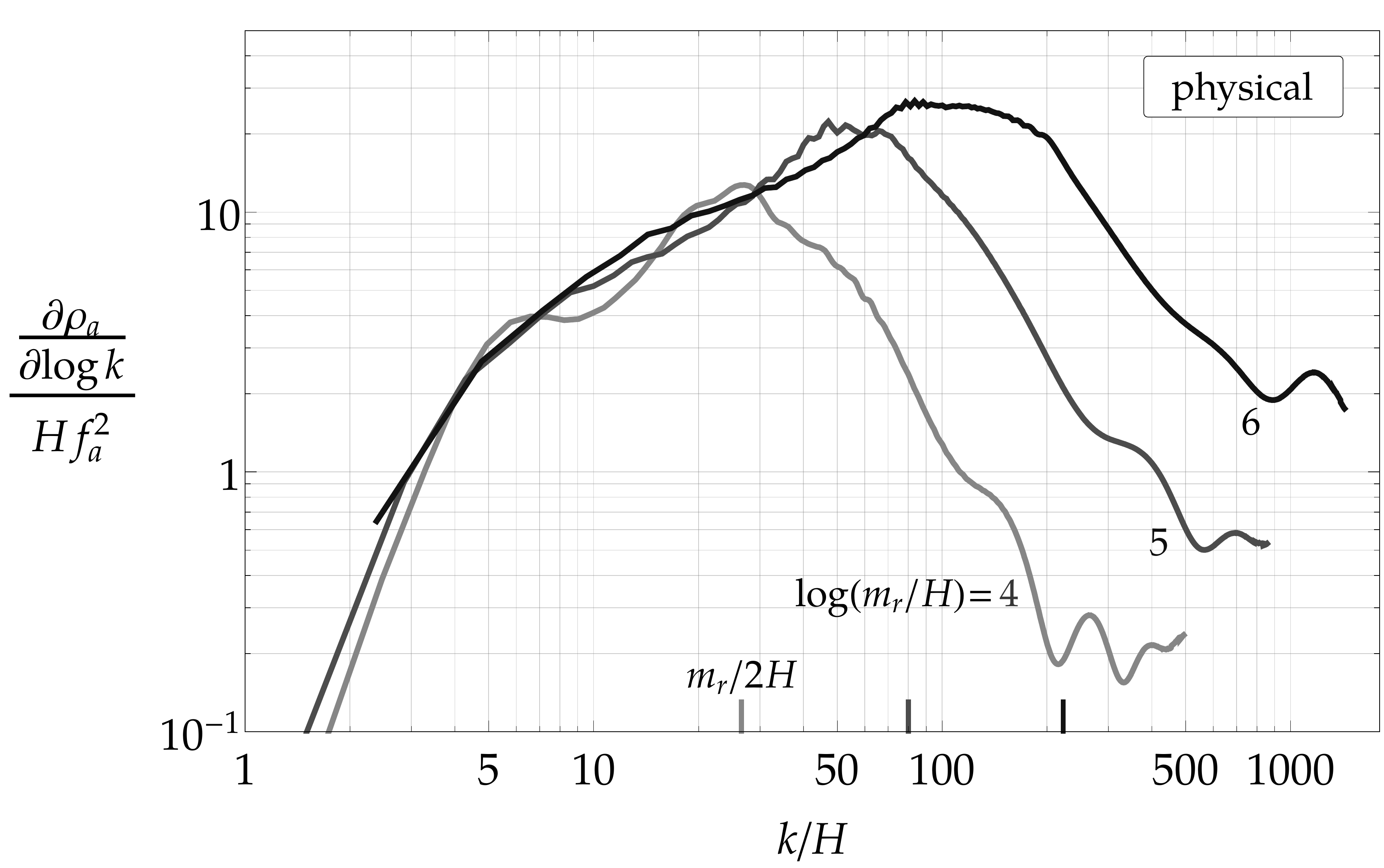}
\caption{
The axion energy density spectrum $\partial \rho_a/\partial \log k$ for the fat string (left) and the physical (right) cases (the data is the same as in Figure~\ref{fig:spectra} but represented on a log scale).
}
\label{fig:spectralog}
 \end{center}
\end{figure}

The fact that $q$ is less than 1 can be more clearly seen from Figure~\ref{fig:xF}, in which we plot $x F(x,y)$. In these plots $q<1$ corresponds to the increase between the IR and UV peaks, which is evident both in the fat string and physical cases. Moreover, most of the area under the curves in Figure~\ref{fig:xF} is at UV momenta $k\sim m_r/2$, which shows that the energy density is dominantly emitted in the form of very high momentum axions, although the axion number density is dominated by low momentum states since $q>0$. In Figure~\ref{fig:spectralog} we plot the total energy spectrum of Figure~\ref{fig:spectra} on a log scale, i.e.  $\partial \rho_a/\partial \log k$. The positive gradients show that the energy density in axions is dominated by UV modes at all times, for both the fat string and the physical case.

\begin{figure}[t]
\begin{center}
\includegraphics[height=10.cm]{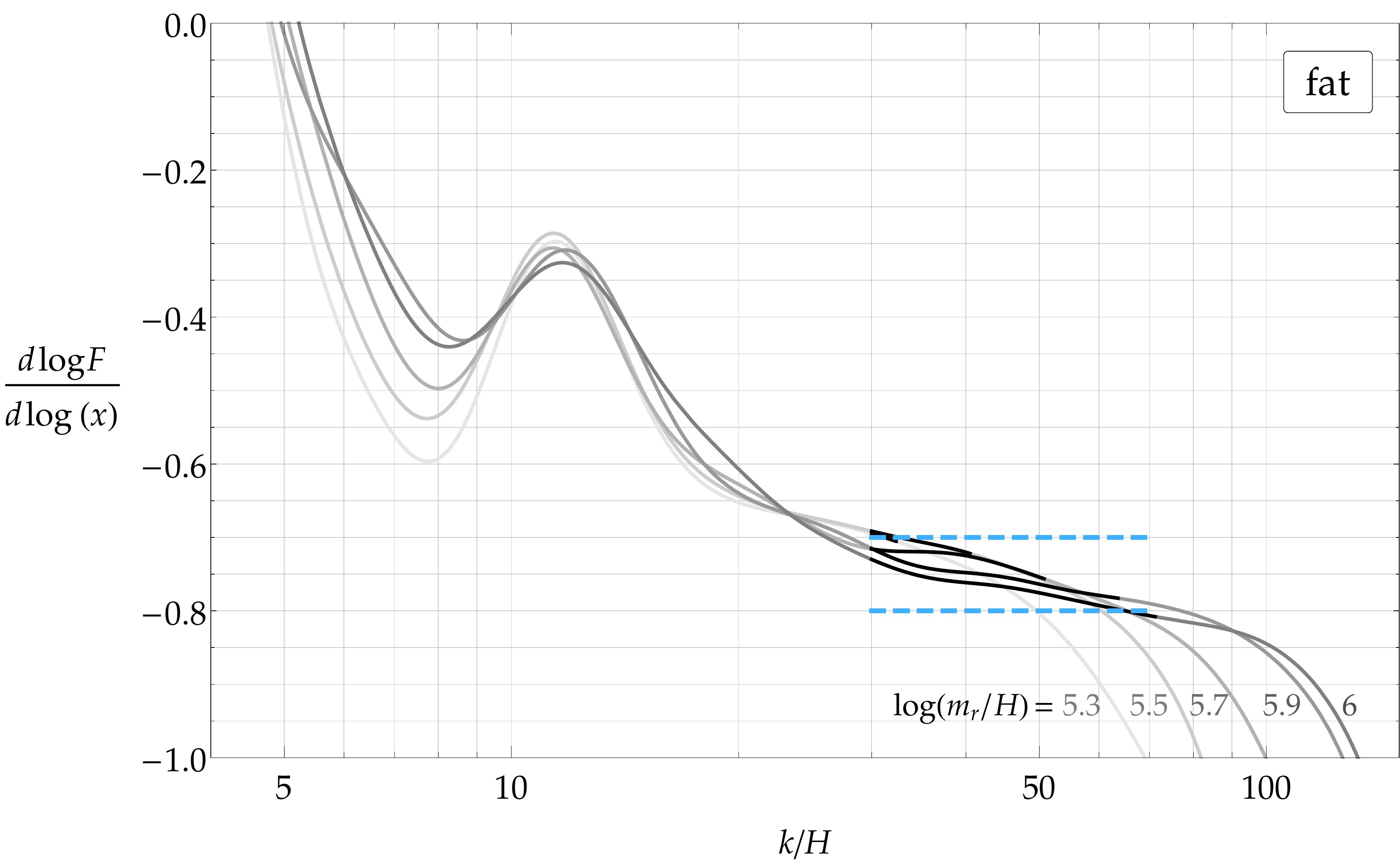}
\caption{
The value of $d\log{F(x,y)}/d\log x$ as a function of the rescaled momentum $x=k/H$, for different time shots, i.e. different values of $y=m_r/H$, for the fat string system in the scaling regime. The black sections corresponds to momenta $30H< k< m_r/6$, which are sufficiently far from the IR and UV peaks to be uncontaminated by effects at these scales. The roughly constant values in these regions correspond to the value of $-q$ in the approximate power law, and the blue dashed lines represent our estimated range for this. 
}
\label{fig:dlogF}
 \end{center}
\end{figure}

Due to the challenges in the physical case, discussed in Section~\ref{sec:instant}, we only attempt a detailed analysis of $q$ in the fat string scenario. To do this, we consider $F$ at late times, $\log(m_r/H)\gtrsim5$, and focus on the region with momenta a factor of $3$ larger than the Hubble peak and a factor of $3$ smaller than the core peak, i.e. $30H\lesssim k\lesssim m_r/6$, which is sufficiently uncontaminated by the cutoffs. 
The power $q$ is then given by $-d\log{F(x,y)}/d\log(x)$.\footnote{Because the derivative $d\log{F(x,y)}/d\log(x)$ is subject to more fluctuations we smooth it by convoluting with a Gaussian $g(x)$ with $\sigma=1/6$, i.e. we consider $\int d\log(z) \ g(\log(x)-\log(z)) \ d\log{F(z,y)}/d\log(z)$.} In Figure~\ref{fig:dlogF} we plot $d\log{F(x,y)}/d\log(x)$ for different values of $y=m_r/H$, corresponding to values of $\log(m_r/H)>5$. The sections of the curves plotted in black indicate momenta in the range $30H< k< m_r/6$, which are safe from contamination from the IR and UV peaks.

$d\log F/d\log(x)$ crosses zero at about $k\sim 5 H$, signalling the position of the Hubble scale peak, and it is smaller than $-1$ at large momenta, meaning that after the UV cutoff the instantaneous spectrum falls off steeply. Moreover, in the intermediate region of interest $- d\log F/d\log(x)$ changes slowly as a function of momentum and across different time shots, ranging from about $0.7$ to $0.8$. We note that in this range $- d\log F/d\log(x)$ (and therefore $q$) shows a tendency to increase at later times. However, given the present uncertainties this apparent change is not significant enough to draw any conclusion, in particular we are not able to assess whether it is due to a residual contamination from the nearby UV peak or it constitutes a genuine feature of the spectrum
(similar analysis with larger grids and more statistics could be beneficial to understand this point).  In the regime $5\lesssim\log(m_r/H)\lesssim6$, we therefore estimate $q=0.75\pm0.05$, where the uncertainly mostly comes from $d\log F/d\log(x)$ not being constant over the momentum range of interest.

In order to extrapolate the axion number density with different power laws, 
we used an analytic form of $F(x,y)$ that closely matches the three approximate power laws visible in Figure~\ref{fig:Fs},
\begin{equation}\label{eq:Fth}
 F(x,y)= N \frac{\left(\frac{x}{x_1}\right)^{q_1} \left[1+\theta (x-x_2) \left(\left(\frac{x_2}{x}\right)^{q_2-q}-1\right)\right]}{\left(\frac{x}{x_1}\right)^{q_1+q}+1} \  \propto \ 
\left \{
\begin{array}{lc}
x^{q_1} & \quad x\ll x_1 \\ & \\
\frac{1}{x^q} & \quad x_1 \ll x\ll x_2  \\ &\\ 
\frac{1}{x^{q_2}} & \quad  x> x_2  \, ,
\end{array}  \right. 
\end{equation}
where $\theta(x)$ is the step function. In this expression $x_1$ and $x_2$ are the positions of the IR and UV cutoffs in units of Hubble, $q_1,q_2$ are the powers that suppress the spectrum in the IR and in the UV respectively, and $q$ is the power between the two cutoffs. The normalisation $N$ is required so that $\int F(x,y) dx =1$. A reasonable fit for the $5$ free parameters is $x_1\sim3, x_2=y/2$ and $q=0.7\div0.8$, $q_1\sim3$, $q_2\sim2$ in the fat string case, and similarly in the physical case 
except that $q_2\sim3$. 
The extrapolations in Figure~\ref{fig:naextrap} are carried out by keeping all parameters
fixed except for $q$. What matters most for the final extrapolation is the parameter $q$, which regulates the hardness of the spectrum, while the remaining parameters only lead to changes of order 1.

\section{Comparison with EFT Estimates} \label{sec:app_eft}

As mentioned in Section~\ref{sec:spectrum}, the dynamics of global string loops can be equivalently described by an effective theory where the fundamental degrees of freedom are the string and the axion radiation~\cite{Dabholkar:1989ju}, with an interaction governed by the Kalb--Ramond action~\cite{Davis:1988rw}. This theory is valid in the regime where the string and the emitted radiation (with frequency $\omega\sim 1/R$) are not strongly coupled, which corresponds to $\log(m_r R)\gg 1$, where $R$ is the loop radius (as shown in~\cite{Dabholkar:1989ju}, the effective coupling of the emitted axion radiation to the string is proportional to $1/\log(m_r R)$). As a result, for loops with a large hierarchy between the radius and the core size, the emitted axion radiation should approximate the one predicted by the Nambu--Goto effective theory. Indeed, using such a theory it was shown~\cite{Dabholkar:1989ju} that a circular loop starting at rest with $\log(m_r R_0)=100$ follows the cosine time-law for the Nambu--Goto strings with percent precision, at least for values of the loop radius such that $\log(m_r R)\gg1$, where the theory is applicable.

\begin{figure}[t]
\begin{center}
\includegraphics[height=7.cm]{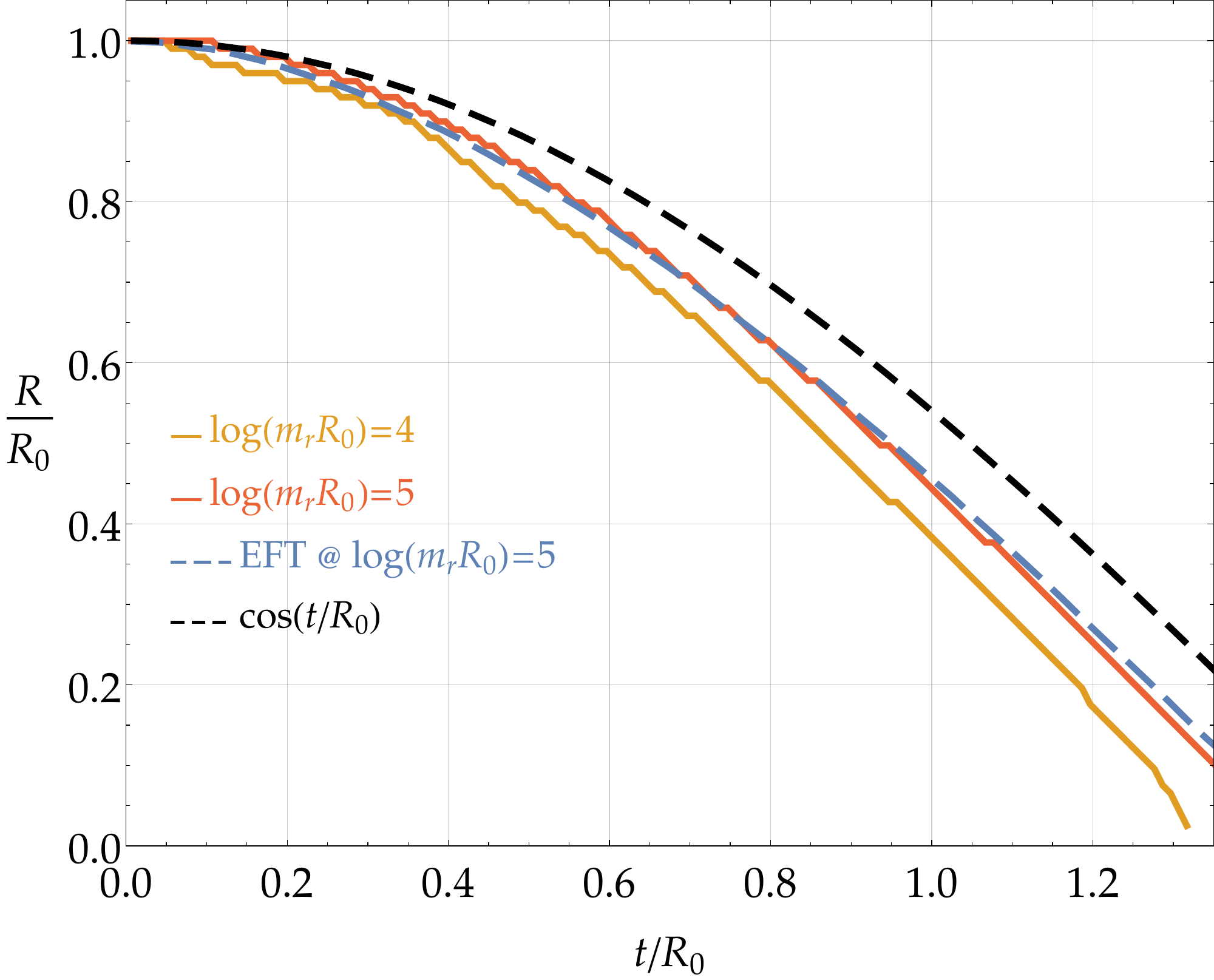} \end{center}
\caption{
Circular loop radius time-law $R(t)$, normalised to the initial radius $R_0$. Green and orange dashed lines correspond to the solution of the field equations for $\log(m_r R_0)=4,5$. The solid blue line is the effective field theory prediction for $\log(m_r R_0)=5$ and the solid black line is the free Nambu--Goto solution $R_{\rm NG}(t)=R_0\cos(t/R_0)$.
}
\label{fig:circular}
\end{figure}

We show now that the evolution of circular loops provided by the solutions of the field equations matches the one predicted by the effective theory of strings, for the values of $\log(m_r R_0)$ reachable in our field theoretic simulations. We solve eq.~\eqref{eq:eom} in Minkowski space, i.e. with a time independent scale factor, $H=0$, and initial conditions $\phi(x)$ and $\dot{\phi}(x)$ that approximately resemble a static circular loop with initial radius $R_0$.\footnote{More precisely, $\dot{\phi}(x)=0$, while $\phi(x)$ is chosen to be cylindrically symmetric around the $z$-axis and in the $y=0$ plane is given by the field generated by the superposition of two point like charges with charge $\pm1$ in the position $(\pm R_0/2,0,0)$. The field generated by a point-like charge $\pm 1$ in the origin is provided by eq.~\eqref{eq:protostring} (with phase $e^{\pm i \theta}$ resp.) and the superposition of fields is defined by their product.} In Fig.~\ref{fig:circular} we plot the time-law for the loop radius $R(t)$ normalised to the initial radius $R_0$ for $\log(m_r R_0)=4$ and  $5$. We also plot the prediction given by the effective theory for $\log(m_r R_0)=5$ and the free Nambu--Goto time law, $R_{\rm NG}(t)=R_0\cos(t/R_0)$. The result of the simulation for $\log(m_r R_0)=5$ matches very well the EFT prediction where this is valid. Moreover, as $\log(m_r R_0)$ increases, the circular loop time law gets closer to the free Nambu--Goto prediction, indeed indicating that at large $\log(m_r R_0)$ global string dynamics converge to that of free Nambu--Goto strings.

Although the convergence $R(t)\to R_{\rm NG}(t)$ is good when the loop radius is sufficiently larger than the core size $m_r^{-1}$, i.e. for $\log(m_r R)\gg 1$, there is a substantial deviation when the string becomes strongly coupled, $\log(m_r R)\lesssim2$,
which prevents the loop from bouncing many times.
This however is not in conflict with the EFT prediction, which correctly  reproduces the
right time law before the loops collapses even for $\log(m_r R_0)$ as small as 5.

Given the limitations in performing direct field theoretic simulations at much larger values of 
$\log(m_r R_0)$ we cannot test whether the EFT expectation that loops will bounce many times (thus emitting an IR dominated spectrum) is correctly reproduced or whether 
core effects when the loops shrink keep inhibiting the rebounce.

\end{document}